\documentclass[11pt,a4paper]{article}
\usepackage{jheppub}
\usepackage[manuscript,readable]{macros}

\preprint{ }

\title{Rank-one 4d \(\mathcal N=3\) SCFTs: Schur index, VOA modules, and modularity}

\author[1]{Zhaoting Guo,}
\author[1]{Satoshi Nawata,}
\author[2]{Yiwen Pan}
\author[1]{and Qituan Zhang}

\affiliation[1]{Department of Physics and Center for Field Theory and Particle Physics, Fudan University, \\
2005 Songhu Road, 200438 Shanghai, China}
\affiliation[2]{Department of Physics, Sun Yat-Sen University, Guangzhou, Guangdong, China}

\emailAdd{24210190011@m.fudan.edu.cn}
\emailAdd{snawata@gmail.com}
\emailAdd{panyw5@mail.sysu.edu.cn}
\emailAdd{24210190037@m.fudan.edu.cn}

\abstract{
We study the representation theory of the vertex operator algebras (VOAs)
associated with rank-one 4d $\mathcal{N} = 3$ superconformal field theories. For the
$\mathbb{Z}_3$ S-fold theory, whose VOA
$\mathcal{W}_{\mathbb{Z}_3}$ has central charge $c_{\mathrm{2d}} = -15$, we use the
$\mathcal{N} = 1$ Lagrangian description to obtain the unflavored Schur index in terms of Dedekind eta functions, while Wilson-loop indices yield
the unflavored non-vacuum characters. These characters all solve a modular linear
differential equation (MLDE) whose solution space also contains a logarithmic
character. Combining flavored MLDEs from null states with Zhu's associative algebra
and a free-field realization, we study four highest-weight modules of
$\mathcal{W}_{\mathbb{Z}_3}$ and their flavored characters in closed form. A
parallel analysis applies to the $\mathcal{N} = 3$ theories obtained by gauging a
discrete $\mathbb{Z}_n$ flavor subgroup of $\mathcal{N} = 4$ $U(1)$ and $SU(2)$
super-Yang--Mills, for which we also obtain closed-form Schur indices and a new
free-field realization of the VOA of the $\mathbb{Z}_4$ quotient.
}

\begin{document}
\maketitle

\section{Introduction and Summary}

Four-dimensional (4d) $\mathcal{N} = 2$ superconformal field theories (SCFTs) have been one of the major subjects of theoretical research, thanks to their rich but relatively tractable strong-coupling dynamics. In particular, genuine $\mathcal{N} = 2$ SCFTs (8 Poincar\'{e} supercharges) and $\mathcal{N} = 4$ (16 Poincar\'{e} supercharges) super-Yang--Mills (SYM) have received the most attention. However, SCFTs with $\mathcal{N} = 3$ supersymmetry occupy an unusual place in the space of 4d SCFTs, even though they in principle have a richer structure than the $\mathcal{N} = 4$ SYMs and are better controlled than their $\mathcal{N} = 2$ cousins \cite{Aharony:2015oyb,Garcia-Etxebarria:2015wns,Aharony:2016kai,Nishinaka:2016hbw}.

$\mathcal{N} = 3$ theories share the relation $a=c$ with the $\mathcal{N} = 4$ SYMs, and genuine ones exist only as isolated, strongly coupled fixed points; an $\mathcal{N} = 3$ theory with a Lagrangian or an exactly marginal coupling always turns out to be an $\mathcal{N} = 4$ SYM \cite{Aharony:2015oyb}. This has a representation-theoretic origin, because the only $\mathcal{N}=3$ multiplet with
spins $s \le 1$ is the vector multiplet, whose CPT completion
coincides with the
$\mathcal{N}=4$ vector multiplet. It was unclear for a long time whether local $\mathcal{N}=3$ theories distinct from $\mathcal{N}=4$ SYM exist at all. Indeed, while supersymmetric field theories in four dimensions have been studied for decades, examples of genuine $\mathcal{N}=3$ SCFTs were constructed only relatively recently
\cite{Aharony:2015oyb,Garcia-Etxebarria:2015wns,Aharony:2016kai,Agarwal:2016rvx,Garcia-Etxebarria:2016erx}.

Conventional field-theoretic methods are of limited use for these theories,
which are constructed in string theory as the worldvolume theories on D3-branes probing an S-fold \cite{Garcia-Etxebarria:2015wns,Aharony:2016kai,Agarwal:2016rvx,Etheredge:2023ler}, a
non-perturbative generalization of an orientifold plane in which a geometric
quotient is combined with a twist by an element of the $SL(2,\mathbb{Z})$
duality group. The twist freezes the holomorphic gauge coupling $\tau$ at a
special algebraic value, so the theory has no marginal coupling, and therefore
no weak-coupling limit or Lagrangian description. The resulting fixed
points are governed by the superconformal algebra $\mathfrak{su}(2,2|3)$ with
$R$-symmetry $SU(3)_R\times U(1)_R$. Their moduli spaces carry a
$3n$-complex-dimensional \emph{triple special K\"{a}hler} geometry \cite{Argyres:2019ngz,Argyres:2019yyb}, which
combines the Coulomb and Higgs branches of an $\mathcal{N}=2$ theory into a
single structure acted on by $SU(3)_R$, and the allowed orbifolds are classified
by the Shephard--Todd theory of complex reflection groups
\cite{Argyres:2019ngz,Argyres:2019yyb,Garcia-Etxebarria:2016erx,Kaidi:2022lyo,Chevalley:1955ub,Shephard:1954oo}.

Besides the moduli space and the central charges, there are other protected quantities that are particularly interesting to study for strongly coupled SCFTs. The first is the superconformal index, and the second is the associated chiral algebra, or vertex operator algebra (VOA). The superconformal index counts BPS operators in the spectrum preserved by some chosen supercharge(s) \cite{Gadde:2011uv}. Under exactly marginal deformations and renormalization group (RG) flows, the spectrum changes continuously while the index remains invariant. This feature makes the superconformal index a useful probe of strongly coupled SCFTs, in combination with $\mathcal{N} = 2$ dualities and RG flows.

Each 4d $\mathcal{N} = 2$ SCFT is associated with a non-unitary VOA that encodes protected information about the 4d theory \cite{Beem:2013sza}. This SCFT/VOA correspondence has been studied extensively for broad classes of theories, including class-$\mathcal{S}$ and Argyres--Douglas theories \cite{Beem:2014rza,Wang:2015mra,Lemos:2014lua,Creutzig:2017qyf,Song:2017oew,Wang:2018gvb,Creutzig:2018lbc,Arakawa:2018egx,Xie:2019zlb,Kiyoshige:2020uqz,Buican:2020moo}. Some of the associated VOAs admit convenient free-field realizations \cite{Beem:2019tfp,Bonetti:2018fqz,Beem:2024fom}. The Schur index is a special limit of the $\mathcal{N} = 2$ superconformal index and can be identified with the vacuum character of the associated VOA \cite{Gadde:2011uv,Beem:2013sza}.\footnote{The vacuum supercharacter is related to the physical Schur index by
$\operatorname{ch}_{\textrm{vac}}=q^{-c_{\mathrm{2d}}/24}\mathcal I_{\mathrm{phys}}$.
Throughout this paper, ``Schur index'' denotes the index corrected by this
factor, so $\mathcal I=\operatorname{ch}_{\textrm{vac}}$.
Full superconformal indices and the formal Wilson-loop integrals retain their
unshifted trace normalization.} Non-vacuum modules are expected to be realized by BPS line or surface defects in four dimensions \cite{Cordova:2015nma,Cordova:2016uwk,Cordova:2017mhb,Nishinaka:2018zwq,Bianchi:2019sxz,Zheng:2022zkm}. Moreover, the associated variety of the VOA is conjecturally identified with the Higgs branch of the 4d SCFT; this relation is closely tied to the quasi-lisse property and the modular linear differential equations (MLDEs) obeyed by ordinary module characters \cite{Gaberdiel:2008pr,Arakawa:2016hkg,Beem:2017ooy}.

The invariance of the superconformal index under RG flow makes it possible to compute the index for theories without Lagrangian descriptions. In particular, $\mathcal{N} = 1$ ultraviolet (UV) Lagrangians for $\mathcal{N} = 2$ infrared (IR) SCFTs have been studied in detail \cite{Gadde:2015xta,Agarwal:2016pjo,Maruyoshi:2016aim,Maruyoshi:2016tqk,Agarwal:2017roi,Benvenuti:2017bpg,Maruyoshi:2018nod,Agarwal:2018ejn}. Most importantly for our study, some $\mathcal{N}=3$ SCFTs arise as
IR fixed points of $\mathcal{N}=1$ Lagrangian gauge theories
\cite{Zafrir:2020epd}, which makes their index accessible by
conventional localization techniques. 

We are interested in the VOAs associated with $\mathcal{N} = 3$ SCFTs and in their representation theory. The enhanced supersymmetry forces the associated VOA to contain an $\mathcal{N}=2$ super-Virasoro subalgebra; the simplest rank-one algebra and protected subsectors of more general $\mathcal{N}=3$ theories have been studied by bootstrap methods \cite{Nishinaka:2016hbw,Lemos:2016xke}, while free-field realizations and superconformal indices provide complementary descriptions \cite{Bonetti:2018fqz,Bourton:2018jwb,Agarwal:2021oyl}. As these VOAs lie outside the better-understood class of $\mathcal{W}$-algebras obtained by Drinfeld--Sokolov reduction of Kac--Moody algebras, their representation theory lacks a systematic mathematical treatment. To gain access to the character space of the associated VOA, we make use of the analytic computation approach discussed in \cite{Pan:2021mrw,Guo:2023mkn,Pan:2025vyu}. Although it has not been proven in full generality, by computing the contour integral (with arbitrary polynomial insertions), one can often extract (some or all of) the characters of the associated VOA \cite{Li:2025nhc}. A second tool is provided by the (flavored) MLDEs, which strongly constrain the characters of a quasi-lisse VOA \cite{PeelaersUnpublished,Beem:2017ooy,Pan:2021ulr,Zheng:2022zkm,Pan:2023jjw,Pan:2024dod}. These equations usually arise from null states in the algebra by applying Zhu's recursion formula and also from the flat connection on the space of conformal blocks \cite{Arakawa:2026nvd}.

\paragraph{Summary of results.}
We focus on the simplest rank-one $\mathcal{N}=3$ theories whose moduli spaces are
$\mathbb{C}^3/\mathbb{Z}_n$. One of them belongs to the class of S-fold theories \cite{Garcia-Etxebarria:2015wns,Aharony:2016kai}, and a few more are constructed by discretely gauging the $\mathcal{N} = 4$ $U(1)$ and $SU(2)$ theories \cite{Bourton:2018jwb}.
\begin{itemize}
  \item \textbf{Closed-form index.} 
  Using the UV superconformal index in its contour integral form \cite{Zafrir:2020epd}, we evaluate the unflavored Schur index of the $\mathbb{Z}_3$ S-fold theory (the vacuum character of the associated VOA $\mathcal{W}_{\mathbb{Z}_3}$) in closed form in terms of Dedekind eta functions. Furthermore, we find that
  inserting formal Wilson-loop characters into the $\mathcal{N}=1$ integral leads directly to the closed-form unflavored non-vacuum characters of $\mathcal{W}_{\mathbb{Z}_3}$. We do not, however, have a direct $\mathcal{N} = 1$ or $\mathcal{N} = 3$ interpretation of the resulting integral. In addition, using flavored MLDEs and the modular structure, we are able to identify all the flavored characters in closed form, which are independently reproduced by a state count in a reduced Poincar\'{e}--Birkhoff--Witt (PBW) basis. See the summary in Table~\ref{tab:WZ3-modules}.

\begin{table}[h]
\centering
\begin{tabular}{c c c c c c c}
\toprule
module & character & $\alpha$ & $h$ & $m$ & unflavored & flavored \\
\midrule
 $M_{\textrm{vac}}$ & $\operatorname{ch}_{\textrm{vac}}$        & $\frac{5}{8}$    & $0$            & $0$  & \eqref{eq:SchurClosedForm}      & \eqref{eq:flavored-Schur-index-WZ3} \\
$M_0$& $\operatorname{ch}_{-\frac{1}{24}}$       & $-\frac{1}{24}$  & $-\frac{2}{3}$ & $0$  & \eqref{eq:nonvac-characters-1} & \eqref{eq:flavored-non-vacuum-characters-WZ3-1} \\
$M_+$ &$\operatorname{ch}_{\frac{1}{8}}^{+}$     & $\frac{1}{8}$    & $-\frac{1}{2}$ & $+1$ & \eqref{eq:nonvac-characters-2} & \eqref{eq:flavored-non-vacuum-characters-WZ3-2} \\
$M_-$ &$\operatorname{ch}_{\frac{1}{8}}^{-}$     & $\frac{1}{8}$    & $-\frac{1}{2}$ & $-1$ & \eqref{eq:nonvac-characters-2} & \eqref{eq:flavored-non-vacuum-characters-WZ3-3} \\
\bottomrule
\end{tabular}
\caption{The four highest-weight modules studied here for $\mathcal{W}_{\mathbb{Z}_3}$, with the
leading exponent $\alpha=h-c_{\mathrm{2d}}/24$ of the character, the conformal
weight $h$ and the $J_0$-charge $m$ of the highest-weight state, and the
equations in which the unflavored and flavored characters are recorded. The
solution space of the MLDE \eqref{eq:MLDE-order4} contains one further element,
the logarithmic character $\operatorname{ch}_{\log}$ of \eqref{eq:chlog-def},
which completes the $\Gamma^0(2)$ orbit but is not the character of an ordinary
module.}
\label{tab:WZ3-modules}
\end{table}

  Although the Schur index of the $\mathcal{N} = 4$ $SU(2)$ theory can be computed by a contour integral, the $\mathbb{Z}_n$ quotient does not commute with the $Q_\text{BRST}$ reduction, making the quotient index less direct to compute. Using the branching rule of \cite{Creutzig:2018ltv}, we nonetheless obtain a closed form \eqref{eq:flavored-Schur-index-Zn-quotient} for all $n = 2,3,4,6$. We also observe that the $\mathbb{Z}_3$ quotient of the $\mathcal{N} = 4$ $U(1)$ theory and the $\mathbb{Z}_3$ S-fold theory share the same unflavored character space but define different VOAs. 
  
  We further find that the leading small-circle divergence of the Schur index matches the regulated residue of the corresponding ABJM $S^3$ partition function.
  
  \item \textbf{Modular structure.} The unflavored vacuum/non-vacuum characters satisfy
  a fourth-order $\Gamma^0(2)$-MLDE. By adding the logarithmic partner of the vacuum character, we construct the space of solutions, which is invariant under the modular group $\Gamma^0(2)$. The non-logarithmic unflavored characters span a three-dimensional subspace, since the charged pair coincides at $a=1$; the logarithmic solution completes it to the four-dimensional MLDE solution space.

  \item \textbf{Null states and flavored MLDEs.} The algebra $\mathcal{W}_{\mathbb{Z}_3}$ is known from the bootstrap approach \cite{Nishinaka:2016hbw,Lemos:2016xke}. Using the free-field realization from \cite{Bonetti:2018fqz}, we study the null states in more detail and construct their corresponding flavored MLDEs. These equations and their null states form a universal quasi-modular structure and provide a set of consistency checks for the flavored character analysis. For the $\mathbb{Z}_n$-quotient theories, we explicitly construct the VOAs by taking quotients of the $\mathcal{N} = 4$ VOAs. In particular, we find a new free-field realization for the VOA obtained from the $\mathbb{Z}_4$-quotient of the small $\mathcal{N} = 4$ superconformal algebra.

  \item \textbf{Highest-weight modules and characters.} Combining the flavored MLDEs with
  Zhu's associative algebra and the free-field realization, we study
  four highest-weight modules of $\mathcal{W}_{\mathbb{Z}_3}$ and obtain
  their flavored characters in closed form, together with the logarithmic
  partner of the vacuum character.

\end{itemize}

\paragraph{Organization.}

This paper is organized as follows. In section~\ref{section:2}, we review basic facts of 4d $\mathcal{N} = 3$ SCFTs, including the S-fold and discrete gauging construction. In section~\ref{section:3}, we focus on the $\mathbb{Z}_3$ S-fold theory, the simplest rank-one S-fold, whose moduli space is $\mathbb{C}^3/\mathbb{Z}_3$. We start by computing its unflavored Schur index in closed form, which is the starting point for our analysis of the representation theory of the associated VOA $\mathcal{W}_{\mathbb{Z}_3}$, from MLDEs, modules to free-field representations and quasi-modularity. Along the way, we observe that the flavored characters take the shape of a Verma-module character and conjecture a reduced PBW basis built from five of the eight strong generators whose state count reproduces them. In section~\ref{sec:discrete-gauging}, we turn to rank-one $\mathcal{N} = 3$ theories obtained from discrete gauging of $\mathcal{N} = 4$ theories, where we also study the closed-form Schur index and the corresponding VOA representation theory. Section~\ref{sec:S1-reduction} then compares the small-circle residues of both the S-fold and the discretely gauged $SU(2)/\mathbb{Z}_n$ theories with rank-one ABJM partition functions. In the appendix, we collect the definitions and useful identities of various special functions and some detailed computation results.

\section{Generalities of \texorpdfstring{$\mathcal{N}=3$}{N=3} SCFTs and associated VOAs}\label{section:2}

In this section, we review some general features of 4d $\mathcal{N} = 3$ SCFTs
and of their associated VOAs, mainly to fix notation and to collect results
used later.

The $\mathcal{N}=3$ superconformal algebra $\mathfrak{su}(2,2|3)$ has three
sets of Poincar\'e and conformal supercharges,
\begin{equation}
  Q^I_\alpha,\quad \widetilde Q_{I\dot\alpha},\quad
  S_I^\alpha,\quad \widetilde S^{I\dot\alpha},
  \qquad I=1,2,3,
\end{equation}
along with the conformal generators and the $U(3)_\mathcal{R}$ generators
$R^I{_J}$.  An $\mathcal{N}=2$ subalgebra can be embedded in several
inequivalent ways; we always take the supercharges with $I=1,2$, for which the
$R$-symmetry Cartan generators are
\begin{equation}
  r=R^1{_1}+R^2{_2},\qquad
  R=\frac12\bigl(R^1{_1}-R^2{_2}\bigr).
\end{equation}
The commutant of $SU(2)_\mathcal{R}\times U(1)_r$ in $U(3)_\mathcal{R}$ is a
$U(1)_f$ flavor symmetry with generator
\begin{equation}
  f=2R^3{_3}+r.
\end{equation}
This embedding is used throughout and lets us treat the moduli space, the
superconformal index, and the associated VOA in $\mathcal{N}=2$ language.

Viewed as an $\mathcal{N}=2$ theory, the additional supersymmetry constrains
the Higgs and Coulomb branch operators.  Higgs branch operators are
$\mathcal{N}=3$ superconformal primaries with
\begin{equation}
  E=2R,\qquad j_1=j_2=r=0,
\end{equation}
while Coulomb branch operators have
\begin{equation}
  E=-r,\qquad R=j_1=j_2=0.
\end{equation}
The relation between the two branches is clearest at rank one.  There, the
Coulomb branch is a one-dimensional cone generated by $u$, and
$U(3)_\mathcal{R}$ forces the Higgs branch to be a hyperK\"ahler cone of
quaternionic dimension one, hence a quotient of $\mathbb{C}^2$ by a finite subgroup of
$SU(2)$.  Requiring $U(1)_f$ to act isometrically leaves only the cyclic case
$\mathbb{C}^2/\mathbb{Z}_\ell$, with coordinate ring
\begin{equation}
  \mathbb{C}[j,w^\pm]/\langle w^+w^- - j^\ell\rangle.
\end{equation}
Here, $j$ and $w^\pm$ correspond to the chiral operators $J$ and $W^\pm$, where
$J$ is the moment map operator for $U(1)_f$ and $\Delta(J)=2=2R(J)$.  Since
$U(3)_\mathcal{R}$ rotates $u,W^+,W^-$ into one another, they all have the same
scaling dimension,
\begin{equation}
  \Delta(u)=\Delta(W^\pm)=\ell\in\mathbb{Z}.
\end{equation}
The Seiberg--Witten geometry is a scale-invariant elliptic fibration over the
Coulomb branch, or equivalently one of Kodaira's $I_0,I_0^*,IV^*,III^*,II^*$
singularities, so integrality of $\Delta(u)$ leaves
\begin{equation}
  \ell=\Delta(u)=1,2,3,4,6.
\end{equation}

For a rank-$n$ theory, the moduli space $\mathcal{M}$ is parametrized by the
vacuum expectation value of the $\mathcal{N}=3$ vector multiplet scalar.  Its
geometry is less constrained than that of the essentially rigid
$\mathcal{N}=4$ moduli space, but much more constrained than in the
$\mathcal{N}=2$ case, which remains largely unexplored.  It has complex
dimension $3n$ and carries a triple special K\"ahler structure, with a
$\mathbb{C}P^2$ worth of metric-compatible complex structures
\cite{Argyres:2019ngz,Argyres:2019yyb}.  Under mild assumptions, $\mathcal{M}$
is an orbifold
\begin{equation}
  \mathcal{M}=\mathbb{C}^{3n}/\mathsf{G},
\end{equation}
where $\mathsf{G}$ is a crystallographic complex reflection group.  The complex-reflection
property follows from assuming that the Coulomb branch chiral ring is freely
generated\footnote{$\mathcal{N} = 3$ theories obtained from discrete gauging $\mathbb{Z}_n$ symmetry of $\mathcal{N} = 4$ generally do not have a freely generated Coulomb branch chiral ring \cite{Bourton:2018jwb}.}, and the crystallographic property from the action of $\mathsf{G}$ on the
electromagnetic charge lattice.

Recall that a complex reflection fixes a hyperplane in $\mathbb{C}^n$ pointwise
and multiplies a complementary direction by a root of unity $\gamma$; in a
suitable basis, it acts as $z_i\mapsto\gamma z_i$ for a single index $i$ and
leaves the other coordinates alone.  For $\gamma=-1$, one gets ordinary
reflections, so all Weyl groups of semisimple Lie algebras are of this type.
Crystallographic complex reflection groups have been completely classified
\cite{Popov1982Utrecht,LehrerTaylor2009}.

We will only need one infinite family, the monomial groups $\mathsf{G}(k,p,n)$ with
$p\mid k$.  These are generated by coordinate permutations, forming a copy of
$S_n$, together with the diagonal phase rotations
\begin{equation}
  (z_1,\ldots,z_n)
  \longmapsto
  \bigl(e^{2\pi i a_1/k}z_1,\ldots,e^{2\pi i a_n/k}z_n\bigr),
  \qquad
  \sum_{i=1}^n a_i\equiv0\pmod p.
\end{equation}
Their invariant rings are freely generated, with degrees
\cite{Aharony:2016kai,Bonetti:2018fqz}
\begin{equation}\label{eq:Gkpn-invariant-degrees}
  d_i=k,2k,\ldots,(n-1)k,\frac{nk}{p}.
\end{equation}
Such groups are crystallographic only for $k=1,2,3,4,6$.  For $k=1,2$, they are
the classical Weyl groups of $ABCD$ type, and a theory with moduli space
$\mathbb{C}^{3n}/\mathsf{G}(k,p,n)$ is then really $\mathcal{N}=4$; genuine
$\mathcal{N}=3$ theories require $k=3,4,6$.  There are two useful constructions
of such theories, which we now describe in turn.

\subsection{S-folds}\label{sec:S-folds}

The S-fold construction realizes some of the above orbifold geometries in
string theory
\cite{Garcia-Etxebarria:2015wns,Aharony:2016kai,Agarwal:2016rvx}.  Consider
F-theory on
\begin{equation}
  \mathbb{R}^{1,3}\times
  \bigl(\mathbb{C}^{3}\times T^{2}\bigr)/\mathbb{Z}_{k} \, ,
\end{equation}
where $T^2$ is the F-theory torus.  The quotient is by a
\begin{equation}
  \mathbb{Z}_{k}\subset Spin(6)\times SL(2,\mathbb{Z})
\end{equation}
symmetry of the background, the first factor acting as the $R$-symmetry and the
second as the S-duality group.  Its generator combines a rotation by $2\pi/k$
in the three complex directions transverse to a D3-brane with an order-$k$
automorphism of the torus, which in type IIB is a non-perturbative duality
transformation; this is why the fixed locus is called an S-fold.  Such an
automorphism exists only for $k=3,4,6$, at the corresponding fixed values of
the axiodilaton.  The diagonal action is arranged so that the two phases cancel
on twelve of the sixteen supercharges, and the theory on D3-branes at the fixed
locus therefore has $\mathcal{N}=3$ supersymmetry.  The same quotient also
explains why a genuine S-fold theory has no exactly marginal gauge coupling:
the axiodilaton is frozen.

Now, put $n$ D3-branes on the S-fold.  At a generic point of the moduli space,
their positions give $n$ triplets of complex scalars.  Permuting the branes and
identifying each one with its S-fold images gives a diagonal action of a
monomial complex reflection group on the three copies of $\mathbb{C}^{n}$.  A
theory in this family is specified by the number $n$ of D3-branes together with
the pair
\begin{equation}
  (k,\ell)\,,\qquad \ell=\frac{k}{p}\,,\qquad p\mid k \, ,
\end{equation}
and has moduli space
\begin{equation}\label{eq:S-fold-moduli-space}
  \mathcal{M}_{k,\ell}^{(n)}
  =\frac{\mathbb{C}^{3n}}{\mathsf{G}(k,p,n)} \, .
\end{equation}
For this family, the crystallographic reflection groups are not just a
classification of possible triple special K\"ahler cones; they are the
identifications acting on the D3-brane positions.  The allowed variants are
determined by the discrete three-form flux \cite{Aharony:2016kai}, and for
$k=3,4,6$, they are
\begin{equation}
  (k,\ell)=(3,1),(3,3),(4,1),(4,4),(6,1) \, .
\end{equation}

With the $\mathcal{N}=2$ embedding above, one of the three copies of
$\mathbb{C}^{n}$ is the Coulomb branch, so the degrees
\eqref{eq:Gkpn-invariant-degrees}, which here read
$d_i=k,2k,\ldots,(n-1)k,n\ell$, are the dimensions of its independent chiral
generators.  At rank one, $\mathsf{G}(k,p,1)=\mathbb{Z}_{\ell}$, so that
\begin{equation}
\mathcal{M}_{k,\ell}^{(1)}=\mathbb{C}^{3}/\mathbb{Z}_{\ell}\,,
  \qquad \Delta(u)=\ell\,,
\end{equation}
with Higgs branch $\mathbb{C}^{2}/\mathbb{Z}_{\ell}$.  This reproduces the
rank-one chiral rings found above and makes manifest how $U(3)_\mathcal{R}$
relates $u$ to $W^{\pm}$.

The $\mathcal{N}=3$ supersymmetry also forces the 4d central charges to agree,
\begin{equation}
  a_{\mathrm{4d}}=c_{\mathrm{4d}},
\end{equation}
in complete analogy with 4d $\mathcal{N}=4$ SYM theories.  For S-fold theories,
the Shapere--Tachikawa condition \cite{Shapere:2008zf} is believed to hold and fixes this common
value in terms of the Coulomb branch spectrum,
\begin{equation}
  4\bigl(2a_{\mathrm{4d}}-c_{\mathrm{4d}}\bigr)
  =\sum_i\bigl(2\Delta(u_i)-1\bigr),
  \qquad
  c_{\mathrm{4d}}
  =\frac14\sum_i\bigl(2\Delta(u_i)-1\bigr),
\end{equation}
where $u_i$ denote the independent Coulomb branch operators.  For
$\mathcal{N}=3$ theories obtained by discrete gauging, the relation fails, and
the central charges are instead inherited from the parent $\mathcal{N}=4$
theory.  The interacting theory studied in section~\ref{section:3} is the
rank-one S-fold with $k=\ell=3$, hence $p=1$, which we refer to throughout as
the $\mathbb{Z}_3$ S-fold theory, and for which
\begin{equation}
  \mathsf{G}(3,1,1)=\mathbb{Z}_{3}\,,\qquad
  \mathcal{M}=\mathbb{C}^{3}/\mathbb{Z}_{3}\,,\qquad
  \Delta(u)=3\,,\qquad
  a_{\mathrm{4d}}=c_{\mathrm{4d}}=\frac54.
\end{equation}

When $p>1$, the S-fold theory has a residual $\mathbb{Z}_{p}$ symmetry that may
itself be gauged \cite{Aharony:2016kai}, and the enlarged orbifold group can
give the same moduli space as another S-fold variant without giving the same
local spectrum.  For example, gauging the residual $\mathbb{Z}_{3}$ symmetry of
the free rank-one $(k,\ell)=(3,1)$ theory produces the same quotient
$\mathbb{C}^{3}/\mathbb{Z}_{3}$ as the $\mathbb{Z}_3$ S-fold theory, with a
Coulomb branch again generated by an operator of dimension $\Delta(u)=3$.  It
nevertheless keeps the free-theory value $a_{\mathrm{4d}}=c_{\mathrm{4d}}=1/4$, rather than the $5/4$
that Shapere--Tachikawa would assign to that spectrum.  The moduli space by
itself therefore does not determine the theory.

\subsection{Discrete gauging}\label{sec:2-discrete-gauging}

The second construction is by discrete gauging an $\mathcal{N}=4$ SYM theory
with gauge group $G$, using a finite symmetry that combines electric--magnetic
duality with an $R$-symmetry rotation
\cite{Garcia-Etxebarria:2015wns,Aharony:2016kai}.  Here, $G$ is understood to
include the global form and the line operator spectrum, and $\tau$ is the
exactly marginal complexified gauge coupling.  On the coupling, the generators act as
\begin{equation}
 T:\tau\longmapsto\tau+1,\qquad
 S:\tau\longmapsto-\frac{1}{\lambda_{\mathfrak g}^{2}\tau},
 \qquad \lambda_{\mathfrak g}=2\cos\frac{\pi}{p_{\mathfrak g}}.
\end{equation}
Here $p_{\mathfrak g}=3$ for $ADE$, $4$ for $B$, $C$, and $F_4$, and $6$
for $G_2$. The generator $S$ exchanges the gauge group with its Langlands
dual, while $T$ can change the discrete theta angle and the line spectrum
\cite{Aharony:2013hda}. Matrix representatives are
\begin{equation}\label{eq:duality-matrices}
 S=\begin{pmatrix}0&-\lambda_{\mathfrak g}^{-1}\\
                   \lambda_{\mathfrak g}&0\end{pmatrix},\qquad
 T=\begin{pmatrix}1&1\\0&1\end{pmatrix},\qquad
 S^2=(ST)^{p_{\mathfrak g}}=-I.
\end{equation}
Their projective action on $\tau$ is the Hecke group. For $ADE$,
$\lambda_{\mathfrak g}=1$, the matrices generate $SL(2,\mathbb Z)$, and
its action on $\tau$ factors through $PSL(2,\mathbb Z)$.
The central element $-I$ fixes $\tau$, but its action on line charges and
its lift to supercharges must still be kept.

A symmetry of a fixed theory must preserve both the coupling and the global
form with its chosen line spectrum. Being of $ADE$ type only identifies the
Lie algebra with its Langlands dual; it does not make every duality a symmetry
of that fixed theory \cite{Aharony:2013hda,Bourton:2018jwb}.
The finite cyclic subgroups of $SL(2,\mathbb Z)$ and their fixed couplings are
listed below. A subgroup can be gauged only when it acts on the chosen theory
and the gauging is anomaly-free.
\begin{center}
  \begin{tabular}{c c}
    \toprule
    cyclic subgroup $\subset SL(2,\mathbb{Z})$ & special value of $\tau$ \\
    \midrule
    $\mathbb{Z}_2 = \langle S_2 = S^2 \rangle$ & any $\tau \in \mathbb H$ \\
    $\mathbb{Z}_3 = \langle S_3 = TS^3 \rangle$ & $\tau = e^{\frac{\pi i}{3}}$ \\
    $\mathbb{Z}_4 = \langle S_4 = S^3 \rangle$ & $\tau = i$ \\
    $\mathbb{Z}_6 = \langle S_6 = TS \rangle$ & $\tau = e^{\frac{\pi i}{3}}$ \\
    \bottomrule
  \end{tabular}
\end{center}
Here $\mathbb H$ is the upper half-plane, and products act from right to left.
We use the lift of $S_k$ to the four $\mathcal N=4$ Poincar\'e supercharges\footnote{Each $Q_k$ denotes a spinor.}
with phases \cite{Bourton:2018jwb}
\begin{equation}
  S_k:(Q_1,Q_2,Q_3,Q_4)\longmapsto
  e^{-\pi i/k}(Q_1,Q_2,Q_3,Q_4).
\end{equation}
To preserve supersymmetry, we combine this lift with the $SU(4)_R$ transformation
$R_k$ acting as\footnote{$R_k$ is denoted as $e^{\frac{2\pi i}{k} r_k}$, and $S_k$ as $e^{\frac{2\pi i}{k}s_k}$ in \cite{Bourton:2018jwb}.}
\begin{equation}
  R_k:(Q_1,Q_2,Q_3,Q_4)\longmapsto
  \left(e^{\pi i/k}Q_1,e^{\pi i/k}Q_2,e^{\pi i/k}Q_3,
  e^{-3\pi i/k}Q_4\right).
\end{equation}
Under the diagonal element $R_kS_k$, twelve of the sixteen supercharges are
invariant, and
\begin{equation}
  \mathbb{Z}_k=\langle R_kS_k\rangle
  \subset Spin(6)\times SL(2,\mathbb{Z})
\end{equation}
preserves an $\mathcal{N}=3$ superconformal subalgebra.  From the point of view
of the $\mathcal{N}=2$ embedding fixed above, it is an extra flavor symmetry.

In this paper, we will encounter $U(1)$ and $SU(2)$ $\mathcal{N} = 4$ SYM; the former theory is just a free $\mathcal{N} = 4$ vector multiplet. Acting on the six real scalars $X^i$ of the free $\mathcal{N}=4$ vector
multiplet, the combined transformation $R_kS_k$ sends
\begin{equation}
  \begin{aligned}
    (z_1,z_2,z_3)&\longmapsto
    \big(e^{2\pi i/k}z_1,e^{2\pi i/k}z_2,e^{-2\pi i/k}z_3\big),\\
    z_1&=X^1+iX^2,\qquad
    z_2=X^3+iX^4,\qquad
    z_3=X^5+iX^6.
  \end{aligned}
\end{equation}
More generally, in any $\mathcal{N}=4$ theory, a Coulomb branch operator $u$ transforms under $R_k S_k$ as\footnote{As noted in \cite{Bourton:2018jwb}, this action is in general not a complex reflection group action. As a result, the Coulomb branch chiral ring is not freely generated.}
\begin{equation}
  u\longmapsto e^{2\pi i\Delta(u)/k}u \ ,
\end{equation}
where $\Delta(u)$ is the scaling dimension of $u$. One may keep this phase as a discrete flavor fugacity in the $\mathcal{N}=2$
superconformal index, or gauge the symmetry, which projects onto invariant local operators and introduces the accompanying twisted sectors.  The Coulomb
and Higgs branch chiral rings then change, but the stress tensor multiplet and
its self-OPE are preserved, so the central charges are those of the parent
theory,
\begin{equation}
  a_{\mathrm{4d}}^{\mathrm{quot}}
  =a_{\mathrm{4d}}^{\mathrm{parent}},\qquad
  c_{\mathrm{4d}}^{\mathrm{quot}}
  =c_{\mathrm{4d}}^{\mathrm{parent}}.
\end{equation}
For example, $\mathcal{N}=4$ $SU(2)$ SYM has
$a_{\mathrm{4d}}=c_{\mathrm{4d}}=3/4$, as do all of its $\mathbb{Z}_k$
quotients.  The central charges therefore distinguish discrete quotients from
S-fold SCFTs even when the two have the same moduli space.

\subsection{Associated VOA}\label{sec:assoc-VOA}

Neither construction gives a Lagrangian description of the resulting fixed
point, so the localization formulas for the superconformal index of a
Lagrangian theory
\cite{Kinney:2005ej,Romelsberger:2005eg,Dolan:2008qi,Gadde:2011uv} do not apply
directly, and no general prescription for the index of an $\mathcal{N}=3$ SCFT
is available.  We use two methods.  One may realize the theory as the IR fixed point of an $\mathcal{N}=1$ Lagrangian gauge theory and compute the index in the UV, where it is protected along the flow \cite{Zafrir:2020epd};
this is the route taken in section~\ref{section:3}.  Alternatively, one may pass
to the protected subsector isolated by the SCFT/VOA correspondence
\cite{Beem:2013sza}, which assigns to any 4d $\mathcal{N}\geq2$ SCFT a vertex
operator algebra whose vacuum character reproduces the Schur limit of the
superconformal index \cite{Gadde:2011ik}.  The latter viewpoint underlies the
rest of the paper, so we briefly recall how the algebra is obtained.

Let $\mathcal{T}$ be a 4d $\mathcal{N}=2$ SCFT and fix a plane
$\mathbb{R}^2\subset\mathbb{R}^4$ with complex coordinate $z$.  Among the
supercharges of $\mathfrak{su}(2,2|2)$, one may form the two nilpotent
combinations
\begin{equation}
  \mathbbm{Q} = \mathcal{Q}^1_{\ -} + \tilde{\mathcal{S}}^{2\dot -} \ , \qquad
  \widetilde{\mathbbm{Q}} = \tilde{\mathcal{Q}}_{2 \dot -} - \mathcal{S}^{\ -}_{1} \ ,
  \qquad
  \{\mathbbm{Q}, \widetilde{\mathbbm{Q}}\} = E - (j_1 + j_2) - 2R \eqqcolon \delta \ .
\end{equation}
Local operators annihilated by $\delta$, equivalently those saturating
\begin{equation}\label{eq:Schur-condition}
  E = 2R + j_1 + j_2 \ , \qquad r = j_2 - j_1 \ ,
\end{equation}
are the \emph{Schur operators}, and they are precisely the operators that
survive in $\mathbbm{Q}$-cohomology.  Translation in $\bar z$ does not act
within the cohomology, so it is replaced by the twisted translation
$\widehat{\bar L}_{-1}=\bar L_{-1}+\mathcal{R}^{-}$, which amounts to
contracting $SU(2)_{\mathcal R}$ indices with the position-dependent
polarization $u_I(\bar z)=(1,\bar z)$.  For the twisted-translated operators,
$\partial_{\bar z}$ acts $\mathbbm{Q}$-exactly, so their cohomology classes
depend on $z$ alone.  With the OPE inherited from four dimensions, these
classes form a vertex operator algebra $\chi[\mathcal{T}]$, graded by
\begin{equation}
  h = \tfrac{1}{2}\left(E + j_1 + j_2\right) = R + j_1 + j_2 \ .
\end{equation}

The entries of the dictionary that we need are
\begin{equation}
  c_{\mathrm{2d}}=-12c_{\mathrm{4d}},\qquad
  k_{\mathrm{2d}}=-\frac12 k_{\mathrm{4d}}.
\end{equation}
The Schur operator in the $SU(2)_{\mathcal R}$ current multiplet maps to the 2d
stress tensor $T$, and the moment map operators of a 4d flavor algebra
$\mathfrak g$ with flavor central charge $k_{\mathrm{4d}}$ generate an affine
subalgebra $\widehat{\mathfrak g}_{k_{\mathrm{2d}}}$.  Since $E-R=h$ on Schur
operators, the Schur index in our convention is the vacuum supercharacter,
\begin{equation}
  \mathcal{I}(q) = q^{-c_{\mathrm{2d}}/24}\operatorname{Tr}(-1)^F q^{E - R}
  = \operatorname{Tr}_{\chi[\mathcal{T}]}(-1)^F q^{L_0-c_{\mathrm{2d}}/24}
  =\operatorname{sch}_{\chi[\mathcal{T}]}(q),
\end{equation}
and gauging a flavor symmetry in four dimensions is implemented by a BRST
reduction in two dimensions, as used in section~\ref{sec:discrete-gauging}.
Finally, the associated variety of $\chi[\mathcal{T}]$ is conjecturally the
Higgs branch of $\mathcal{T}$.  In the cases at hand, it is a symplectic variety
with finitely many symplectic leaves, hence quasi-lisse, which is what
guarantees that the characters of ordinary modules satisfy an MLDE
\cite{Arakawa:2016hkg}.

For $\mathcal{N}\geq3$, one chooses an $\mathcal{N}=2$ subalgebra as above.  The
third pair of supercharges $Q^3_\alpha,\widetilde Q_{3\dot\alpha}$ is then not
used in building $\mathbbm{Q}$ and survives in cohomology, contributing two
fermionic currents of weight $h=3/2$.  Together with $T$ and the current $J$
associated with the $U(1)_f$ commutant, these generate a 2d $\mathcal{N}=2$
super-Virasoro subalgebra \cite{Nishinaka:2016hbw}, which organizes all of the
algebras appearing below.

For the S-fold theories, this structure was made explicit in
\cite{Bonetti:2018fqz}.  To a complex reflection group $\mathsf{G}$ of rank $n$, one
associates an $\mathcal{N}=2$ vertex operator algebra $\mathcal{W}_{\mathsf{G}}$,
obtained by extending the super-Virasoro algebra by generators in
correspondence with the fundamental invariants of $\mathsf{G}$: for each invariant
of degree $d_\ell$, $\ell=1,\ldots,n$, one adds a pair of $\mathcal{N}=2$
chiral and anti-chiral multiplets whose primaries have conformal weight
$h=d_\ell/2$.  Consistency of the resulting OPEs fixes the central charge to a
single negative value,
\begin{equation}\label{eq:c2d-degrees}
  c_{\mathrm{2d}}=-12c_{\mathrm{4d}}
  =-3\sum_{\ell=1}^{n}\bigl(2d_\ell-1\bigr),
\end{equation}
in agreement with the Shapere--Tachikawa value of $c_{\mathrm{4d}}$ once the
Coulomb branch dimensions are identified with the degrees $d_\ell$.  The
algebra $\mathcal{W}_{\mathsf{G}}$ admits a free-field realization in terms of $n$
copies of a $bc\beta\gamma$ ghost system, with the quantum numbers recorded in
Table~\ref{tab:ghost-qnums}, generalizing Adamovi\'c's realization of the small
$\mathcal{N}=4$ algebra \cite{Adamovic:2014lra}.  When $\mathsf{G}$ is
crystallographic, $\mathcal{W}_{\mathsf{G}}$ is conjectured to be the VOA
associated with the corresponding S-fold SCFT.

The restriction to S-folds is essential here, since the quotient
$\mathbb{C}^{3n}/\mathsf{G}$ by itself does not determine the chiral algebra.
Equation \eqref{eq:c2d-degrees} builds in the S-fold value of the central
charge, whereas a discrete quotient inherits $c_{\mathrm{4d}}$ from its parent
and is realized as a finite quotient of the parent VOA.  Two SCFTs with the
same moduli space can therefore have different associated VOAs.  The case
relevant for section~\ref{section:3} is $\mathsf{G}=\mathsf{G}(3,1,1)=\mathbb{Z}_3$, with a
single invariant of degree $d_1=3$ and hence $c_{\mathrm{2d}}=-15$, while the
$\mathbb{Z}_3$ quotient of the free $U(1)$ theory, which has the same moduli
space, gives $c_{\mathrm{2d}}=-3$.

\section{The \texorpdfstring{$\mathbb{Z}_3$}{Z3} S-fold theory}\label{section:3}

In this section, we focus on the $\mathbb{Z}_3$ S-fold theory, the rank-one $\mathcal{N} = 3$ theory associated with the complex reflection group $\mathsf{G}(3,1,1) = \mathbb{Z}_3$. We will study in detail the representation theory of the associated VOA $\mathcal{W}_{\mathbb{Z}_3}$. We will work out its null states at low weight and the corresponding flavored/unflavored MLDEs, vacuum/non-vacuum flavored/unflavored characters in closed form, derive alternatively the prediction of module highest (or lowest, depending on the preferred language) weights from Zhu's associative algebra, propose the free-field realization for the non-vacuum modules, conjecture a reduced PBW basis whose state count reproduces the flavored characters, and confirm the quasi-modularity structure among the null states.

The theory has one Coulomb branch operator $u$ of dimension $\Delta(u) = 3$, hence
\begin{equation}
    a_{\mathrm{4d}} = c_{\mathrm{4d}} = \frac{2 \Delta(u) - 1}{4} = \frac{5}{4} \ .
\end{equation}
Although the $\mathcal{N} = 3$ theory does not admit $\mathcal{N}=2$-preserving exactly marginal deformations, it was shown in \cite{Zafrir:2020epd} that the theory sits on the same $\mathcal{N} = 1$ conformal manifold of the IR fixed point of an $\mathcal{N} = 1$ Lagrangian theory. The Lagrangian theory in the UV is an $SU(2) \times SU(2)$ gauge theory with four chiral multiplets $A, B, C, F$. The theory has a $U(1)_v \times U(1)_x \times U(1)_R$ global symmetry, where the two flavor symmetries can mix with $U(1)_R$ in the superconformal $R$-charge. Relevant superpotentials were introduced sequentially along the renormalization-group flow to break $U(1)_v \times U(1)_x$ while preserving $U(1)_R$, so that the $U(1)_R$ symmetry visible in the UV coincides with the $\mathcal{N} = 3$ superconformal $R$-charge. This Lagrangian description makes the protected quantities accessible to direct computation.

\subsection{Unflavored index in closed form}\label{sec:unflavored_WZ3}

The $\mathcal{N} = 2$ superconformal index of the $\mathbb{Z}_3$ S-fold theory is computed by the index of the UV gauge theory, written as a contour integral,
\begin{equation}
\label{eq:full-index}
\begin{aligned}
  \mathcal{I} = \frac{\kappa^{2}\,\Gamma\!\left(x^{-8}(pq)^{\frac{2}{3}};p,q\right)}{4}
  \oint \frac{dz}{2\pi i z}\,\frac{dy}{2\pi i y}\;
  &\frac{\Gamma\!\left(z^{\pm 2}vx^{-2}(pq)^{\frac{1}{3}};p,q\right)
         \Gamma\!\left(z^{\pm}y^{\pm 2}x\,(pq)^{\frac{1}{3}};p,q\right)}
        {\Gamma(z^{\pm 2};p,q)\,\Gamma(y^{\pm 2};p,q)}\\
  &\times
  \Gamma\!\left(z^{\pm}x^{13}v^{-4}(pq)^{\frac{1}{3}};p,q\right)
  \Gamma\!\left(z^{\pm}y^{\pm}x^{-4}(pq)^{\frac{1}{6}};p,q\right),
\end{aligned}
\end{equation}
where
\begin{equation}
  \kappa \;\equiv\; (p;p)(q;q) \;=\; p^{-\frac{1}{24}}q^{-\frac{1}{24}}\,\eta(p)\,\eta(q).
\end{equation}
Here, $v$ and $x$ are the fugacities for $U(1)_v$ and $U(1)_x$, respectively. We are interested in the Schur limit given by $p \to \sqrt{q}$. We also set $x, v \to 1$. The integral simplifies to
\begin{align}
\label{eq:Schur-integral}
  q^{-5/8}\mathcal{I} = & \ \frac{\eta\!\left(\frac{\tau}{2}\right)^{5}}{4q\,\eta(\tau)}
  \oint \frac{dz}{2\pi i z}\,\frac{dy}{2\pi i y}\;
  \frac{\vartheta_{1}\!\left(2\mathfrak{y}\big|\frac{\tau}{2}\right)
        \vartheta_{1}\!\left(2\mathfrak{y}\big|\tau\right)
        \vartheta_{1}\!\left(2\mathfrak{z}\big|\tau\right)^{2}
        \vartheta_{1}\!\left(\mathfrak{z}\pm 2\mathfrak{y}\big|\tau\right)}
       {\vartheta_{1}\!\left(\mathfrak{z}+\frac{\tau}{2}\big|\tau\right)^{2}
        \vartheta_{1}\!\left(\mathfrak{z}\pm 2\mathfrak{y}\big|\frac{\tau}{2}\right)
        \vartheta_{1}\!\left(\mathfrak{z}\pm\mathfrak{y}-\frac{\tau}{4}\big|\frac{\tau}{2}\right)} \cr
        = & \ \oint \frac{dz}{2\pi i z} \frac{dy}{2\pi i y} \mathcal{Z}(\mathfrak{z}, \mathfrak{y}) \ .
\end{align}
It is straightforward to check that the integrand $\mathcal{Z}(\mathfrak{z}, \mathfrak{y})$ is doubly elliptic as a function of both $\mathfrak{z}, \mathfrak{y}$,
\begin{equation}
    \mathcal{Z}(\mathfrak{z} + 1, \mathfrak{y})
    = \mathcal{Z}(\mathfrak{z} + \tau, \mathfrak{y})
    = \mathcal{Z}(\mathfrak{z}, \mathfrak{y} + 1)
    = \mathcal{Z}(\mathfrak{z}, \mathfrak{y} + \tau) = \mathcal{Z}(\mathfrak{z}, \mathfrak{y}) \ .
\end{equation}
Applying the integration method of \cite{Pan:2021mrw}, the integral gives the corrected Schur index in the closed form
\begin{equation}
\label{eq:SchurClosedForm}
  \mathcal{I} = \frac{1}{3}\!\left(
    \frac{\eta(q)^{4}}{\eta(\sqrt{q})^{2}}
    - \frac{\eta(\sqrt{q})\,\eta(q^{3})^{2}}{\eta(q)^{2}\,\eta(q^{3/2})}
  \right)
  = q^{\frac{5}{8}}\!\left(1+q+q^{2}+2q^{3}-2q^{\frac{7}{2}}+3q^{4}
    -2q^{\frac{9}{2}}+\cdots\right).
\end{equation}
A series of identities between Jacobi theta functions and the Dedekind eta function are used to bring the result of the integration into the simple form above; see appendix~\ref{app:identities}. The resulting $q$-series expansion agrees with the result of \cite{Zafrir:2020epd}.

The Schur index is identified with the vacuum character of the associated VOA $\mathcal{W}_{\mathbb{Z}_3}$ \cite{Beem:2013sza} of the $\mathcal{N} = 3$ theory. Besides the vacuum module, we are also interested in its full representation theory. In general, the non-vacuum modules of the algebra correspond to line or surface defects of the 4d superconformal theory, and the corresponding defect Schur indices are expected to reproduce the non-vacuum characters \cite{Cordova:2015nma,Cordova:2017mhb,Nishinaka:2018zwq,Bianchi:2019sxz,Zheng:2022zkm}. Much less is known, however, about these defects in the $\mathcal{N} = 3$ theory.

The UV $\mathcal{N} = 1$ Lagrangian description also provides a way to access the \emph{unflavored} non-vacuum characters of $\mathcal{W}_{\mathbb{Z}_3}$. A 4d $\mathcal{N} = 1$ vector multiplet does not contain an adjoint scalar, so a standard Wilson line operator does not preserve the supercharges used to define the superconformal index. The $\mathcal{N} = 3$ theory is not a gauge theory at all, and hence there is no Wilson line operator to speak of either. Nevertheless, one can still formally write down a ``Wilson-loop index'' in analogy with those of an $\mathcal{N} = 2$ theory, by inserting a gauge group character into the integral. Although this object has no clear origin in either the $\mathcal{N} = 1$ or the $\mathcal{N} = 3$ description (or rather the $\mathcal{N} = 2$ description, since we are interested in the chiral algebra, which is intrinsically an $\mathcal{N} = 2$ structure), it does produce the desired characters.

The character of the spin-$j$ representation of $SU(2)$ is given by
\begin{equation}\label{spin-j}
  \chi^{\mathrm{spin}}_{j}(z) = \frac{z^{2j+1}-z^{-(2j+1)}}{z-z^{-1}} \ .
\end{equation}
We insert the spin-$j_z$ and spin-$j_y$ characters of the $SU(2)_z \times SU(2)_y$ gauge group into the contour integral that computes the Schur index,
\begin{equation}
  \langle W_{(j_{z},j_{y})}\rangle
  = \oint \frac{dz}{2\pi i z}\,\frac{dy}{2\pi i y}\;
    \chi^{\mathrm{spin}}_{j_{z}}(z)\,\chi^{\mathrm{spin}}_{j_{y}}(y)\,\mathcal{Z}(z,y),
\end{equation}
where $\mathcal{Z}(z,y)$ is the unshifted integrand in \eqref{eq:Schur-integral}. The integral can be easily carried out using the integration formula in \cite{Pan:2021mrw,Guo:2023mkn}.

\paragraph{Representations $(j,0)$ and $(0,j)$.}
With a single non-trivial spin, the answer involves, besides the Schur index
$\operatorname{ch}_{\textrm{vac}}\coloneqq\operatorname{ch}_{\frac{5}{8}}\coloneqq\mathcal{I}$ itself, only the two
$q$-series
\begin{align}\label{eq:nonvac-characters-1}
  \operatorname{ch}_{-\frac{1}{24}}
  &= \frac{\eta(q^{\frac{3}{2}})^{2}}{\eta(q)\,\eta(q^{3})}
  = q^{-\frac{1}{24}}\!\left(1 + q - 2q^{\frac{3}{2}} + 2q^{2} - 2q^{\frac{5}{2}}
    + 3q^{3} - 4q^{\frac{7}{2}} + \cdots\right),\\
      \operatorname{ch}_{\frac{1}{8}}
  &= \frac{\eta(q^{\frac{1}{2}})\,\eta(q^{3})^{2}}{\eta(q)^{2}\,\eta(q^{\frac{3}{2}})}
  = q^{\frac{1}{8}}\!\left(1 - q^{\frac{1}{2}} + q - q^{\frac{3}{2}}
    + 2q^{2} - 3q^{\frac{5}{2}} + 3q^{3} - 4q^{\frac{7}{2}} + \cdots\right)~.\label{eq:nonvac-characters-2}
\end{align}
As the labels indicate, these are the unflavored non-vacuum characters of the
associated VOA; they are re-derived from the MLDE in section~\ref{sec:MLDE}.
Half-integral spin contributes nothing,
\begin{equation}
  \langle W_{(j,0)}\rangle = \langle W_{(0,j)}\rangle = 0 \ ,
  \qquad j \in \mathbb{Z}+\tfrac{1}{2} \ ,
\end{equation}
while for $j\in\mathbb{Z}$,
\begin{equation}
\label{eq:Wilson-jz}
\begin{aligned}
  \langle W_{(j,0)}\rangle
  &= q^{-5/8}\operatorname{ch}_{\textrm{vac}}
    + \sum_{\substack{n=-2j\\ n\neq 0}}^{2j}
      \bigl(a_{n}\,\operatorname{ch}_{-\frac{1}{24}} + b_{n}\,\operatorname{ch}_{\frac{1}{8}}\bigr) \ , \\
  \langle W_{(0,j)}\rangle
  &= q^{-5/8}\operatorname{ch}_{\textrm{vac}}
    + \sum_{\substack{n=-2j\\ n\neq 0}}^{2j}
      \bigl(c_{n}\,\operatorname{ch}_{-\frac{1}{24}} + d_{n}\,\operatorname{ch}_{\frac{1}{8}}\bigr) \ .
\end{aligned}
\end{equation}
The coefficients are rational functions of $q$.  Abbreviating
$\delta_{a}\coloneqq\delta_{n\equiv a\,(3)}$ and
$\Sigma_{n}\coloneqq 2(q^{n}-1)(1+q^{n})^{2}$,
\begin{equation}
\label{eq:Wilson-jz-coeff}
\begin{aligned}
  a_{n} &= \frac{q^{-\frac{5}{4}+\frac{2n}{3}}}{2\,(q^{4n}-1)}
           \Bigl[\delta_{2}-\delta_{1}\,q^{\frac{8n}{3}}\Bigr] \ , \\[1mm]
  b_{n} &= \frac{q^{-\frac{5}{4}+\frac{2n}{3}}}{2\,(q^{4n}-1)}
           \Bigl[\delta_{0}\bigl(1-q^{\frac{8n}{3}}\bigr)
                +\bigl(\delta_{2}-\delta_{1}\bigr)q^{\frac{4n}{3}}\Bigr] \ , \\[1mm]
  c_{n} &= -\frac{q^{-\frac{5}{4}+\frac{5n}{6}}}{\Sigma_{n}}
           \Bigl[\delta_{1}\bigl(1-2q^{n}\bigr)
                +\delta_{2}\,q^{\frac{n}{3}}\bigl(2-q^{n}\bigr)\Bigr] \ , \\[1mm]
  d_{n} &= \frac{q^{-\frac{5}{4}+\frac{n}{2}}}{\Sigma_{n}}
           \Bigl[\bigl(\delta_{2}-\delta_{1}\bigr)\bigl(1-q^{n}+q^{2n}\bigr)
                +\delta_{0}\,q^{\frac{n}{3}}\bigl(1-q^{\frac{n}{3}}\bigr)
                 \bigl(1+q^{\frac{n}{3}}\bigr)^{3}\Bigr] \ .
\end{aligned}
\end{equation}

\paragraph{General case.}
Since the $SU(2)$ character is a polynomial in $z$ or $y$, for general $j_z,j_y$, it is convenient to separate the spin labels from the
monomial exponents.  We define
\begin{equation}
  \mathcal{I}_{r,s}
  \equiv
  \oint \frac{dz}{2\pi i z}\,\frac{dy}{2\pi i y}\;
  z^{r}y^{s}\,\mathcal{Z}(z,y),
\end{equation}
where $r,s\in\mathbb Z$ are the exponents of the monomials inserted in the
integrand.  With the convention \eqref{spin-j}, the weights appearing in
$\chi^{\mathrm{spin}}_{j_z}(z)\chi^{\mathrm{spin}}_{j_y}(y)$ are
\[
  r=-2j_z,-2j_z+2,\ldots,2j_z,\qquad
  s=-2j_y,-2j_y+2,\ldots,2j_y .
\]
The integrals with $r=0$ or $s=0$ were already covered by the two cases above,
so we now focus on $r,s\ne 0$.

Evaluating the residues explicitly (see appendix~\ref{app:residues} for the
definitions of $S_i$ and $W_i$), one finds
\begin{equation}
\label{eq:Imn}
\begin{aligned}
  \mathcal{I}_{r,s}
  = {}&\delta_{s,r}\,(S_{2}-W_{1}-W_{2}+W_{3})
       + \delta_{s,-r}\,(S_{2}-W_{1}-W_{2}+W_{4})\\
      &+\delta_{s,\frac{r}{2}}\,(-S_{1}-W_{2}+W_{3}+W_{4})
       + \delta_{s,-\frac{r}{2}}\,(-S_{1}-W_{1}+W_{3}+W_{4}) .
\end{aligned}
\end{equation}
The ``Wilson-loop index'' in a general representation $(j_z,j_y)$ is then
obtained by summing over the character weights,
\begin{equation}
\label{eq:Wilson-general}
  \langle W_{(j_z,j_y)}\rangle
  =
  \sum_{\substack{r=-2j_z\\ r\equiv 2j_z\!\!\!\pmod 2}}^{2j_z}
  \sum_{\substack{s=-2j_y\\ s\equiv 2j_y\!\!\!\pmod 2}}^{2j_y}
  \mathcal{I}_{r,s}.
\end{equation}
The term with $r=s=0$ is the unshifted index without insertion, $\mathcal I_{0,0}=q^{-5/8}\mathcal I$.

\subsection{Modular linear differential equation}\label{sec:MLDE}

Via the SCFT/VOA correspondence, the Schur index is identified with the vacuum character of the associated VOA. The Higgs branch of the 4d theory is identified with the associated variety of the algebra. The fact that the Higgs branch, here $\mathbb{C}^2/\mathbb{Z}_3$, is a symplectic variety with finitely many symplectic leaves implies that the VOA is quasi-lisse. The analysis of \cite{Arakawa:2016hkg} then guarantees that the unflavored character of any ordinary module satisfies an MLDE. The presence of half-integer powers of $q$ in the unflavored Schur index suggests that this equation should be $\Gamma^0(2)$-modular. Such an equation can be obtained by making the following ansatz:
\begin{equation}
\label{eq:MLDE-general}
  \left[D_{q}^{(n)}+\sum_{k = 1}^{n}\sum_{a,b\geq 0}^{4a+2b=2k}
  c_{k,a,b}\,E_{4}^{a}\,\Em{2}{1}^{\!b}\,
  D_{q}^{(n-k)}\right]\mathcal{I}=0 .
\end{equation}
The equation is homogeneous in modular weight, and $D_q^{(n)} \coloneqq \partial_{(2n-2)} \circ \cdots \circ \partial_{(2)}\circ \partial_{(0)}$ denotes the modular differential operator built out of the Serre derivative $\partial_{(k)} \coloneqq q \partial_q + k E_2(\tau)$. Note that the equation contains only $E_4(\tau)$ and $\Em{2}{1}$ (see \eqref{eq:Epm-shorthand} for the definition), which generate all the $\Gamma^0(2)$-modular forms; for example,
\begin{equation}
    E_6(\tau) = \frac{8}{35}\Em{2}{1}^3 - \frac{6}{7}E_4(\tau) \Em{2}{1} \ .
\end{equation}

We find that the minimal-order unflavored MLDE with Wronskian index zero is of \emph{order four},
\begin{align}\label{eq:MLDE-order4}
  0 = \Big(&D_{q}^{(4)}
    + 4\Em{2}{1}D_{q}^{(3)}
    - 10\Em{2}{1}^{2}D_{q}^{(2)}
    - 50E_{4}(\tau)D_{q}^{(2)}
    +100E_{4}(\tau)\Em{2}{1}D_{q}^{(1)}
    \nonumber\\
    &
    + 150E_{4}(\tau)\Em{2}{1}^{2}
    - 60\Em{2}{1}^{3}D_{q}^{(1)}
    -375E_{4}(\tau)^{2}
    - 15\Em{2}{1}^{4}\Big)\mathcal{I} \ .
\end{align}
We look for solutions as $q$-series of the form
\begin{equation}
\label{eq:Frobenius}
  \operatorname{ch}(\tau) = q^{\alpha}\sum_{n=0}^{\infty} a_{n}\,q^{n/2}, \qquad a_{0}\neq 0 \ .
\end{equation}
The resulting indicial equation has four roots, listed here with multiplicity:
\begin{equation}
\label{eq:indicial-roots}
  \alpha\;=\;\frac{5}{8}\,,\qquad -\frac{1}{24}\,,\qquad \left[\frac{1}{8}\right]_2 \ .
\end{equation}
The repeated root at $\alpha = \frac{1}{8}$ implies the existence of exactly one\footnote{This fact will be useful for analyzing the closed form of the flavored Schur index.} logarithmic solution. It is straightforward to identify the characters for these indicial roots with \eqref{eq:SchurClosedForm}, \eqref{eq:nonvac-characters-1}, \eqref{eq:nonvac-characters-2},  respectively. 
In particular, the solution with leading exponent $\alpha = \frac{5}{8}$ is precisely the unflavored Schur index computed in \eqref{eq:SchurClosedForm}.

The last unflavored solution must be logarithmic and can be determined from the modular transformations of the solutions already found. Explicitly, we compute the $STS$ transformation for all non-logarithmic solutions,
\begin{equation}
\label{eq:S-chm124}
  S\,\operatorname{ch}_{-\frac{1}{24}}
  = \frac{2}{\sqrt{3}}\,\frac{\eta(q^{\frac{2}{3}})^{2}}{\eta(q^{\frac{1}{3}})\,\eta(q)}
  = \frac{2}{\sqrt{3}}\!\left(
    \frac{\eta(q^{2})\,\eta(q^{3})^{2}}{\eta(q)^{2}\,\eta(q^{6})}
    + \frac{\eta(q^{6})^{2}}{\eta(q)\,\eta(q^{3})}
  \right),
\end{equation}
where the second equality is particularly convenient for iterating further modular transformations, leading to
\begin{align}
      STS\,\operatorname{ch}_{-\frac{1}{24}}
  &= \frac{1}{\sqrt{3}}\,e^{\frac{\pi i}{6}}\,\operatorname{ch}_{-\frac{1}{24}}
    + \frac{2}{\sqrt{3}}\,e^{-\frac{\pi i}{6}}\,\operatorname{ch}_{\frac{1}{8}},
\end{align}
and similarly
\begin{equation}
\label{eq:STS-on-etaquotients}
\begin{aligned}
  STS\,\operatorname{ch}_{\frac{1}{8}} &= \frac{1}{\sqrt{3}}\,e^{-\frac{\pi i}{6}}\,\operatorname{ch}_{-\frac{1}{24}}
    + \frac{i}{\sqrt{3}}\,\operatorname{ch}_{\frac{1}{8}}.
\end{aligned}
\end{equation}

The transformation of $\operatorname{ch}_{\textrm{vac}} = \mathcal{I}$ is qualitatively different: under $S$, it acquires a term linear in $\tau$,
signaling the logarithmic sector predicted by the repeated indicial root \eqref{eq:indicial-roots}. Explicitly,
\begin{equation}
\label{eq:S-ch58}
  S\,\operatorname{ch}_{\textrm{vac}}
  = -\frac{1}{3\sqrt{3}}\!\left(
    \frac{\eta(q^{2})\,\eta(q^{3})^{2}}{\eta(q)^{2}\,\eta(q^{6})}
    - \frac{2\,\eta(q^{6})^{2}}{\eta(q)\,\eta(q^{3})}
  \right)
  - \frac{i\tau}{6}\,\frac{\eta(q)^{4}}{\eta(q^{2})^{2}}.
\end{equation}
Further $T$, $S$ transformation leads to the logarithmic solution to the MLDE,
\begin{equation}
\label{eq:chlog-def}
  \operatorname{ch}_{\log} \coloneqq STS\,\operatorname{ch}_{\textrm{vac}}
  = \frac{1-\tau}{3}\,\frac{\eta(q)^{4}}{\eta(\sqrt{q})^{2}}
    -\frac{i}{3\sqrt{3}}\,\operatorname{ch}_{\frac{1}{8}}
    -\frac{e^{-\frac{i\pi}{6}}}{3\sqrt{3}}\,\operatorname{ch}_{-\frac{1}{24}} .
\end{equation}
The solutions $\operatorname{ch}_{-\frac{1}{24}}$, $\operatorname{ch}_{\frac{1}{8}}$, $\operatorname{ch}_{\textrm{vac}}$, and $\operatorname{ch}_{\log}$ form a basis of the solution space, which closes under the $\Gamma^0(2)$ modular group:
\begin{equation}
\label{eq:STS-on-chlog}
  STS\,\operatorname{ch}_{\log}
  = -\operatorname{ch}_{\textrm{vac}} + 2\,\operatorname{ch}_{\log}
    + \tfrac{1}{3}\,e^{-\frac{\pi i}{3}}\,\operatorname{ch}_{-\frac{1}{24}}
    + \tfrac{2}{3}\,e^{\frac{2\pi i}{3}}\,\operatorname{ch}_{\frac{1}{8}}.
\end{equation}
In the ordered basis $\bigl(\operatorname{ch}_{\textrm{vac}},\,\operatorname{ch}_{\log},\,
\operatorname{ch}_{-\frac{1}{24}},\,\operatorname{ch}_{\frac{1}{8}}\bigr)$, the transformation $STS$ is represented by the matrix
\begin{equation}
\label{eq:STS-matrix}
  STS =
  \begin{pmatrix}
    0 & 1 & 0 & 0 \\
    -1 & 2 & \frac{1}{3}e^{-\frac{\pi i}{3}} & \frac{2}{3}e^{\frac{2\pi i}{3}} \\
    0 & 0 & \frac{1}{\sqrt{3}}e^{\frac{\pi i}{6}} & \frac{2}{\sqrt{3}}e^{-\frac{\pi i}{6}} \\
    0 & 0 & \frac{1}{\sqrt{3}}e^{-\frac{\pi i}{6}} & \frac{i}{\sqrt{3}}
  \end{pmatrix}.
\end{equation}

\subsection{\texorpdfstring{$\mathcal{W}_{\mathbb{Z}_3}$}{WZ3} algebra, its null states, and flavored MLDE}\label{sec:WZ3-VOA}

We now wish to understand the representation theory of the associated VOA $\mathcal{W}_{\mathbb{Z}_3}$.
To this end, we first take a closer look at the algebra itself, which was studied in detail in \cite{Nishinaka:2016hbw} and later in \cite{Bonetti:2018fqz}.

\begin{table}[!ht]
\centering
\begin{tabular}{c c c c c c}
    \toprule
                            & $h$           & $m$            & $r$            & $R$           &$(-1)^F$\\
    \midrule
    $J$           & $1$           & $0$            & $0$            & $1$           & $1$    \\
    $T$           & $2$           & $0$            & $0$            & $1$           &$1$     \\
    $W$           & $\frac{3}{2}$ & $3$  & $0$            & $\frac{3}{2}$ & $1$    \\
    $\tilde W$          & $\frac{3}{2}$ & $-3$ & $0$            & $\frac{3}{2}$ &  $1$   \\
    $G$           & $\frac{3}{2}$ & $-1$ & $\frac{1}{2}$  & $1$           &$-1$    \\
    $\tilde{G}$   & $\frac{3}{2}$ & $1$  & $-\frac{1}{2}$ & $1$           &$-1$    \\
    $G_W$         & $2$           & $2$            & $\frac{1}{2}$  & $\frac{3}{2}$ &$-1$    \\
    $\tilde G_{\tilde{W}}$ & $2$           & $-2$           & $-\frac{1}{2}$ & $\frac{3}{2}$ &$-1$    \\
    \bottomrule
\end{tabular}
\caption{The quantum numbers of the generators of $\mathcal{W}_{\mathbb{Z}_3}$. Here, $r$ and $R$ are the quantum numbers of the corresponding Schur operators in four dimensions.}
\label{tab:WZ3-generator-quantum-numbers}
\end{table}

The algebra has central charge $c_{\mathrm{2d}} = -15$, following from the 4d central charge $c_{\mathrm{4d}} = \frac{5}{4}$. It is strongly generated by $T, J, G, \tilde G, W, \tilde W, G_W, \tilde G_{\tilde W}$, which are all Virasoro primaries (besides $T$ itself) with conformal weights $2, 1, \frac{3}{2}, \frac{3}{2}, \frac{3}{2}, \frac{3}{2}, 2, 2$; their quantum numbers are collected in Table~\ref{tab:WZ3-generator-quantum-numbers}. As the theory is a genuine $\mathcal{N} = 3$ SCFT, the associated VOA contains the 2d $\mathcal{N} = 2$ superconformal algebra as a subalgebra, generated by
\begin{equation}
    T, \quad  J, \quad G, \quad \tilde{G} \ .
\end{equation}
In particular, the current $J$ generates a $U(1)_{-5}$ subalgebra, under which $G, \tilde G$ carry charge $-1, +1$. The 2d $\mathcal{N} = 2$ supersymmetry descends from the third pair of supercharges $Q^3_\alpha, \tilde Q^3_{\dot \alpha}$ from four dimensions, which act on the bosonic Schur operators to create fermionic ones. The additional strong generators\footnote{Strong generators generate all operators in the VOA through normal ordered product and derivatives, which themselves cannot be written as normal ordered product or derivative of other operators.} organize into (anti-)chiral multiplets with $J$-charges $(3,2)$ and $(-3, -2)$ of the superconformal algebra,
\begin{equation}
    \{W, G_W\}, \qquad \{\tilde W, \tilde G_{\tilde W}\} \ , \quad G(z)W(w) \sim \frac{G_W(w)}{z - w}, \quad \tilde G(z)\tilde W(w) \sim \frac{\tilde G_{\tilde W}(w)}{z - w} \ ,
\end{equation}
and trivial $\tilde G(z) W(w)$ and $G(z)\tilde W(w)$ OPEs. The absence of second or higher-order poles in the OPEs of $W, \tilde W$ with $G, \tilde G$ implies that they are $\mathcal{N} = 2$ superconformal primaries, while the triviality of the $\tilde G(z)W(w)$ and $G(z)\tilde W(w)$ OPEs implies that $W$ and $\tilde W$ are $\mathcal{N} = 2$ chiral and anti-chiral operators, respectively. Besides the standard OPEs fixed by the conformal weights and $J$-charges, the remaining non-trivial OPEs are
\begin{align}
    W(z)\tilde W(w) \sim & \ \frac{-20/9}{(z - w)^3} + \frac{4}{3} \frac{J(w)}{(z - w)^2} + \frac{\frac{2}{3}T(w) - \frac{1}{3}(JJ)(w) + \frac{2}{3}J'(w)}{z - w}\ ,\cr
    G(z) \tilde G(w) \sim & \ \frac{5 }{(z - w)^3} + \frac{J(w)}{(z - w)^2} + \frac{-T(w) + \frac{1}{2}J'(w)}{z - w} \ ,\cr
    \tilde G(z) G_W(w) \sim & \frac{-3 W(w)}{(z - w)^2} - \frac{W'(w)}{z - w} \ ,\qquad
    G(z) \tilde G_{\tilde W}(w) \sim  \frac{-3\tilde W(w)}{(z - w)^2} - \frac{\tilde W'(w)}{z - w} \ ,\cr
    \tilde W(z) G_W(w) \sim & \ \frac{\frac{4}{3}G(w)}{(z - w)^2} + \frac{\frac{2}{3}(JG)(w) + \frac{2}{3}G'(w)}{z - w} \ ,\cr
    W(z) \tilde G_{\tilde W}(w) \sim & \ \frac{ -\frac{4}{3}\tilde G(w) }{(z - w)^2}
    + \frac{\frac{2}{3}(J \tilde G)(w) - \frac{2}{3} \tilde G'(w)}{z - w}\ ,\cr
    G_W(z) \tilde G_{\tilde W}(w) \sim & \ \frac{- \frac{20}{3}}{(z - w)^4}
    + \frac{8}{3} \frac{J(w)}{(z - w)^3}
    + \frac{2T(w) - \frac{1}{3}(JJ)(w) + \frac{4}{3}J'(w)}{(z - w)^2}\cr
    & \ + \frac{\frac{2}{3}(G\tilde G)(w) - \frac{2}{3}(JT)(w) - \frac{1}{3}(JJ')(w)+ \frac{4}{3}T'(w) + \frac{1}{3} J''(w)}{z - w} \ .
\end{align}
The algebra $\mathcal{W}_{\mathbb{Z}_3}$ admits a free-field realization in terms of a single $bc\beta \gamma$ system \cite{Bonetti:2018fqz},
\begin{align}
    J = & \ 2 (bc) + 3 (\beta \gamma)\ ,
    & T = & \ -2(bc') - (b'c) - \frac{3}{2} (\beta \gamma') - \frac{1}{2} \beta' \gamma \ ,\cr
    G = & \ (\gamma b)\ , \qquad
    & \tilde G = & \ 3 (\beta c') + 2 (\beta' c) \ ,\cr
    W = & \ \beta, \qquad
    & \tilde W = & \ \beta(\beta(\gamma (\gamma \gamma))) + 2 (\beta (\gamma (bc))) - 4 (\beta (\gamma' \gamma)) - \frac{4}{3} (\gamma(b c'))\cr
    & & & + \frac{2}{3} (\gamma (b'c))
    + \frac{2}{3} (\beta'(\gamma \gamma)) - \frac{8}{3} (\gamma'(bc)) + \frac{10}{9}\gamma''\ ,\cr
    G_W = & \ b\ ,
    & \tilde G_{\tilde W} = & \ - \frac{8}{3} (b (c''c)) - 3 (\beta (\beta(\gamma(\gamma c')))) + 4 (\beta (\gamma (b(c'c))))
    + 4 (\beta (\gamma c'')) \cr
    & & & + 4 (\beta (\gamma' c'))
    + \frac{2}{3}(b' (c'c))
    - 2 (\beta'(\beta (\gamma(\gamma c))))
    + \frac{8}{3} (\beta' (\gamma'c)) \cr
    & & & - \frac{2}{3} (\beta'' (\gamma c))
    - \frac{10}{9}c''' \ . 
\end{align}
The OPEs between the ghosts are chosen to be
\begin{equation}
\beta(z)\,\gamma(w)\sim - \frac{1}{z-w},\qquad b(z)\,c(w)\sim \frac{1}{z-w},
\end{equation}
and their quantum numbers are listed in Table~\ref{tab:ghost-qnums}.

\PaperBeginGhostTable
  \PaperGhostTableSpacing
  \centering
  \begin{tabular}{c c c c c}
    \toprule
      & $h$ & $\mathsf{G}$ & $r$ & $R$ \\
    \midrule
    $\beta_{\ell}$   & $\frac{1}{2} d_{\ell}$        & $d_{\ell}$        & $0$              & $\frac{1}{2} d_{\ell}$ \\
    $b_{\ell}$       & $\frac{1}{2}(d_{\ell}+1)$     & $(d_{\ell}-1)$     & $+\frac{1}{2}$   & $\frac{1}{2}d_{\ell}$ \\
    $c_{\ell}$       & $-\frac{1}{2}(d_{\ell}-1)$    & $-(d_{\ell}-1)$    & $-\frac{1}{2}$   & $1-\frac{1}{2}d_{\ell}$ \\
    $\gamma_{\ell}$  & $1-\frac{1}{2}d_{\ell}$       & $-d_{\ell}$        & $0$              & $1-\frac{1}{2}d_{\ell}$ \\
    $\partial$       & $1$                           & $0$                           & $0$              & $0$ \\
    \bottomrule
  \end{tabular}
  \caption{Quantum numbers for the $\beta_\ell\gamma_\ell b_\ell c_\ell$ ghost systems realizing the VOA associated to a complex reflection group $\mathsf{G}$.
  Here, $d_\ell$ are the degrees of the fundamental invariants of $\mathsf{G}$; for $\mathbb Z_k$, there is a single invariant, so $\ell = 1$ and $d_1=k$.}
  \label{tab:ghost-qnums}
\end{table}

In four dimensions, the Schur operators corresponding to $J, W, \tilde W$ are Higgs-branch generators, obeying the chiral ring relation $W \tilde W \sim J^3$. The associated variety is $\mathbb{C}^2/\mathbb{Z}_3$, with $\mathbb{Z}_3$ acting on the two complex planes with opposite phases, $x \to e^{2\pi i /3} x$, $y \to e^{-2\pi i /3}y$. Its coordinate ring is generated by $w = x^3$, $\tilde w = y^3$, and $j = xy$, subject to the relation $w \tilde w = j^3$. At the level of the VOA, this relation is realized by a null state of conformal weight three \cite{Bonetti:2018fqz}:
\begin{equation}\label{eq:N3}
    \mathcal{N}_3 = (W \tilde W) - \frac{1}{27}(J(JJ)) + \frac{4}{9} (TJ) + \frac{1}{3}(J'J) - \frac{2}{9}T' - \frac{13}{27} J'' + \frac{2}{9} (G\tilde G) \ .
\end{equation}
The null state is a singular vector, annihilated by all the positive modes,
\begin{align}
    L_{n \ge 1}\mathcal{N}_3 = & \ J_{n \ge 1} \mathcal{N}_3 = G_{n > 0} \mathcal{N}_3 = \tilde G_{n > 0} \mathcal{N}_3 \cr
    = & \ W_{n > 0} \mathcal{N}_3 = \tilde W_{n > 0}\mathcal{N}_3 = (G_W)_{n \ge 1} \mathcal{N}_3 = (\tilde G_{\tilde W})_{n \ge 1} \mathcal{N}_3 = 0 \ .
\end{align}
This null state can be completely fixed by its leading terms $W \tilde W - \frac{1}{27}(J(JJ))$ together with the superconformal primary condition, and it vanishes identically in the $bc\beta \gamma$ realization. It is expected to generate the maximal ideal of $\mathcal{W}_{\mathbb{Z}_3}$: further null states of the VOA are obtained by taking normally ordered products, derivatives, and OPE poles of $\mathcal{N}_3$ with elements of the VOA. There are two more weight-three null states \cite{Bonetti:2018fqz}, given by $\mathcal{N}_{3,+2} = (JG_W) - 3(GW) + G_W'$ and $\mathcal{N}_{3,-2} = (J\tilde G_{\tilde W})  + 3(\tilde G \tilde W) - \tilde G_{\tilde W}'$, where the second subscript records the $J_0$-charge. The three null states are related by taking OPE poles with $G_W$ and $\tilde G_{\tilde W}$,
\begin{equation}
  \bigl\{G_{W}\, \mathcal{N}_3\bigr\}_{2}
  \;=\;\frac{- 2}{9}\,\mathcal{N}_{3,+2}\,,
  \qquad
  \bigl\{\tilde G_{\tilde W}\,\mathcal{N}_3\bigr\}_{2}
  \;=\;\frac{4}{9}\,\mathcal{N}_{3,-2}\, .
\end{equation}

To analyze the full space of (super-)characters of $\mathcal{W}_{\mathbb{Z}_3}$, we construct the flavored MLDEs from null states of higher conformal weight using Zhu's recursion formula \cite{Gaberdiel:2008pr,zhu1996modular,Beem:2017ooy,Pan:2021ulr,Pan:2023jjw,Zheng:2022zkm,Pan:2024dod}. The first candidate comes from the singular vector $\mathcal{N}_3$ itself,
\begin{align}\label{eq:D3}
\mathcal{D}_3 = {}& D_{a} D_q^{(1)} + \Big[\frac{1}{2} \Em{1}{a} + \frac{3}{2} \Em{1}{a^3}\Big]D_q^{(1)} -\frac{1}{12} D_{a}^3 -\frac{3}{4} \Em{1}{a^3} D_{a}^2 \nonumber\\
& + \Big[\frac{9}{4} E_2(\tau) + \frac{1}{2} \Em{2}{a} + 3 \Em{2}{a^3}\Big]D_{a} \nonumber\\
& + \Big[\frac{15}{4} E_2(\tau) \Em{1}{a^3} - \frac{5}{2} \Em{3}{a} - 5 \Em{3}{a^3}\Big] \ .
\end{align}
Here, $D_a = a \partial_a$, and $D_a^n$ means $(a\partial_a)^n$. The next candidates are the ten null states at conformal weight four, four of which are neutral under $J$. All of them vanish identically in the $bc \beta \gamma$ realization. They are descendants of $\mathcal{N}_3$, spanned by $(J \mathcal{N}_3)$, $\mathcal{N}_3'$, $\{\tilde G \{G \mathcal{N}_3\}_1\}_1$, $\{\tilde W \{W\mathcal{N}_3\}_1\}_1$,
\begin{align}
    \mathcal{N}^{(1)}_{4} = (J \mathcal{N}_3), \quad
    \mathcal{N}_{4}^{(2)} = - \frac{3}{2} \mathcal{N}'_3 - \frac{9}{4} \{\tilde G \{G \mathcal{N}_3\}_1\}_1 - \frac{9}{8} \{\tilde W \{W \mathcal{N}_3\}_1\}_1 \ , \cr
    \mathcal{N}_4^{(3)} = - 9 \mathcal{N}'_3 - \frac{9}{2} \{\tilde G \{G \mathcal{N}_3\}_1\}_1 - \frac{27}{4} \{\tilde W \{W \mathcal{N}_3\}_1\}_1 \ , \cr
    \mathcal{N}_4^{(4)} = - \frac{9}{2} \{\tilde G \{G \mathcal{N}_3\}_1\}_1 - \frac{27}{4} \{\tilde W \{W \mathcal{N}_3\}_1\}_1 \ .
\end{align}
Among the four neutral null states at conformal weight four, the
particular combination needed in the ordinary Zhu-algebra analysis
below is
\begin{equation}
\begin{aligned}
\mathcal N_4
&:=
3\mathcal N_4^{(1)}
-\frac{1}{3}\mathcal N_4^{(4)}
\\
&=
3(J\mathcal N_3)
+\frac{3}{2}
 \bigl\{\tilde G\{G\mathcal N_3\}_{1}\bigr\}_{1}
+\frac{9}{4}
 \bigl\{\tilde W\{W\mathcal N_3\}_{1}\bigr\}_{1}.
\end{aligned}
\label{eq:N4-descendant-combination}
\end{equation}
Substituting \eqref{eq:N3} for $\mathcal N_3$ and reducing
all iterated products to the normal-ordering convention used throughout
this section gives
\begin{align}
\mathcal N_4={}&
-\frac{1}{9}\bigl(J(J(JJ))\bigr)
+3\bigl(J(W\tilde W)\bigr)
+\bigl(T(JJ)\bigr)
+\frac{7}{6}\bigl(J'(JJ)\bigr)
\nonumber\\
&-2(T'J)
-\frac{22}{9}(J''J)
+\frac{23}{18}J'''
-\bigl(G_W\tilde G_{\tilde W}\bigr)
-(TJ')
\nonumber\\
&-3(W\tilde W')
-\frac{5}{6}(J'J')
+T'' .
\label{eq:N4-explicit-normal-ordered}
\end{align}
This is the representative whose ordinary Zhu projection is denoted
by $[\mathcal N_4]$ below.

The above weight-four null states contribute three flavored MLDEs $\mathcal{D}_{4}^{(i)}\operatorname{ch}=0$ ($i=1,2,3$), where
\begin{align}
\mathcal{D}_{4}^{(1)} = {}& D_{a}^2 D_q^{(1)} + \Big[\frac{1}{2} \Em{1}{a} + \frac{3}{2} \Em{1}{a^3}\Big]D_{a} D_q^{(1)} \nonumber\\
& + \Big[-5 E_2(\tau) - \Em{2}{a} - 9 \Em{2}{a^3} - \frac{1}{2} \Em{1}{a}^2 - \frac{9}{2} \Em{1}{a^3}^2\Big]D_q^{(1)} -\frac{1}{12} D_{a}^4 \nonumber\\
& -\frac{3}{4} \Em{1}{a^3} D_{a}^3 + \Big[\frac{9}{2} E_2(\tau) + \frac{1}{2} \Em{2}{a} + \frac{15}{2} \Em{2}{a^3} + \frac{9}{4} \Em{1}{a^3}^2\Big]D_{a}^2 \nonumber\\
& + \Big[\frac{1}{2} E_2(\tau) \Em{1}{a} + \frac{51}{4} E_2(\tau) \Em{1}{a^3} - 4 \Em{3}{a} - 32 \Em{3}{a^3} \nonumber\\
& \qquad - \frac{1}{2} \Em{2}{a} \Em{1}{a} - 9 \Em{2}{a^3} \Em{1}{a^3}\Big]D_{a} \nonumber\\
& + \Big[\frac{35}{2} E_4(\tau) + 10 \Em{4}{a} + 60 \Em{4}{a^3} - \frac{45}{4} E_2(\tau)^2 - \frac{5}{2} E_2(\tau) \Em{2}{a} \nonumber\\
& \qquad - \frac{75}{2} E_2(\tau) \Em{2}{a^3} - \frac{45}{4} E_2(\tau) \Em{1}{a^3}^2 + \frac{5}{2} \Em{3}{a} \Em{1}{a} \nonumber\\
& \qquad + 15 \Em{3}{a^3} \Em{1}{a^3}\Big] \ ,
\end{align}
\begin{align}
\mathcal{D}_{4}^{(2)} = {}& \Big[\frac{2}{3} \Em{1}{a} - \frac{4}{3} \Ep{1}{a^2}\Big]D_{a} D_q^{(1)} + D_q^{(2)} \nonumber\\
& + \Big[-\frac{2}{3} \Em{1}{a}^2 + \frac{4}{3} \Ep{1}{a^2} \Em{1}{a} - 2 \Em{2}{a} + 4 \Ep{2}{a^2}\Big]D_q^{(1)} \nonumber\\
& + \Big[\frac{2}{3} \Em{2}{a} - \frac{2}{3} \Ep{2}{a^2}\Big]D_{a}^2 \nonumber\\
& + \Big[\frac{2}{3} E_2(\tau) \Em{1}{a} - \frac{4}{3} E_2(\tau) \Ep{1}{a^2} - \frac{20}{3} \Em{3}{a} + \frac{16}{3} \Ep{3}{a^2} \nonumber\\
& \qquad - \frac{2}{3} \Em{2}{a} \Em{1}{a} + \frac{4}{3} \Em{2}{a} \Ep{1}{a^2}\Big]D_{a} \nonumber\\
& + \Big[- 10 E_4(\tau) + \frac{70}{3} \Em{4}{a} - \frac{40}{3} \Ep{4}{a^2} - \frac{10}{3} E_2(\tau) \Em{2}{a} + \frac{10}{3} E_2(\tau) \Ep{2}{a^2} \nonumber\\
& \qquad + \frac{10}{3} \Em{3}{a} \Em{1}{a} - \frac{20}{3} \Em{3}{a} \Ep{1}{a^2}\Big] \ ,
\end{align}
\begin{align}
\mathcal{D}_{4}^{(3)} = {}& D_{a}^2 D_q^{(1)} + \Big[2 \Em{1}{a} - 2 \Ep{1}{a^2}\Big]D_{a} D_q^{(1)} \nonumber\\
& + \Big[-5 E_2(\tau) - 2 \Em{1}{a}^2 + 2 \Ep{1}{a^2} \Em{1}{a} - 4 \Em{2}{a} + 6 \Em{2}{a^3} + 6 \Ep{2}{a^2}\Big]D_q^{(1)} \nonumber\\
& + \Big[2 \Em{2}{a} - 3 \Em{2}{a^3} - \Ep{2}{a^2}\Big]D_{a}^2 \nonumber\\
& + \Big[2 E_2(\tau) \Em{1}{a} - 2 E_2(\tau) \Ep{1}{a^2} - 16 \Em{3}{a} + 24 \Em{3}{a^3} + 8 \Ep{3}{a^2} \nonumber\\
& \qquad - 2 \Em{2}{a} \Em{1}{a} + 2 \Em{2}{a} \Ep{1}{a^2}\Big]D_{a} \nonumber\\
& + \Big[2 D_{a}^2 E_2(\tau) + 15 E_4(\tau) + 40 \Em{4}{a} - 60 \Em{4}{a^3} - 20 \Ep{4}{a^2} - 5 E_2(\tau)^2 \nonumber\\
& \qquad - 10 E_2(\tau) \Em{2}{a} + 15 E_2(\tau) \Em{2}{a^3} + 5 E_2(\tau) \Ep{2}{a^2} + 10 \Em{3}{a} \Em{1}{a} \nonumber\\
& \qquad - 10 \Em{3}{a} \Ep{1}{a^2}\Big] \ .
\end{align}
The fourth null gives the same equation as the third.

We can also go up in conformal weight to find more null states with zero $J$-charge using the free-field realization, to generate flavored MLDEs. We find 14 null states at weight 5, 37 null states at weight 6, and 133 null states at weight 7. For instance, one useful weight-6 null state is
\begin{equation}
\begin{aligned}
\mathcal{N}_6 ={}&
\frac{2}{3}\bigl(\partial G\,\partial^2\tilde{G}\bigr)
+\frac{2}{3}\bigl(\partial G\,((\partial\tilde{G})J)\bigr)
+\frac{2}{9}\bigl(\partial G\,(\tilde{G}\,\partial J)\bigr)
\\
&+\frac{1}{9}\bigl(\partial G\,(\tilde{G}(JJ))\bigr)
-\frac{2}{3}\bigl(\partial G\,(\tilde{G}T)\bigr)
+\bigl(\partial G\,(\tilde{G}_{\tilde{W}}W)\bigr) \ .
\end{aligned}
\end{equation}
In total, they lead to a set of flavored MLDEs, which impose strong constraints on the space of characters.

In the end, we obtain the following non-logarithmic $q$-series solutions, organized according to their indicial roots $\alpha = \frac{5}{8}, [\frac{1}{8}]_2, - \frac{1}{24}$. First, we have for the indicial root $\alpha = \frac{5}{8}$, \emph{i.e.}, the vacuum character,
\begin{equation}
    \operatorname{ch}_{\textrm{vac}}(q,a)
    =
    q^{5/8}
    \big(
        \sum_{n=0}^{8} V_n(a)q^{n/2}
        +\cO(q^{9/2})
    \big)\ ,
\end{equation}
where the coefficients $V_n(a)$ are written in terms of the $SU(2)$ characters $\chi_\ell(a) = \chi^{\mathrm{spin}}_{\ell/2}(a) = a^\ell+a^{\ell-2}+\cdots+a^{-\ell}$,
\begin{align*}
V_0 &= \chi_0,                 & V_1 &= 0,                     & V_2 &= \chi_0,\\
V_3 &= \chi_3-2\chi_1,          & V_4 &= 4\chi_0-\chi_2,        & V_5 &= 2\chi_3-4\chi_1,\\
V_6 &= \chi_6-2\chi_4-\chi_2+8\chi_0,
& V_7 &= -\chi_5+6\chi_3-10\chi_1,\\
V_8 &= 2\chi_6-5\chi_4-\chi_2+17\chi_0 \ .
\end{align*}
This solution is manifestly the character of the vacuum module of the associated VOA and is expected to coincide with the Schur index.

The next one is $\alpha = - \frac{1}{24}$,
\begin{equation}
    \operatorname{ch}_{-\frac{1}{24}}(q,a)
    =
    q^{-1/24}
    \big(
        \sum_{n=0}^{8} U_n(a)q^{n/2}
        +\cO(q^{9/2})
    \big) \ ,
\end{equation}
where
\begin{align*}
U_0 &= \chi_0,\\
U_1 &= \chi_3-2\chi_1,\\
U_2 &= \chi_6-2\chi_4+4\chi_0,\\
U_3 &= \chi_9-2\chi_7+4\chi_3-6\chi_1,\\
U_4 &=
\chi_{12}-2\chi_{10}+4\chi_6-6\chi_4+\chi_2+10\chi_0,\\
U_5 &=
\chi_{15}-2\chi_{13}+4\chi_9-6\chi_7
+\chi_5+11\chi_3-16\chi_1,\\
U_6 &=
\chi_{18}-2\chi_{16}+4\chi_{12}-6\chi_{10}
+\chi_8+11\chi_6-17\chi_4
+2\chi_2+25\chi_0,\\
U_7 &=
\chi_{21}-2\chi_{19}+4\chi_{15}-6\chi_{13}
+\chi_{11}+11\chi_9-17\chi_7\\
&\hspace{1.5cm}
+2\chi_5+28\chi_3-38\chi_1,\\
U_8 &=
\chi_{24}-2\chi_{22}+4\chi_{18}-6\chi_{16}
+\chi_{14}+11\chi_{12}-17\chi_{10}\\
&\hspace{1.5cm}
+2\chi_8+28\chi_6-41\chi_4
+5\chi_2+57\chi_0.
\end{align*}
There are two $\alpha = \frac{1}{8}$ non-logarithmic solutions:
\begin{equation}
    \operatorname{ch}_{\frac{1}{8}}^{\pm}(q,a)
    =
    q^{1/8}
    \left(
        \sum_{n=0}^{8} P_n^{\pm}(a)q^{n/2}
        +\cO(q^{9/2})
    \right),
    \qquad
    P_n^{-}(a)=P_n^{+}(a^{-1}).
\end{equation}
In particular,
\begin{align*}
P_0^{+} &= a,\\
P_1^{+} &= a^4-a^2 - 1,\\
P_2^{+} &= 2a-a^3-a^5+a^7,\\
P_3^{+} &=
a^{10}-a^8-a^6+3a^4-2a^2-2+a^{-2},\\
P_4^{+} &=
a^{13}-a^{11}-a^9+3a^7-3a^5-2a^3+6a-a^{-3},\\
P_5^{+} &=
a^{16}-a^{14}-a^{12}+3a^{10}-3a^8-2a^6
+8a^4-5a^2-6+3a^{-2},\\
P_6^{+} &=
a^{19}-a^{17}-a^{15}+3a^{13}-3a^{11}-2a^9
+9a^7-7a^5
-6a^3+13a-a^{-1}-3a^{-3}+a^{-5},\\
P_7^{+} &=
a^{22}-a^{20}-a^{18}+3a^{16}-3a^{14}-2a^{12}
+9a^{10}\\
&\hspace{1.5cm}
-8a^8-6a^6+19a^4-10a^2-12+8a^{-2}-a^{-6},\\
P_8^{+} &=
a^{25}-a^{23}-a^{21}+3a^{19}-3a^{17}-2a^{15}
+9a^{13} -8a^{11}-6a^9+21a^7-17a^5-13a^3+29a\\
&\hspace{1.5cm}
-2a^{-1}-8a^{-3}+3a^{-5}.
\end{align*}
From this computation, we therefore predict four highest-weight modules of $\mathcal{W}_{\mathbb{Z}_3}$, whose quantum numbers are summarized as in Table~\ref{tab:WZ3-modules}.
The fact that $\alpha = 1/8$ is a double indicial root implies the existence of at least one logarithmic character. We will construct such a solution momentarily.

\subsection{Flavored index from modularity}

The four non-logarithmic character functions found above have finite
$a\to1$ limits:
\begin{equation}
    \operatorname{ch}_{-\frac{1}{24}} (a \to 1, q) = \frac{\eta(\frac{3}{2}\tau)^2}{\eta(\tau)\eta(3\tau)}, \qquad
    \operatorname{ch}_{\frac{1}{8}}^\pm(a \to 1, q) = \frac{\eta(3\tau)^2}{\eta(\tau)^2} \frac{\eta(\frac{1}{2}\tau)}{ \eta(\frac{3}{2}\tau)} \ .
\end{equation}
Matching the $q$-expansions of the previous subsection against ratios of theta
and eta functions, we identify the three non-vacuum characters in closed form:
\begin{align}\label{eq:flavored-non-vacuum-characters-WZ3-1}
        \operatorname{ch}_{-\frac{1}{24}}(a;q)
        &=q^{-\frac{1}{3}}\frac{\eta(3\tau)^3}{\eta(\tau)^2}
        \frac{\vartheta_4\left(\mathfrak a|\tau\right)}
        {\vartheta_4\left(3\mathfrak a-\tau|3\tau\right)
         \vartheta_4\left(-3\mathfrak a-\tau|3\tau\right)},\\
        \operatorname{ch}_{\frac{1}{8}}^+(a;q)
        &=q^{-\frac{1}{6}}a\frac{\eta(3\tau)^3}{\eta(\tau)^2}
        \frac{\vartheta_4\left(\mathfrak a|\tau\right)}
        {\vartheta_4\left(3\mathfrak a-\tau|3\tau\right)
         \vartheta_4\left(3\mathfrak a|3\tau\right)},\label{eq:flavored-non-vacuum-characters-WZ3-2}\\
        \operatorname{ch}_{\frac{1}{8}}^-(a;q)
        &=q^{-\frac{1}{6}}a^{-1}\frac{\eta(3\tau)^3}{\eta(\tau)^2}
        \frac{\vartheta_4\left(\mathfrak a|\tau\right)}
        {\vartheta_4\left(-3\mathfrak a-\tau|3\tau\right)
         \vartheta_4\left(-3\mathfrak a|3\tau\right)},\label{eq:flavored-non-vacuum-characters-WZ3-3}
\end{align}
where $a = e^{2\pi i \mathfrak{a}}$.  The vacuum character itself is not of this
form, since it is the one solution whose $\Gamma^0(2)$ orbit contains a
logarithmic partner.  In the rest of this subsection, we determine the vacuum character, together
with that partner, and then work out the modular orbit of the full set of
characters.  Our strategy is to constrain the answer by modularity, using three
observations.

The first concerns the number of logarithmic solutions.  In the unflavoring
limit, one obtains exactly one logarithmic character starting from the vacuum,
namely $STS\operatorname{ch}_{\textrm{vac}} = \operatorname{ch}_{\log}$ given in
\eqref{eq:chlog-def}, and it is unique up to the addition of non-logarithmic
solutions.  Further $STS$ and $T^2$ transformations produce no new logarithmic
direction: they stay within the span of $\operatorname{ch}_{\log}$ and the
non-logarithmic characters.  The most natural assumption is
that this remains true in the flavored case: there is only one logarithmic
element in the $\Gamma^0(2)$-orbit of the flavored vacuum character.

The second comes from known Lagrangian indices.  The fact that the unflavored
characters $\operatorname{ch}_\alpha$ were obtained by contour integration
suggests that the flavored Schur index likewise originates from some integral
of elliptic functions.  The closed form of such an integral is always a sum of
residues of the integrand multiplied by polynomials in Eisenstein series.
Moreover, in all known cases, the residues themselves solve the flavored MLDEs
and are therefore linear combinations of module characters.

The third concerns the origin of the flavored MLDEs.  They are the result of
applying Zhu's recursion relations, so the $a$ in the Eisenstein series
$E_k^\pm (a^m)$ appearing there encodes the charges of the
generators of the associated VOA.  For the flavored Schur index
$\operatorname{ch}_{\textrm{vac}}$ to solve such a set of equations, it is natural
to assume that the Eisenstein series occurring in it carry the same generator
charges.

Combining the three observations, we assume that the flavored Schur index takes
the form
\begin{equation}
    \operatorname{ch}_{\textrm{vac}}(a, q)
    = \sum_{\alpha = - \frac{1}{24}, [\frac{1}{8}]^\pm} \operatorname{ch}_{\alpha}(a, q) P_{\alpha} (a, q)\ , \qquad
    P_\alpha (a, q) =  \sum_{m = 1}^3\sum_{\pm} \lambda_{\alpha m\pm} E_1^\pm(a^m) \ ,
\end{equation}
where $[\frac{1}{8}]^\pm$ denotes the pair
$\operatorname{ch}_{\frac{1}{8}}^\pm$, the upper characteristic runs
over both spin structures, and the $\lambda_{\alpha m\pm}$ are numerical
coefficients to be determined.  Each ingredient traces back to one of the
observations above: the expansion in module characters with Eisenstein
coefficients is the structure predicted by the second; the restriction to
$E_1$, rather than higher $E_k$, follows from the single logarithmic partner
assumed in the first; and $m = 1,2,3$ runs over the possible charges under $J$
as dictated by the third.  Here, we use the symmetry of the Eisenstein series to
take the charges positive.

Imposing the flavored MLDEs fixes all the $\lambda_{\alpha m\pm}$ uniquely.  The
coefficients multiplying $\operatorname{ch}_{-\frac{1}{24}}$ turn out to vanish, so
that only the pair $\operatorname{ch}^{\pm}_{\frac{1}{8}}$ contributes, and one obtains
the closed form
\begin{align}\label{eq:flavored-Schur-index-WZ3}
\operatorname{ch}_{\textrm{vac}}(a, q) = & \ \frac{
a\,\eta(3\tau)^3\,
\vartheta_4(\mathfrak{a})\,
}{
6q^{1/6}\eta(\tau)^2\,
\vartheta_4(3\mathfrak{a}|3\tau)\,
\vartheta_4(-\tau+3\mathfrak{a}|3\tau)
}\left(
-1
+2\Em{1}{a}
+2\Ep{1}{a^2}
\right)
\cr
&\quad
-\frac{
\eta(3\tau)^3\,
\vartheta_4(\mathfrak{a})\,
}{
6q^{1/6}a\eta(\tau)^2\,
\vartheta_4(3\mathfrak{a}|3\tau)\,
\vartheta_4(\tau+3\mathfrak{a}|3\tau)
}
\left(
1
+2\Em{1}{a}
+2\Ep{1}{a^2}
\right) \ .
\end{align}
As a check, the first few terms in the $q$-expansion of
\eqref{eq:flavored-Schur-index-WZ3} match the prediction of
\cite{Agarwal:2021oyl}.

It remains to compute the modular orbit of the characters.  The expressions
above involve theta functions at modulus $3\tau$, and rewriting their
transforms in the original basis requires the decomposition identities
\begin{equation}
    \vartheta_3(\mathfrak{z}|\tau) = \sum_{r \in \{\pm 1, 0\} } q^{\frac{r^2}{2}} z^r \vartheta_3(3 \mathfrak{z} + 3r \tau | 9\tau) \ , \quad \prod_{r \in \{\pm 1, 0\}} \frac{\vartheta_3(\mathfrak{z} + r(\frac{1}{3} + \tau)|\tau)}{\vartheta_3(3 \mathfrak{z} + 3r \tau|9\tau)}
    = e^{\frac{2\pi i}{3}} \frac{\eta(\tau)^3}{\eta(9\tau)^3} \ .
\end{equation}
Written out, the three cases of the first identity that we need are
\begin{align}
    \vartheta_3(\mathfrak{z} - (\frac{1}{3} + \tau)|\tau)
    = & \ \vartheta_3(3 \mathfrak{z} - 3 \tau|9\tau)
    + q^{- \frac{1}{2}} e^{\frac{4\pi i}{3}} z \vartheta_3(3 \mathfrak{z}|9\tau)
    + e^{\frac{2\pi i}{3}} z^2 \vartheta_3(3\mathfrak{z} + 3 \tau|9\tau) \ , \\
    \vartheta_3(\mathfrak{z}|\tau)
    = & \ q^{\frac{1}{2}} z^{-1}\vartheta_3(3 \mathfrak{z} - 3 \tau|9\tau)
    + \vartheta_3(3 \mathfrak{z}|9\tau)
    + q^{\frac{1}{2}} z \vartheta_3(3\mathfrak{z} + 3 \tau|9\tau) \ , \\
    \vartheta_3(\mathfrak{z} + (\frac{1}{3} + \tau)|\tau)
    = & \ e^{\frac{2\pi i}{3}} z^{-2} \vartheta_3(3 \mathfrak{z} - 3 \tau|9\tau)
    + q^{- \frac{1}{2}} e^{\frac{4\pi i}{3}} z^{-1} \vartheta_3(3 \mathfrak{z}|9\tau)
    + \vartheta_3(3\mathfrak{z} + 3 \tau|9\tau) \ .
\end{align}
With these formulas, the three non-vacuum characters are found to close among
themselves under $STS$,
\begin{align}
    STS\operatorname{ch}_{\frac{1}{8}}^+ = & \ 
      \frac{1}{\sqrt{3}}(e^{\frac{\pi i}{6}} \operatorname{ch}_{\frac{1}{8}}^+ +
      e^{-\frac{\pi i}{6}} \operatorname{ch}_{- \frac{1}{24}}
      - e^{- \frac{\pi i}{6}} \operatorname{ch}_{\frac{1}{8}}^- ) \ ,
      \\
    STS\operatorname{ch}_{-\frac{1}{24}} = & \ \frac{1}{\sqrt{3}}(e^{-\frac{\pi i}{6}} \operatorname{ch}_{\frac{1}{8}}^+ 
    + e^{\frac{\pi i}{6}} \operatorname{ch}_{- \frac{1}{24}}
    + e^{-\frac{\pi i}{6}} \operatorname{ch}_{\frac{1}{8}}^-
    ) \ ,\\
    STS\operatorname{ch}_{\frac{1}{8}}^- = & \ \frac{1}{\sqrt{3}}(- e^{-\frac{\pi i}{6}} \operatorname{ch}_{\frac{1}{8}}^+ 
    + e^{-\frac{\pi i}{6}} \operatorname{ch}_{- \frac{1}{24}}
    + e^{\frac{\pi i}{6}} \operatorname{ch}_{\frac{1}{8}}^-
    ) \ .
\end{align}
The vacuum character, on the other hand, does not remain in this
three-dimensional pure-character space: its $STS$ transform acquires terms
linear in $\tau$ and $\mathfrak{a}$.  All of the new structure is carried by a
single additional function.  Writing
\begin{equation}
  \mathcal{A}
  \coloneqq 3\operatorname{ch}_{\textrm{vac}}
  + \frac{\operatorname{ch}_{\frac18}^{+}+\operatorname{ch}_{\frac18}^{-}}{2}\,,
  \qquad
  \Psi_{\log}
  \coloneqq \tau\,\mathcal{A}
  + 3\mathfrak{a}\left(\operatorname{ch}_{\frac18}^{+}-\operatorname{ch}_{\frac18}^{-}\right),
 \label{eq:flavored-logarithmic-companion}
\end{equation}
the closed form \eqref{eq:flavored-Schur-index-WZ3} gives
\begin{equation}
 STS\operatorname{ch}_{\textrm{vac}}
 = \operatorname{ch}_{\textrm{vac}}
   - \frac{1}{3}\,\Psi_{\log}
   + c_\omega\, u_\omega \,,
 \qquad
 u_\omega \coloneqq
 \operatorname{ch}_{\frac18}^{+}
 -\operatorname{ch}_{-\frac1{24}}
 +\operatorname{ch}_{\frac18}^{-} \,,
 \qquad
 c_\omega \coloneqq \frac{e^{-\frac{i\pi}{6}}}{3\sqrt3} \,.
 \label{eq:flavored-vacuum-STS}
\end{equation}
Since $\Psi_{\log}$ is itself invariant,
\begin{equation}
 STS\,\Psi_{\log}=\Psi_{\log} \,,
 \label{eq:Psi-log-invariant}
\end{equation}
the orbit closes once this one function is adjoined.  The smallest space that
contains the four characters and is stable under $\Gamma^0(2)$ is therefore
five-dimensional,
\begin{equation}
 \mathcal{V}=\operatorname{span}\left\{
 \operatorname{ch}_{\frac18}^{+},\;
 \operatorname{ch}_{-\frac1{24}},\;
 \operatorname{ch}_{\frac18}^{-},\;
 \operatorname{ch}_{\textrm{vac}},\;
 \Psi_{\log}
 \right\}.
\end{equation}
Note that $\Psi_{\log}$ depends explicitly on $\tau$ and $\mathfrak{a}$ and is
not a graded trace; it is the logarithmic companion of the vacuum character
rather than the character of a fifth module.  It is the flavored counterpart of
the logarithmic solution of section~\ref{sec:MLDE}: in the unflavored limit the
second term of \eqref{eq:flavored-logarithmic-companion} drops out and
\begin{equation}
 \Psi_{\log}\Big|_{\mathfrak{a}\to0}
 =\tau\left(3\operatorname{ch}_{\textrm{vac}}+\operatorname{ch}_{\frac18}\right)
 =\tau\,\frac{\eta(\tau)^4}{\eta(\frac{\tau}{2})^2}\,,
\end{equation}
so that \eqref{eq:flavored-vacuum-STS} reduces to \eqref{eq:chlog-def}.

We can now read off the $\Gamma^0(2)$ representation carried by $\mathcal{V}$.
On the three-dimensional block of pure characters, $STS$ acts with order three
and is diagonalized by
\begin{equation}
 u_1=\operatorname{ch}_{\frac18}^{+}+\operatorname{ch}_{-\frac1{24}},
 \qquad
 u_2=\operatorname{ch}_{\frac18}^{-}-\operatorname{ch}_{\frac18}^{+},
 \qquad
 u_\omega \,,
\end{equation}
with eigenvalues $1$, $1$ and $\omega=e^{2\pi i/3}$, respectively.  The
remaining two directions cannot be diagonalized.  Indeed, the $u_\omega$
admixture in \eqref{eq:flavored-vacuum-STS} can be absorbed by mixing the
vacuum character with $\operatorname{ch}_{\frac18}^{\pm}$, and the combination
that does so is precisely $\mathcal{A}$ (unique up to adding the invariant
characters $u_1$, $u_2$),
\begin{equation}
 STS\,\mathcal{A}=\mathcal{A}-\Psi_{\log} \,.
 \label{eq:flavored-logarithmic-transform}
\end{equation}
Together with \eqref{eq:Psi-log-invariant}, this exhibits
$(\mathcal{A},-\Psi_{\log})$ as a Jordan pair of size two, and in the ordered
basis $(u_1,u_2,u_\omega,\mathcal{A},-\Psi_{\log})^{\mathsf T}$ the
transformation takes the Jordan form
\begin{equation}
 STS\sim
 \begin{pmatrix}
  1&0&0&0&0\\
  0&1&0&0&0\\
  0&0&\omega&0&0\\
  0&0&0&1&1\\
  0&0&0&0&1
 \end{pmatrix}.
 \label{eq:full-Jordan-form}
\end{equation}
A single Jordan block of size two
is exactly the consequence of one logarithmic solution.

\subsection{Zhu's algebra and modules}

The simple modules of a VOA $\mathbb{V}$ are in one-to-one correspondence with the modules of Zhu's associative algebra $A(\mathbb{V})$. The latter describes how the zero modes of $\mathbb{V}$ act on the highest weight space of a simple module. Let us first review the definition of Zhu's algebra $A(\mathbb{V})$ and then use it to constrain the highest weights considered for $\mathcal{W}_{\mathbb{Z}_3}$.

This associative algebra should not be confused with the $C_2$-algebra
$R_{\mathbb{V}} \coloneqq \mathbb{V}/C_2(\mathbb{V})$, which is a commutative
Poisson algebra.  The former controls the top spaces of modules, whereas the
latter controls the associated variety
$X_{\mathbb{V}} = \operatorname{Specm} R_{\mathbb{V}}$.  We write
$R_{\mathbb{V}}^{\mathrm{red}} \coloneqq
R_{\mathbb{V}}/\sqrt{(0)}$ for its reduction.

Consider a $\frac{1}{2}\mathbb{Z}$-graded super-VOA $\mathbb{V} = \mathbb{V}_0 \oplus \mathbb{V}_{1/2}$, where $\mathbb{V}_0$ and $\mathbb{V}_{1/2}$ denote states with integral or half-integral conformal weights. One can define two products, a star product $*$ and a circle product $\circ$. Denoting $a(z) = \sum_n a_n z^{-n - h_a}$,
\begin{equation}
    a * b = \begin{cases}
    \sum_{\ell \ge 0} \binom{h_a}{\ell} a_{-h_a + \ell} b & a, b \in \mathbb{V}_0 \\
    0 & a \in \mathbb{V}_{1/2} ~ \text{or} ~ b \in \mathbb{V}_{1/2}
    \end{cases} \ ,
\end{equation}
and
\begin{equation}
    a \circ b =  \begin{cases}
    \displaystyle\sum_{\ell \ge 0}\binom{h_a}{\ell} a_{-h_a - 1 + \ell} b & a \in \mathbb{V}_0 \\
    \displaystyle\sum_{\ell=0}^{h_ a-\frac12}
    \binom{ h_a-\frac12}{\ell} a_{-h_a + \ell} b = (ab) + \sum_{\ell = 1}^{h_a - \frac{1}{2}} \binom{h_a - \frac{1}{2}}{\ell} \{a b\}_\ell \ , & a \in \mathbb{V}_{1/2}
    \end{cases} \ .
\end{equation}
The linear span $O(\mathbb{V}) = \operatorname{span}\{a \circ b \ \mid a, b \in \mathbb{V}\}$ forms a two-sided ideal with respect to the star product,
\begin{equation}
    a * (b \circ c), (b \circ c)* a \in O(\mathbb{V}) \ .
\end{equation}
Moreover, the star product is associative up to terms in $O(\mathbb{V})$,
\begin{equation}
    (a * b) * c = a * (b * c) \mod O(\mathbb{V}) \ ,
\end{equation}
rendering the quotient $A(\mathbb{V}) = \mathbb{V}/O(\mathbb{V})$ an associative algebra,
\begin{equation}
    [a * b] = [a] * [b] \ .
\end{equation}

Passing from $\mathbb{V}$ to $A(\mathbb{V})$ simplifies the structure in two ways. First,
derivatives are reduced to lower representatives.  For $a\in\mathbb{V}_0$,
\[
  a\circ \mathbf 1=\partial a+h_a a,
\]
and therefore, in $A(\mathbb V)$,
\[
  [\partial a]=[-h_a a],\qquad
  [\partial^r a]=(-1)^r h_a(h_a+1)\cdots(h_a+r-1)[a].
\]
Second, in the ordinary Zhu algebra relevant for NS ordinary modules, an
individual state of half-integral conformal weight has trivial Zhu class:
if $a\in\mathbb V_{1/2}$, then $a\circ\mathbf 1=a$, hence $[a]=0$.
This does not mean that half-integral fields can be ignored completely:
their OPEs and normally ordered products may have integral conformal weight
and can contribute nontrivially to $A(\mathbb V)$.

The projection of a normally ordered product is obtained from the definition
of the star and circle products.  In the conventions above,
\begin{equation}\label{eq:AV-representative-of-NO}
    [(ab)] = \begin{cases}
        \displaystyle [a*b] - \sum_{\ell \ge 1}\binom{h_a}{\ell}
        [\{ab\}_{\ell}] \ , & a \in \mathbb{V}_0,\\[3mm]
        \displaystyle - \sum_{\ell \ge 1} \binom{h_a - \frac{1}{2}}{\ell}
        [\{ab\}_{\ell}] \ , & a \in \mathbb{V}_{1/2}.
    \end{cases}
\end{equation}

We denote by $o(a)=a_0$ the zero mode of an integral-weight state $a$; it
carries zero conformal weight and maps the lowest weight space $M_h$ of an
ordinary module $M$ to itself.  When acting on $M_h$,
\begin{equation}
    o(a*b)M_h=o(a)o(b)M_h .
\end{equation}
Thus, the ordinary Zhu algebra action on the lowest weight space is generated
by integral-weight representatives, while half-integral fields enter only
through integral composites and OPE coefficients.
The irreducible modules of $A(\mathbb V)$ are in one-to-one correspondence
with irreducible ordinary $\mathbb V$-modules; hence, we now describe
$A(\mathbb V)$ and find its modules.

The relevant $A(\mathbb{V})$ generators are
$j=[J]$, $t=[T]$, $g_w=[G_W]$, and
$\tilde g_{\tilde w}=[\tilde G_{\tilde W}]$, where $[~\cdot~]$ denotes the
projection to $A(\mathbb V)$.  States involving normally ordered products with
$G,\tilde G,W,\tilde W$ can be treated using
\eqref{eq:AV-representative-of-NO}.  For instance, taking
$a=G$ with $h_G=\frac32$ and $b=\tilde G$, the half-integral formula gives
\begin{equation}
\begin{aligned}
  [(G\tilde G)]
  &=
  -\sum_{\ell\ge1}\binom{h_G-\frac12}{\ell}[\{G\tilde G\}_{\ell}]
  =
  -[\{G\tilde G\}_{1}] 
  =
  \left[T-\frac{J'}{2}\right]
  =
  t+\frac12 j .
\end{aligned}
\end{equation}
We collect the projection of several key null states at lower weights,
\begin{align}
    [\mathcal{N}_3] = & \ \frac{2}{9}(\frac{7}{6}j - \frac{1}{6} j*j*j + 2t * j) \ ,\\
    [\mathcal{N}_{3,+2}] = & \ j* g_w - g_w, \qquad
    [\mathcal{N}_{3,-2}] = j * \tilde g_{\tilde w} + \tilde g_{\tilde w} \ ,\\
    [\mathcal{N}_4] =& \ g_w * \tilde g_{\tilde w}- \frac{5}{9} j + \frac{11}{18} j* j + \frac{1}{18}j * j * j - \frac{1}{9} j* j * j * j - 2j * t + t*j + t*j*j\ ,\\
    [\mathcal{N}_6] = & \ \tilde g_{\tilde w}* g_w - \frac{4}{3} j + \frac{1}{6} j* j * j + \frac{2}{9} j*t + \frac{4}{9} t - \frac{23}{9} t* j  - \frac{1}{9} t* j * j + \frac{2}{3}t * t \ .
\end{align}

We can now analyze the irreducible modules of $A(\mathbb{V})$, with the lowest (or highest, depending on the preferred terminology) weight space $V$ supporting the modules. Acting on $V$, $(j - 1)* g_w = 0$ and $(j + 1)* \tilde g_{\tilde w} = 0$ implies the images of $g_w$ and $\tilde g_{\tilde w}$ sit in the $j$-eigenspace with eigenvalues $1$ and $-1$ respectively. On the other hand, $t$ is central having eigenvalue $h$, and $[\mathcal{N}_3] = 0$ implies $j(12h - j^2 + 7) = 0$. Therefore,
\begin{equation}
    (12h + 6) g_w = (12 h + 6) \tilde g_{\tilde w} = 0\ ,
\end{equation}
since $j = \pm 1$ on the image of $g_w$ and $\tilde g_{\tilde w}$.
Hence there are only two possibilities:
\begin{equation}
    g_w = \tilde g_{\tilde w} = 0, \qquad \text{or} \qquad
    h = - \frac{1}{2}\ .
\end{equation}

The former scenario implies a one-dimensional lowest weight space as an irreducible module of $A(\mathbb{V})$, since only $t, j$ act non-trivially.

The latter scenario actually gives rise to a one-dimensional lowest weight space as well. When $h = - \frac{1}{2}$, $[\mathcal{N}_3] = 0$ reduces to $j(j^2 - 1) = 0$, restricting the $j$-eigenvalues to be $0, \pm 1$. One can decompose the lowest weight space $V$ into $j$-eigenspaces, $V = V_{-1} \oplus V_0 \oplus V_{+1}$, where $g_w$ and $\tilde g_{\tilde w}$ map between $V_{\pm 1}$ and act as zero on $V_0$, since $g_w, \tilde g_{\tilde w}$ carry $j$-charge $\pm 2$. The equations $[\mathcal{N}_6] = [\mathcal{N}_4] = 0$ imply $g_w \tilde g_{\tilde w} = 0$, $\tilde g_{\tilde w} g_w = \frac{1 - j^2}{18}$ since $j^3 = j$ when acting on either $V_0$ or $V_{\pm1}$. However, $\tilde g_{\tilde w}g_w |_{V_0} = 0$ while $\frac{1 - j^2}{18}|_{V_0} \ne 0$; this is possible only if $V_0$ is zero-dimensional. We are left with $V_{\pm 1}$: $g_w \tilde g_{\tilde w} = \tilde g_{\tilde w} g_w = 0$ on both $V_{\pm 1}$. As an irreducible representation of $A(\mathbb{V})$, $V$ can only be $V_{+1}$ or $V_{-1}$, and $g_w = \tilde g_{\tilde w} = 0$. Hence only $t, j$ act non-trivially on $V$, and as an irreducible representation, $V$ must be one-dimensional.

Now that we have established that the lowest weight space is one-dimensional, the null equations lead to the following equations on the quantum numbers $(h,m)$, the eigenvalues of the Zhu elements $t$ and $j$,
\begin{equation}
    m(12h- m^2 + 7) = m(m-1)(18h - 2m^2 - m + 10) = 12h^2 - 2hm^2 - 42hm + 8h + 3m^3 - 24 m = 0 \ .
\end{equation}
They give rise to four solutions:
\begin{equation}
    (h, m) = (0, 0), \quad  (- \frac{1}{2}, \pm1), \quad (-\frac{2}{3}, 0) \ .
\end{equation}
The first solution corresponds to the vacuum module, the second and third to $\operatorname{ch}_{1/8}^\pm$, and the last to $\operatorname{ch}_{- \frac{1}{24}}$. We will denote these modules as $M_\textrm{vac}$, $M_{\pm }$ and $M_0$, respectively. In fact, as we will see below, the charged modules $M_\pm$ are reducible and their irreducible quotients $N_\pm$ are relevant in the following free-field realization.

The free-field construction below realizes the neutral module $M_0$ and images of
the charged quotients $N_\pm$, rather than the reducible modules $M_\pm$. The $bc \beta\gamma$ system realizing $\mathcal{W}_{\mathbb{Z}_3}$ can be bosonized as (we keep the necessary cocycle structure for the fermionic nature in $b, c$ implicit)
\begin{equation}
    \beta = e^{\chi + \phi}, \qquad
    \gamma = \partial \chi e^{-\chi - \phi}, \qquad
    b = e^{-\sigma}, \qquad
    c = e^\sigma \ .
\end{equation}
Here,
\begin{equation}
    \chi(z)\chi(w) \sim \ln(z - w), \qquad
    \phi(z)\phi(w) \sim - \ln(z - w), \qquad
    \sigma(z)\sigma(w) \sim \ln(z - w) \ .
\end{equation}
This naturally gives a free boson realization of all the strong generators of $\mathcal{W}_{\mathbb{Z}_3}$. In particular,
\begin{equation}
    J = -3 \partial \phi - 2 \partial \sigma, \qquad
    T = \frac{1}{2}(\partial \chi\partial\chi) - \frac{1}{2}(\partial\phi\partial\phi) + \frac{1}{2}(\partial\sigma\partial\sigma) - \frac{1}{2}\partial^2\chi
    + \partial^2 \phi + \frac{3}{2}\partial^2 \sigma \ .
\end{equation}

Consider the operator $v = e^{r \chi + s \phi + u \sigma}$ as the candidate for the highest weight state of a module. Then
\begin{equation}
    J_0 v = (3s - 2u)v, \qquad
    L_0 v = \frac{1}{2}(r^2 - s^2 + r + 2s + u^2 -3u)v \ .
\end{equation}
Furthermore, we require the vertex operator $v$ to be local, meaning that its OPE with the ghosts involves only integral powers; this imposes the conditions
\begin{equation}
    r - s \in \mathbb{Z}, \qquad
    u \in \mathbb{Z} \ .
\end{equation}
Finally, all positive modes of the strong generators must annihilate $v$. Together, these constraints leave the following possibilities\footnote{The state $v_{- \frac{1}{24}}$ will be renamed to $v_0$ in the next subsection.}
\begin{equation}\label{eq:free-boson-realization}
    v_{- \frac{1}{24}} = e^{- \frac{1}{3} \chi + \frac{2}{3}\phi + \sigma}, \quad
    v_+ = e^{\phi + \sigma}, \quad
    v_- = e^{- \frac{1}{3}\chi - \frac{1}{3}\phi} \ ,
\end{equation}
corresponding to $M_0$ and the free-field images of $N_+$ and $N_-$, respectively. The action of the negative modes of the generators is straightforward to work out; here, we list the simplest entries. For example, on $v_{-\frac{1}{24}}$, there are four states at level $\frac{1}{2}$,
\begin{align}
    G_{-1/2}v_{- \frac{1}{24}} = & \ - \frac{1}{3} e^{- \frac{4}{3}\chi - \frac{1}{3}\phi}, \quad
    & \tilde G_{-1/2}v_{- \frac{1}{24}} = & \ e^{\frac{2}{3}\chi + \frac{5}{3}\phi + 2\sigma} \ , \\
    W_{- \frac{1}{2}}v_{- \frac{1}{24}}
    = & \ e^{\frac{2}{3}\chi + \frac{5}{3}\phi + \sigma}, \quad
    & \tilde W_{- \frac{1}{2}} v_{- \frac{1}{24}} = & \ (\frac{8}{27} \partial\chi - \frac{4}{27}\partial \phi - \frac{2}{9}\partial \sigma)e^{- \frac{4}{3}\chi - \frac{1}{3}\phi + \sigma} \ . \nonumber
\end{align}
These four states precisely reproduce the term
\begin{equation}
    q^{- \frac{1}{24} + \frac{1}{2}}(\chi_3 - 2 \chi_1) = q^{\frac{11}{24}}(a^3 - a - a^{-1} + a^{-3}) \in \operatorname{ch}_{- \frac{1}{24}} \ .
\end{equation}

Similarly, for the free-field image of $N_+$ with $(h, m) = (- \frac{1}{2}, + 1)$, the free boson realization implies
\begin{equation}
    G_{- \frac{1}{2}}v_+ = \tilde W_{- \frac{1}{2}}v_+ = 0, \quad
    \tilde G_{- \frac{1}{2}}v_+ = e^{\chi + 2\phi + 2 \sigma}, \quad
    W_{- \frac{1}{2}}v_+ = e^{\chi + 2 \phi + \sigma} \ .
\end{equation}
The two nonzero states in this realization contribute the subleading term $q^{\frac{1}{8} + \frac{1}{2}}(a^4 - a^2)$ to the character of the free-field image of $N_+$.
By contrast, $\operatorname{ch}_{1/8}^{+}$ has coefficient $a^4-a^2-1$
at this level. The proposed addition of $\operatorname{ch}_{\rm vac}$ in
\eqref{eq:charged-irreducible-character-conjecture} removes the extra $-1$. In the universal highest-weight module, $G_{-\frac{1}{2}}v_+$ is a singular vector with $(h,m)=(0,0)$; it vanishes in the free-field realization of the quotient, as shown above. Its only potentially nonzero positive-mode descendant is
\begin{equation}\label{FF-vanish}
    \tilde G_{\frac{1}{2}} G_{-\frac{1}{2}}v_+
    = - \frac{1}{2} (J_0 + 2L_0) v_+ = 0 \ ,
\end{equation}
confirming that $G_{-\frac{1}{2}}v_+$ is singular. This singular vector is retained in $M_+$ and vanishes in its free-field image; we return to this point at the end of section~\ref{subsec:five-generator-PBW}.

\subsection{A reduced PBW basis}
\label{subsec:five-generator-PBW}

The non-vacuum characters $\operatorname{ch}_{- \frac{1}{24}}$ and $\operatorname{ch}_{\frac{1}{8}}^\pm$ in \eqref{eq:flavored-non-vacuum-characters-WZ3-1}--\eqref{eq:flavored-non-vacuum-characters-WZ3-3} are ratios of Jacobi theta and Dedekind $\eta$ functions. Written in terms of the infinite-product representations of these functions, they display the characteristic shape of a Verma-module character,
\begin{equation}
 \operatorname{ch}_\alpha(a;q)=\operatorname{str}_M a^{J_0}q^{L_0-c_{2d}/24},
 \qquad
 \alpha=h-\frac{c_{2d}}{24} ,
 \label{eq:pbwchar-supercharacter}
\end{equation}
namely a graded count of states in a certain PBW basis. In this subsection, we conjecture the existence of such a basis and provide preliminary evidence for it.

The VOA $\mathcal W_{\mathbb Z_3}$ is strongly generated by the following eight operators:
\begin{equation}
    T,\quad J,\quad G,\quad \tilde G,\quad W,\quad \tilde W,\quad
    G_W,\quad \tilde G_{\tilde W}.
\end{equation}
Highest-weight modules of $\mathcal{W}_{\mathbb{Z}_3}$ are constructed by acting with the non-positive modes of these operators on a highest-weight state. We first focus on the module $M_0$ with quantum numbers $(h, m) = (- \frac{2}{3}, 0)$. For the purpose of counting states, we argue that the negative modes of $T$, $G_W$ and $\tilde G_{\tilde W}$ can be absorbed into those of $J, G, \tilde G, W, \tilde W$; that is, $M_0$ admits a reduced PBW basis built from only five generators.

Concretely, we choose the
canonical ordering
\begin{equation}
    J\prec G\prec\tilde G\prec W\prec\tilde W ,
    \label{eq:five-generator-order}
\end{equation}
with modes within each species ordered by decreasing excitation number, and
consider formal monomials subject to
\begin{equation}
    a_n\in\{0,1,2\},
    \qquad
    \epsilon_r,\tilde\epsilon_r\in\{0,1\},
    \label{eq:five-letter-occupancies}
\end{equation}
where $a_n$ is the occupation number of $J_{-n}$, and $\epsilon_r,\tilde\epsilon_r$ those of the fermionic modes $G_{-r-\frac12},\tilde G_{-r-\frac12}$. The bosonic modes of $W,\tilde W$ carry unrestricted occupation numbers, but are subject to the exclusion of
\begin{equation}
    W_{-r-\frac12},\qquad
    \tilde W_{-r-\frac12},
    \qquad
    r\equiv1\pmod3 .
    \label{eq:five-letter-W-restriction}
\end{equation}
The resulting states are the monomials
\begin{equation}
 \prod_{n\ge1}J_{-n}^{a_n}
 \prod_{r\ge0}G_{-r-\frac12}^{\epsilon_r}
 \prod_{r\ge0}\tilde G_{-r-\frac12}^{\tilde\epsilon_r}
 \prod_{\substack{r\ge0\\r\not\equiv1\ ({\rm mod}\ 3)}}
 W_{-r-\frac12}^{p_r}
 \prod_{\substack{r\ge0\\r\not\equiv1\ ({\rm mod}\ 3)}}
 \tilde W_{-r-\frac12}^{\tilde p_r}\,v_0 ,
\end{equation}
where $p_r,\tilde p_r\in\mathbb Z_{\ge0}$, and $v_0\equiv v_{-\frac{1}{24}}$ denotes the highest-weight state of $M_0$. Treating these monomials as linearly independent, their graded supertrace reads
\begin{align}
\chi_{\rm PBW}(a;q)
={}&
q^{-\frac1{24}}
\prod_{n\ge1}(1+q^n+q^{2n})
\prod_{r\ge0}
(1-a^{-1}q^{r+\frac12})
(1-aq^{r+\frac12})
\nonumber\\
&\times
\prod_{\substack{r\ge0\\r\not\equiv1\ ({\rm mod}\ 3)}}
\frac{1}{
(1-a^3q^{r+\frac12})
(1-a^{-3}q^{r+\frac12})}.
\label{eq:five-letter-character-counting}
\end{align}
The three groups of factors mirror the nature of the five generators (see Table~\ref{tab:WZ3-generator-quantum-numbers}): the $\mathbb Z_3$-truncated occupancies of the current $J$ give $\prod_{n\ge1}(1+q^n+q^{2n})$, the fermionic $G,\tilde G$ of $J$-charge $\mp1$ give the numerator, and the bosonic $W,\tilde W$ of $J$-charge $\pm3$ give the denominator, thinned out by the exclusion \eqref{eq:five-letter-W-restriction}.
This counting reproduces the closed-form character \eqref{eq:flavored-non-vacuum-characters-WZ3-1},
\begin{equation}
    \chi_{\rm PBW}(a;q)
    =
    \operatorname{ch}_{-\frac1{24}}(a;q) \ .
    \label{eq:PBW-character-agreement}
\end{equation}

This agreement is our main evidence for the reduction. It does not, however, show that the monomials above are linearly independent in $M_0$, and in fact they are not. We therefore examine the two lowest levels at which the question has content. At level one, the reduction can be verified in closed form. Already at level $\frac52$, on the other hand, the five generators still appear to span the module, while the canonical ordering \eqref{eq:five-generator-order} ceases to be a valid normal form.

\paragraph{Level $N = 1$: how the reduction works.} Consider the level-one subspace, of conformal dimension $h = - \frac{2}{3} + 1$. The eight strong generators give twelve states,
\begin{align}
&J_{-1}v_0,\qquad
T_{-1}v_0,\qquad
(G_W)_{-1}v_0,\qquad
(\tilde G_{\tilde W})_{-1}v_0,
\nonumber\\
&G_{-\frac12}\tilde G_{-\frac12}v_0,\qquad
G_{-\frac12}W_{-\frac12}v_0,\qquad
G_{-\frac12}\tilde W_{-\frac12}v_0,
\tilde G_{-\frac12}W_{-\frac12}v_0,\qquad
\tilde G_{-\frac12}\tilde W_{-\frac12}v_0,
\nonumber\\
&W_{-\frac12}^{2}v_0,\qquad
W_{-\frac12}\tilde W_{-\frac12}v_0,\qquad
\tilde W_{-\frac12}^{2}v_0 .
\label{eq:level1-eight-generator-ambient}
\end{align}
The weight-three null multiplet of the vacuum module imposes three relations among them,
\begin{align}
0&=
\left(
G_{-\frac12}W_{-\frac12}
-\frac23(G_W)_{-1}
\right)v_0 ,
\label{eq:level1-relation-GW}\\
0&=
\left(
\tilde G_{-\frac12}\tilde W_{-\frac12}
-\frac23(\tilde G_{\tilde W})_{-1}
\right)v_0 ,
\label{eq:level1-relation-GWt}\\
0&=
\left(
-\frac8{27}J_{-1}
-\frac29T_{-1}
+\frac29G_{-\frac12}\tilde G_{-\frac12}
+W_{-\frac12}\tilde W_{-\frac12}
\right)v_0 .
\label{eq:level1-relation-T}
\end{align}
These can be solved for precisely the three states built on $G_W$, $\tilde G_{\tilde W}$ and $T$, in terms of the remaining ones,
\begin{align}
(G_W)_{-1}v_0
& = 
\frac32G_{-\frac12}W_{-\frac12}v_0,
\label{eq:level1-rewrite-GW}\\
(\tilde G_{\tilde W})_{-1}v_0
& = 
\frac32\tilde G_{-\frac12}\tilde W_{-\frac12}v_0,
\label{eq:level1-rewrite-GWt}\\
T_{-1}v_0
& = 
-\frac43J_{-1}v_0
+G_{-\frac12}\tilde G_{-\frac12}v_0
+\frac92W_{-\frac12}\tilde W_{-\frac12}v_0 .
\label{eq:level1-rewrite-T}
\end{align}
The three generators $T, G_W, \tilde G_{\tilde W}$ are thereby eliminated, and the nine surviving reduced monomials form a basis of the level-one subspace,
\begin{align}
\bigl\{&
J_{-1}v_0,\,
G_{-\frac12}\tilde G_{-\frac12}v_0,\,
G_{-\frac12}W_{-\frac12}v_0,\,
G_{-\frac12}\tilde W_{-\frac12}v_0,
\nonumber\\
&
\tilde G_{-\frac12}W_{-\frac12}v_0,\,
\tilde G_{-\frac12}\tilde W_{-\frac12}v_0,\,
W_{-\frac12}^{2}v_0,\,
W_{-\frac12}\tilde W_{-\frac12}v_0,\,
\tilde W_{-\frac12}^{2}v_0
\bigr\}.
\label{eq:level1-five-generator-basis}
\end{align}

\paragraph{Level $N = 5/2$: failure of the canonical ordering.}
We label a graded sector by its level $N$, its $J_0$-charge $m$ and its fermion parity $F$. The lowest sector in which the naive prescription breaks down is
\begin{equation}
    \lambda=\left(N,m,F\right)
    =\left(\frac52,+1,\mathrm{odd}\right).
\end{equation}
Subject to the constraints \eqref{eq:five-letter-occupancies}--\eqref{eq:five-letter-W-restriction}, the reduced monomials in this sector are the eight states
\begin{align}
Y_0&=
\tilde G_{-\frac12}
W_{-\frac12}^{2}
\tilde W_{-\frac12}^{2}v_0 ,
\nonumber\\
X_1&=\tilde G_{-\frac52}v_0,\qquad
X_2=J_{-1}\tilde G_{-\frac32}v_0,\qquad
X_3=J_{-2}\tilde G_{-\frac12}v_0,
\nonumber\\
X_4&=J_{-1}^{2}\tilde G_{-\frac12}v_0,\qquad
X_5=
G_{-\frac12}\tilde G_{-\frac12}
\tilde G_{-\frac32}v_0,
\nonumber\\
X_6&=
\tilde G_{-\frac32}
W_{-\frac12}\tilde W_{-\frac12}v_0,\qquad
X_7=
J_{-1}\tilde G_{-\frac12}
W_{-\frac12}\tilde W_{-\frac12}v_0 .
\label{eq:level52-eight-candidates}
\end{align}
These are not linearly independent, however. Evaluating them in the free boson realization, one finds the single relation
\begin{equation}
0=
Y_0
+\frac{1000}{729}X_1
-\frac{1400}{729}X_2
+\frac{500}{729}X_3
+\frac{500}{729}X_4
+\frac{200}{243}X_5
+\frac{200}{27}X_6
-\frac{50}{9}X_7 .
\label{eq:first-five-letter-relation}
\end{equation}
Their span inside the module subspace $M_{0, \lambda}$ is therefore only seven-dimensional, one less than the number of monomials.

To state this precisely, we distinguish the formal reduced PBW symbols $e_{Y_0}, e_{X_i}$ from their actual realizations $Y_0, X_i$ in the module $M_0$. Let
\begin{equation}
    C_\lambda
    =
    \operatorname{Span}_{\mathbb C}
    \{e_{Y_0},e_{X_1},\ldots,e_{X_7}\}, \qquad
    \dim C_\lambda = 8 \ ,
\end{equation}
be the formal eight-dimensional candidate space, and define the evaluation map
\begin{equation}
    \Phi_\lambda:C_\lambda\longrightarrow M_{0,\lambda},
    \qquad
    e_{Y_0}\mapsto Y_0,\qquad
    e_{X_i}\mapsto X_i .
\label{eq:evaluation-map}
\end{equation}
The relation \eqref{eq:first-five-letter-relation} corresponds to the kernel
vector
\begin{align}
s_{\frac52,+}
={}&
e_{Y_0}
+\frac{1000}{729}e_{X_1}
-\frac{1400}{729}e_{X_2}
+\frac{500}{729}e_{X_3}
+\frac{500}{729}e_{X_4}
+\frac{200}{243}e_{X_5}
+\frac{200}{27}e_{X_6}
-\frac{50}{9}e_{X_7} \ ,
\label{eq:level52-kernel-vector}
\end{align}
so that
\begin{equation}
    \ker\Phi_\lambda
    =
    \mathbb C\,s_{\frac52,+},
    \qquad
    \operatorname{dim}\operatorname{Im}\Phi_\lambda=7.
\label{eq:level52-kernel-result}
\end{equation}

The character \eqref{eq:PBW-character-agreement}, on the other hand, assigns to this sector the dimension
\begin{equation}
    \dim M_{0,\lambda}=8 .
\end{equation}
The canonically ordered monomials thus fail to reach one direction of $M_{0,\lambda}$, which may be represented by the state
\begin{equation}
    Z_{\frac52,+}
    =
    T_{-1}\tilde G_{-\frac32}v_0 .
\label{eq:level52-missing-state}
\end{equation}
Although written using $T$, this state does lie in the space generated by the five reduced generators; it merely fails to be canonically ordered,
\begin{equation}
    T_{-1}\tilde G_{-\frac32}v_0
    ={}
    -G_{-\frac12}\tilde G_{-\frac12}
     \tilde G_{-\frac32}v_0 -
    \tilde G_{-\frac12}G_{-\frac12}
     \tilde G_{-\frac32}v_0 .
    \label{eq:level52-nonordered-rewrite}
\end{equation}
The two monomials on the right differ only by the exchange of $G_{-\frac12}$ and $\tilde G_{-\frac12}$, so their sum is the anticommutator of these two modes, which returns $T_{-1}$. Reordering the second term into the canonical order \eqref{eq:five-generator-order} therefore regenerates the very mode of $T$ that one set out to eliminate. This is the sense in which the five-generator spanning statement and the naive canonical normal form must be distinguished.

The two defects found in this sector are of equal size: one linear relation among the eight monomials, and one module direction outside their image. They cancel against each other in the graded character, which is why \eqref{eq:PBW-character-agreement} can hold even though linear independence fails. We have checked this cancellation sector by sector in the free field realization up to relative level $\frac92$, and conjecture that it persists for all $\lambda$, so that the formal span always carries the correct dimension,
\begin{equation}
    \dim C_\lambda = \dim M_{0, \lambda} \ ,
\end{equation}
even though the image of $\Phi_\lambda$ may be a proper subspace of $M_{0,\lambda}$. The restricted monomials \eqref{eq:five-letter-occupancies}--\eqref{eq:five-letter-W-restriction} are thus best regarded as a character-counting set rather than as a literal PBW basis at all levels; we leave the systematic construction of the corrected normal forms to future work.

\paragraph{The charged modules.} We close with the charged modules of quantum numbers
\begin{equation}
    (h,m)=\left(-\frac12,\pm1\right) \ ,
\end{equation}
whose reducible modules and free-field quotients must be distinguished. Let $v_\pm$ denote the corresponding highest-weight states, and let $M_\pm$ be the highest-weight modules generated from them by the non-positive modes of the strong generators. These modules contain the singular vectors
\begin{equation}
    s_+=G_{-\frac12}v_+,
    \qquad
    s_-=\tilde G_{-\frac12}v_- .
    \label{eq:charged-module-singular-vectors}
\end{equation}
This follows by acting on $s_\pm$ with the positive modes. For $s_+$, for instance, the only potentially non-vanishing case is
\begin{equation}
    \tilde G_{\frac{1}{2}} s_+
    = - \frac{1}{2} (J_0 + 2L_0) v_+ = 0 \ ,
\end{equation}
because $J_0 v_+ = v_+$ and $L_0 v_+ = - \frac{1}{2}v_+$. If $v_\pm$ is bosonic, then $s_\pm$ is fermionic, and it generates a proper submodule
$K_\pm=U \left((\mathcal W_{\mathbb Z_3})_{<0}\right)s_\pm$ of $M_\pm$, which is isomorphic to the Fermion-parity reversal $\Pi M_\textrm{vac}$ of the vacuum module $M_\textrm{vac}$. As seen in \eqref{FF-vanish}, $s_\pm$ vanishes identically in the free boson realization. Defining the quotient
\begin{equation}
    N_\pm
    =
    M_\pm/
    K_\pm ,
    \label{eq:charged-module-quotient}
\end{equation}
one obtains the short exact sequence
\begin{equation}
    0\longrightarrow
    K_\pm
    \longrightarrow
    M_\pm
    \longrightarrow
    N_\pm
    \longrightarrow0 .
    \label{eq:charged-vacuum-exact-sequence}
\end{equation}
For the reducible highest-weight modules $M_\pm$, we use a reduced counting ansatz analogous to that of $M_0$, with the exclusion rules
\begin{equation}
\begin{array}{c|cc}
 & W_{-r-\frac12} & \tilde W_{-r-\frac12} \\ \hline
 M_+ & r\not\equiv2\pmod3 & r\not\equiv0\pmod3 \\
 M_- & r\not\equiv0\pmod3 & r\not\equiv2\pmod3
\end{array}
\qquad (r\ge0).
\label{eq:charged-five-letter-W-restrictions}
\end{equation}
This count retains the odd singular vector $s_\pm$ and supports the
identification of $\operatorname{ch}_{1/8}^{\pm}$ \eqref{eq:flavored-non-vacuum-characters-WZ3-2} and \eqref{eq:flavored-non-vacuum-characters-WZ3-3} with the supercharacters
of the reducible modules $M_\pm$. Under the isomorphism
$K_\pm\cong\Pi M_\textrm{vac}$, additivity of supercharacters gives
\begin{equation}
    \operatorname{ch} N_\pm
    =
    \operatorname{ch}_{\frac18}^{\pm}
    +
    \operatorname{ch}_{\rm vac}.
    \label{eq:charged-irreducible-character-conjecture}
\end{equation}

\subsection{Quasi-modularity of the null-state equations}

In the above, we have constructed a space of character functions for $\mathcal{W}_{\mathbb{Z}_3}$, on which the modular group $\Gamma^0(2)$ acts. One easily verifies that the logarithmic partner $STS\operatorname{ch}_{\textrm{vac}}$ also solves the full set of flavored MLDEs constructed from the null states. This points to a modular structure of the set of equations. In the unflavored case, where the coefficients of the MLDE are either the standard Eisenstein series $E_{2n}(\tau)$ or the twisted Eisenstein series $E^{\pm}_{n}\!\left(\pm 1\right)$, which are modular with respect to an appropriate modular group $\Gamma$ (depending on the chosen spin structure), the unflavored equation transforms covariantly under $\Gamma$.

Now, with flavor fugacity $a$ turned on, the coefficients $E^{\pm}_{k}\!\left(a^n\right)$ are quasi-Jacobi instead of modular under $\Gamma$: each $E^{\pm}_{k}\!\left(a^n\right)$ transforms into a linear combination of lower-weight twisted Eisenstein series. Therefore, the flavored MLDEs are not modular, but ``quasi-modular'': one flavored MLDE transforms into a linear combination of other flavored MLDEs with equal or lower modular weight. The relevant modular group for our discussion is $\Gamma^0(2)$ generated by $STS, T^2$. To analyze the modular structure of the equations, we introduce an additional fugacity $y = e^{2\pi i \mathfrak{y}}$ conjugate to the level $k$ of the $U(1)$ affine Lie subalgebra of $\mathcal{W}_{\mathbb{Z}_3}$, and consider the extended characters
\begin{equation}
    \operatorname{ch}_\alpha(y, a, q) = y^k \operatorname{ch}_\alpha(a, q) \ , \qquad
    k = - 5 \ .
\end{equation}
The prefactor $y^k$ is often omitted in the physics literature. Under the $S, T$ transformations,
\begin{equation}
    S: (\mathfrak{y}, \mathfrak{a}, \tau) \to (\mathfrak{y} - \frac{1}{2\tau} \mathfrak{a}^2, \frac{\mathfrak{a}}{\tau}, - \frac{1}{\tau}), \qquad
    T: (\mathfrak{y}, \mathfrak{a}, \tau) \to (\mathfrak{y}, \mathfrak{a}, \tau + 1) \ .
\end{equation}
The derivatives also transform accordingly,
\begin{equation}
    q \partial_q \to \tau^2 q \partial_q + \tau \mathfrak{a} a \partial_a + \frac{k}{2} \mathfrak{a}^2, \qquad
    D_a \coloneqq a \partial_a \to \tau a \partial_a + k \mathfrak{a} \ .
\end{equation}
With the modular property of the Eisenstein series, we can write down the $STS$ transformation of all the flavored MLDEs.

The simplest example is the weight-three equation. It is easy to check that the operator \eqref{eq:D3} transforms with only an overall prefactor $(\tau - 1)^3$:
\begin{align}
\mathcal{D}_3 \xrightarrow{\ STS\ } (\tau-1)^3\,\mathcal{D}_3 \ .
\end{align}
This is because there is a single weight-three flavored MLDE and no equation of lower weight.

The weight-four equations transform less trivially,
\begin{align}
    \mathcal{D}_4^{(1)} 
    \xrightarrow{STS} & \ (\tau - 1)^4 \mathcal{D}_4^{(1)}  - 5(\tau - 1)^3 \mathfrak{a} \mathcal{D}_3 \ , \cr
    \mathcal{D}_4^{(2)} 
    \xrightarrow{STS} & \ (\tau - 1)^4 \mathcal{D}_4^{(2)} \ , \cr
    \mathcal{D}_4^{(3)}  \xrightarrow{STS} & \ (\tau - 1)^4 \mathcal{D}_4^{(3)}  + 12 (\tau - 1)^3 \mathfrak{a} \mathcal{D}_3  \ .
\end{align}

All the higher-weight flavored MLDEs have similar modular transformations. As noted above, each of these flavored MLDEs originates from a null state $\mathcal{N}$, and the modular transformation of each equation may be reformulated in terms of that null state,
\begin{equation}
    \mathcal{N} \xrightarrow{STS} \sum_{i = 0}^{h[\mathcal{N}] - 3} \sum_{n = 0}^{h[\mathcal{N}]} \tau^n \mathfrak{a}^i (-1)^{n + h[\mathcal{N}] - i} \begin{pmatrix}
        h[\mathcal{N}] - i\\n
    \end{pmatrix} (J_1)^i \mathcal{N} \ .
\end{equation}
This is identical to the (quasi) modular structure discussed in \cite{Zheng:2022zkm,Pan:2023jjw,Pan:2024dod} in the context of affine Lie algebras. All these results indicate a universal modular structure among the null states of a quasi-lisse VOA.

\medskip

\section{\texorpdfstring{$\mathcal{N}=3$}{N=3} theories from discrete gauging}
\label{sec:discrete-gauging}

In this section, we consider another series of rank-one $\mathcal{N}=3$ SCFTs, obtained by discrete gauging a $\mathbb{Z}_n$ flavor subgroup of the $\mathcal{N} = 4$ SYM \cite{Argyres:2016yzz,Bourton:2018jwb}. There are essentially two choices: the $\mathcal{N} = 4$ $U(1)$ SYM and the $SU(2)$ SYM. The former is simply the theory of a free $\mathcal{N} = 4$ vector multiplet, and hence trivially superconformal. We will establish the structure of their associated VOAs, use the null states and the flavored MLDEs to construct their representation theory, and work out the characters in closed form. The associated VOA of the former theory is relatively straightforward, involving only the $\mathbb{Z}_n$-quotient of the free $bc\beta \gamma$ system. The analysis of the Schur index of the latter theory requires more care, since the usual BRST-reduction implementing 4d $SU(2)$ gauging does not commute with the $\mathbb{Z}_n$-refinement of the Schur index. Nonetheless, the branching rule of \cite{Creutzig:2018ltv} gives the correct decomposition and allows us to write the index in closed form.

\subsection{Gauging \texorpdfstring{$\mathbb{Z}_n$}{Zn} in the \texorpdfstring{$U(1)$}{U(1)} theory}

We start with gauging a $\mathbb{Z}_n$ subgroup of the $SU(4)_R\times SL(2,\mathbb{Z})$
global symmetry of the $U(1)$ $\mathcal{N}=4$ theory.
At the level of the superconformal index, discrete gauging amounts to
projecting onto $\mathbb{Z}_n$-invariant operators, which is implemented by
averaging over the $n$ discrete holonomies $\epsilon\in\mathbb{Z}_n$:
\begin{equation}
\label{eq:Zn-gauged-index-def}
  \mathcal{I}_{\mathbb{Z}_{n}}(p,q,t;a)
  = \frac{1}{n}\sum_{\epsilon\in\mathbb{Z}_n}
    \mathcal{I}_{\mathcal{N}=4}(p,q,t;a,\epsilon),
\end{equation}
where $a$ is the fugacity for the surviving continuous flavor symmetry and
$\epsilon$ denotes the fugacity associated to the $\mathbb{Z}_n$ flavor symmetry.

Since the theory is free, the Schur limit is relatively simple and can be written as
\begin{equation}
\label{eq:Zn-Schur-PE}
  \mathcal{I}_{\mathbb{Z}_{n}}(q;a)
  = \frac{q^{1/8}}{n}\sum_{\epsilon\in\mathbb{Z}_n}
    \mathrm{PE}\!\left[
      \frac{\sqrt{q}\,(a\epsilon+(a\epsilon)^{-1})}{1-q}
      -\frac{q\,(\epsilon+\epsilon^{-1})}{1-q}
    \right],
\end{equation}
where the plethystic exponential is defined by
\begin{equation}
  \mathrm{PE}[f(q,a,\epsilon)] \equiv \exp\!\left(\sum_{m\geq 1}\frac{f(q^{m}, a^m, \epsilon^m)}{m}\right).
\end{equation}
The Schur index can also be rewritten in terms of Jacobi theta functions,
\begin{equation}
    \mathcal{I}_{\mathbb{Z}_n}(a, q) = \frac{1}{n} \sum_{\epsilon} \frac{a^{\frac{1}{2}}q^{-1/8}\vartheta_1(\mathfrak{e})}{(1 - \epsilon^{-1})\vartheta_1(\mathfrak{a} + \mathfrak{e} - \frac{\tau}{2})} \ .
\end{equation}

\subsubsection{\texorpdfstring{$n = 1$}{n=1}}

The associated VOA of a free vector and a free hypermultiplet is given by the small $bc$ ghost system and a pair of symplectic bosons $\beta, \gamma$, with conformal weights
\begin{equation}
    h[b] = h[\partial c] = 1, \quad h[c] = 0, \quad h[\beta] = h[\gamma] = \frac{1}{2} \ .
\end{equation}
The hypermultiplet has an $SU(2)$ flavor symmetry rotating the two scalars $q, \tilde q$, which descend to $\beta, \gamma$, respectively, at the level of the associated VOA. The Schur index of this theory is given by
\begin{equation}
    \operatorname{ch} = q^{- \frac{c_{\mathrm{2d}}}{24}} \frac{(q;q)(q;q)}{(aq^{\frac{1}{2}};q)(a^{-1}q^{\frac{1}{2}};q)} = \frac{\eta(\tau)^3}{\vartheta_4(\mathfrak{a})} \ , \qquad c_{\mathrm{2d}} = c_{bc} + c_{\beta \gamma} = -3.
\end{equation}
This free VOA contains the small 2d $\mathcal{N} = 4$ superconformal algebra as a sub-VOA, which we will discuss in the $n = 2$ case. Although the VOA is strongly generated by the minimal set $b, \partial c, \beta, \gamma$, we may artificially include the generators in the small 2d $\mathcal{N} = 4$ superconformal algebra. The operators $\beta, b, c, \gamma'$ constitute a short $\mathfrak{psl}(2|2)$ multiplet $\mathfrak{G}_{\frac{1}{2}}$, while $\gamma, - b, - c, \beta'$ form another.

\subsubsection{\texorpdfstring{$n = 2$}{n=2}}

The Schur index of this theory is given by
\begin{equation}
    \operatorname{ch} = \frac{1}{2}\frac{\eta(\tau)^3}{\vartheta_4(\mathfrak{a})} + \frac{\vartheta_2(0)}{4 \vartheta_3(\mathfrak{a})} \ .
\end{equation}
The unflavored index in $q$-series is
\begin{equation}
    \operatorname{ch} = q^{1/8}+3 q^{9/8}-4 q^{13/8}+10 q^{17/8}-16 q^{21/8}+29 q^{25/8} + \cdots \ ,
\end{equation}
which is easily shown to solve the unflavored MLDE
\begin{align}
0 = \Big( & D_q^{(4)} + 6\Em{2}{1} D_q^{(3)} + \Big[-100 E_4(\tau) + 12\Em{2}{1}^2\Big]D_q^{(2)} \nonumber\\
& + \Big[-134\Em{2}{1}^3 + 270 E_4(\tau)\Em{2}{1}\Big]D_q^{(1)} \nonumber\\
& + \Big[-1125 E_4(\tau)^2 + 450 E_4(\tau)
\Em{2}{1}^2 - 45\Em{2}{1}^4\Big]\Big)\operatorname{ch} \ .
\end{align}
The indicial roots are $\alpha = - \frac{1}{8}, [\frac{1}{8}]_2, \frac{5}{8}$, where the double degeneracy of $\frac{1}{8}$ implies the existence of a logarithmic solution. Concretely, the four solutions are given by
\begin{align}
    \operatorname{ch}_{\frac{1}{8}} = & \ \frac{\eta(\tau)^3}{2\vartheta_4(0)} + \frac{\vartheta_2(0)}{4\vartheta_3(0)},
    & \operatorname{ch}_{\frac{5}{8}} = & \ \frac{\eta(\tau)^3}{2\vartheta_4(0)} - \frac{\vartheta_2(0)}{4\vartheta_3(0)} \ , \cr
    \operatorname{ch}_{-\frac{1}{8}} = & \ \frac{\vartheta_3(0)}{\vartheta_2(0)} \ , 
    & \operatorname{ch}_{\log} = & \ \tau \frac{\eta(\tau)^3}{\vartheta_4(0)} \ .
\end{align}

To further understand the full representation theory, we look at the VOA in a bit more detail. The associated VOA is given by the $\mathbb{Z}_2$-quotient of the $bc \beta \gamma$ ghost system with OPE
\begin{equation}
    \beta(z) \gamma(w) \sim \frac{-1}{z - w}, \qquad
    b(z) c(w) \sim \frac{1}{z - w} \ ,
\end{equation}
and the following quantum number assignments.
\begin{table}[ht]
    \centering
    \begin{tabular}{c|c| c}
            & $h$ & $m$ \\
            \hline
        $b$ & $1$ & $1$\\
        $c$ & $0$ & $-1$\\
        $\beta$ & $1/2$ & $1$ \\
        $\gamma$ & $1/2$ & $-1$ 
    \end{tabular}
    \caption{Quantum numbers of the $bc\beta\gamma$ system realizing the VOA of the $\mathcal{N} = 4$ $U(1)$ theory.}
    \label{tab:bcbg-qnums-u1}
\end{table}
The quotient projects the $bc \beta \gamma$ VOA onto the subspace with $m \in 2 \mathbb{Z}$. The subspace is the 2d small $\mathcal{N} = 4$ superconformal algebra at 2d central charge $c_{\mathrm{2d}} = -3$, with strong generators given by
\begin{align}
  T = & \ - (b c') - \frac{1}{2} (\beta \gamma') + \frac{1}{2} (\beta ' \gamma), \quad
  J = (\beta \gamma), \quad
  J^+ = (\beta \beta), \quad
  J^- = (\gamma \gamma) \ , \cr
  G^+ = & \ (\beta b), \quad
  G^- = (\gamma b), \quad
  \tilde G^+ = (\beta c'), \quad
  \tilde G^- = (\gamma c') \ .
\end{align}
The currents $J, J^\pm$ generate an $SU(2)_{-\frac{1}{2}}$ subalgebra, and the stress tensor $T$ is an independent generator,
\begin{equation}
    T = T_\text{Sug} - (b \partial c) \ , \quad
    c_{\mathrm{2d}} - c_\text{Sug} = c_{bc} = -2 \ .
\end{equation}
Note that this $bc \beta \gamma$ system differs from the one used in \cite{Bonetti:2018fqz}: there, one has $m(\beta) - m(b) = 1$, whereas here we have $m(\beta) = m(b)$ instead. Null states can nonetheless be constructed out of the free fields; we list a few of low conformal weight. The singular vector occurs at weight three,
\begin{align}
    \mathcal{N}_3^+ = & \ (G^+ \tilde G^+) - \frac{1}{5}(J(JJ^+)) + \frac{9}{10}(J \partial J^+)
    + \frac{1}{5} (J^+(J^+J^-)) \cr
    & \ + (TJ^+) + \frac{3}{10} (\partial J J^+)
    - \frac{3}{2} \partial^2 J^+ \ ,
\end{align}
which generates two further weight-three null states $\mathcal{N}_3^0$ and $\mathcal{N}_3^-$ under the action of $J_0^-$, forming an $SU(2)$ triplet. The three null states imply
\begin{equation}
    j^a \bigl(t - (j^2 - j^+ j^-)\bigr) = 0 \ ,
\end{equation}
in the commutative Poisson algebra $R_{\mathbb{V}}$, where $j^a$ denotes
$j,j^\pm$ collectively, and $j,j^\pm,t$ are the images of $J,J^\pm,T$,
respectively.  In addition, at weight six, there is a null state
$\mathcal{N}_T = L_{-2}^3|0\rangle + c_2$, with $c_2 \in C_2(\mathbb{V})$.
Its image in $R_{\mathbb{V}}$ gives $t^3=0$; hence $t=0$ only after passing
to $R_{\mathbb{V}}^{\mathrm{red}}$.  The radical of the resulting triplet
relations is generated by $j^2-j^+j^-$, and therefore
\begin{equation}
    R_{\mathbb{V}}^{\mathrm{red}}
    \cong \frac{\mathbb{C}[j,j^+,j^-]}{(j^2-j^+j^-)} \ ,
    \qquad
    X_{\mathbb{V}} \cong \mathbb{C}^2/\mathbb{Z}_2 \ .
\end{equation}

The uncharged null $J_0^- \mathcal{N}_3^+$ gives rise to a flavored MLDE ($D_a \coloneqq a\partial_a$)
\begin{align}
0 = \Big( & D_{a} D_q^{(1)} + \Em{1}{a} D_q^{(1)} -\frac{1}{5}\, D_{a}^3 -\frac{4}{5}\, \Ep{1}{a^{2}} D_{a}^{2} \nonumber\\
& + \Big[\frac{8}{5}\, E_2(\tau)\, + \Em{2}{a} + 4\, \Ep{2}{a^{2}} + \frac{8}{5}\, \Ep{1}{a^{2}}^{2}\Big]D_{a} \nonumber\\
& + \Big[-\Em{3}{a} - \frac{14}{5}\, \Ep{3}{a^{2}} + \frac{4}{5}\, E_2(\tau)\, \Ep{1}{a^{2}} - \frac{4}{5}\, \Ep{2}{a^{2}}
  \Ep{1}{a^{2}}\Big]\Big)\operatorname{ch} \ .
\end{align}
At weight-4, there are 9 uncharged null states,
\begin{align}
    (J^- \mathcal{N}_3^+), \quad (J^0\mathcal{N}_3^0), 
    \quad (J^+ \mathcal{N}_3^-),
    \quad
    \partial\mathcal{N}_3^0,
    \quad
    \{G^- \{\tilde G^- \mathcal{N}_3^+\}_1\}_1 \ ,
    \cr
    \{ G^+ \{ G^- \mathcal{N}_3^0\}_1\}_1,\quad
    \{ G^+ \{ \tilde G^- \mathcal{N}_3^0\}_1\}_1,\quad
    \{ \tilde G^+ \{ G^- \mathcal{N}_3^0\}_1\}_1,\quad
    \{ \tilde G^+ \{\tilde G^- \mathcal{N}_3^0\}_1\}_1 \ .
\end{align}
and at weight-5, there are 33 such states. They give rise to a set of flavored MLDEs that determine the space of characters as $q$-series,
\begin{align}
    \operatorname{ch}_{\frac{5}{8}} = & \ (a + a^{-1})q^{\frac{5}{8}} - 2 q^{\frac{9}{8}} + (a^3 + 2a + 2a^{-1} + a^{-3})q^{\frac{13}{8}} + \cdots \ ,\cr
    \operatorname{ch}_{\frac{1}{8}} = & \ q^{\frac{1}{8}} + (1 + a^2 + a^{-2})q^{9/8} - 2(a + a^{-1})q^{\frac{13}{8}} + \cdots \ ,\cr
    \operatorname{ch}^{(1)}_{- \frac{1}{8}} = &
    \frac{a^{\frac{1}{2}}}{1 - a^2}q^{- \frac{1}{8}}
    - \frac{2a^{3/2}}{1- a^2} q^{\frac{3}{8}}
    + \frac{2a^{\frac{1}{2}} + a^{\frac{5}{2}}}{1 - a^2}q^{\frac{7}{8}}
    - \frac{2a^{-\frac{1}{2}} + 4 a^{\frac{3}{2}}}{1 - a^2}q^{\frac{11}{8}} + \cdots \ ,\cr
    \operatorname{ch}^{(2)}_{- \frac{1}{8}}
    = & \ \frac{a^{3/2}}{1 - a^2}q^{-\frac{1}{8}} - \frac{2a^{\frac{1}{2}}}{1 - a^2} q^{\frac{3}{8}} + \frac{a^{-\frac{1}{2}} + 2a^{\frac{3}{2}}}{1 - a^2}q^{\frac{7}{8}} - \frac{2 a^{\frac{1}{2}}(2 + a^2)}{1 - a^2} q^{\frac{11}{8}} + \cdots \ .
\end{align}
The non-logarithmic solution with $\alpha = \frac{1}{8}$ is the Schur index of the theory. The root $\alpha = - \frac{1}{8}$ admits an extra solution that is singular as $a \to 1$ and is therefore invisible in the unflavored MLDE. The remaining solutions can also be generated by acting with the modular transformations $STS$ and $T^2$ on the Schur index,
\begin{align}
    \operatorname{ch}_{\frac{5}{8}}
    = & \ \frac{\eta(\tau)^3}{2\vartheta_4(\mathfrak{a})} - \frac{\vartheta_2(0)}{4 \vartheta_3(\mathfrak{a})},
    & \operatorname{ch}_{\frac{1}{8}}
    = & \ \frac{\eta(\tau)^3}{2 \vartheta_4(\mathfrak{a})} + \frac{\vartheta_2(0)}{4 \vartheta_3(\mathfrak{a})} \ ,
    \cr
    \operatorname{ch}^{(1)}_{- \frac{1}{8}}
    = & \ \frac{1}{2}\bigg(
    \frac{i \vartheta_4(0)}{\vartheta_1(\mathfrak{a})}
    + \frac{\vartheta_3(0)}{\vartheta_2(\mathfrak{a})}
    \bigg),
    & \ \operatorname{ch}^{(2)}_{- \frac{1}{8}}
    = & \
    \frac{1}{2}\bigg(
    \frac{i \vartheta_4(0)}{\vartheta_1(\mathfrak{a})}
    - \frac{\vartheta_3(0)}{\vartheta_2(\mathfrak{a})}
    \bigg) \ .
\end{align}
There is one logarithmic solution obtained by applying the $STS$ transformation to a linear combination of $\operatorname{ch}_{\frac{5}{8}}$ and $\operatorname{ch}_{\frac{1}{8}}$,
\begin{align}
    \operatorname{ch}_{\log} = \frac{\tau \eta(\tau)^3}{\vartheta_4(\mathfrak{a})} \ .
\end{align}
The Schur index with $n = 1$ appears as the module character
\begin{equation}
    \operatorname{ch}_{n = 1} = \operatorname{ch}_{\frac{1}{8}} + \operatorname{ch}_{\frac{5}{8}}  \ .
\end{equation}
This is because the small $bc\beta \gamma$ system forms a reducible highest weight module of the small $\mathcal{N} = 4$ superconformal algebra. The highest weight state of the small $bc\beta \gamma$ system is given by the standard vacuum $|0\rangle$, while the states $\beta = \beta_{-\frac{1}{2}}|0\rangle, \gamma = \gamma_{- \frac{1}{2}}|0\rangle$ form the highest weight states of a $\mathbb{Z}_2$-odd submodule with conformal weight $h = \frac{1}{2}$. The small $bc\beta \gamma$ system is a direct sum of the irreducible $h = 0$ and $h = \frac{1}{2}$ modules.

The fact that the modular transform of a solution remains a solution of the flavored MLDEs is again a manifestation of quasi-modularity. Concretely, the flavored MLDE associated with a null state $\mathcal{N}$ transforms into a linear combination of equations of equal or lower modular weight. Written in terms of the corresponding null states,
\begin{equation}
    \mathcal{N} \xrightarrow{STS} \sum_{n \ge 0} \sum_{i \ge 0} \tau^n \mathfrak{a}^i(-1)^{n+ h[\mathcal{N}] - i} \begin{pmatrix}
        h[\mathcal{N}] - i\\
        n
    \end{pmatrix} (J_1)^i \mathcal{N} \ .
\end{equation}
This is identical to the relation we encounter in the $\mathcal{W}_{\mathbb{Z}_3}$ case.

\subsubsection{\texorpdfstring{$n = 3$}{n=3}}
Next, we turn to $n = 3$. The associated VOA is given by the $\mathbb{Z}_3$-quotient of one $bc \beta \gamma$ system with central charge $c_{\mathrm{2d}} = -3$. It is strongly generated by $T, J, G, \tilde G, W, Y, V, P, Q, R, S$. The generators can be realized explicitly in terms of the small $bc \beta \gamma$ system associated to the $\mathcal{N} = 4$ $U(1)$ SYM. Explicitly,
\begin{align}
    T = & \ - (b c') - \frac{1}{2}(\beta \gamma') + \frac{1}{2} (\beta'\gamma) \ ,
    \qquad J = (\beta \gamma) \ ,\qquad
    G = (b \gamma), \qquad
    \tilde G = (\beta c'), \cr
    W = & \ \beta^3, \qquad Y = \gamma^3, \qquad
    V = 12 (b c') - 12 (\beta \gamma') + 12 (\beta'\gamma) \ , \cr
    P = & \ (\beta (\beta b)), \qquad
    Q = (\gamma(\gamma c'))\,, \qquad
    R = - \frac{1}{3} (\beta c'') + \frac{2}{3}(\beta'c'), \qquad
    S = (\beta (\gamma(\gamma b))) \ .
\end{align}
The first line constitutes the 2d $\mathcal{N} = 2$ superconformal algebra.

Rewriting the Schur index \eqref{eq:Zn-Schur-PE} in terms of Jacobi theta functions
yields, for $n=3$,
\begin{equation}
\label{eq:I_Z3_flavored}
\begin{aligned}
  \mathcal{I}_{\mathbb{Z}_{3} ~ \text{gauged}}(q;a)
  &= \frac{1}{3}\left(
      \frac{\eta(3\tau)}{\vartheta_{3}\!\left(\mathfrak{a}-\tfrac{1}{6}\,\middle|\,\tau\right)}
      +\frac{\eta(3\tau)}{\vartheta_{4}\!\left(\mathfrak{a}-\tfrac{1}{3}\,\middle|\,\tau\right)}
      +\frac{\eta(\tau)^{3}}{\vartheta_{4}\!\left(\mathfrak{a}\,\middle|\,\tau\right)}
    \right)\\[4pt]
  &= q^{\frac{1}{8}}
    + q^{\frac{9}{8}}
    + \frac{(-1+a^{2})^{2}(1+a^{2})}{a^{3}}\,q^{\frac{13}{8}}
    + \left(4-a^{-2}-a^{2}\right)q^{\frac{17}{8}}
    + \cdots,
\end{aligned}
\end{equation}
where $q = e^{2\pi i\tau}$ and we use additive notation $a = e^{2\pi i\mathfrak{a}}$.

In the unflavored limit $a\to 1$, the first two theta-quotients in
\eqref{eq:I_Z3_flavored} coincide, and the index simplifies to
\begin{equation}
\label{eq:I_Z3_unflavored}
  \mathcal{I}_{\mathbb{Z}_{3}}(q;1)
  = \frac{1}{3}\left(
      2\,\frac{\eta(3\tau)}{\vartheta_{3}\!\left(\tfrac{1}{6}\,\middle|\,\tau\right)}
      +\frac{\eta(\tau)^{3}}{\vartheta_{4}(0\,|\,\tau)}
    \right)
  = q^{\frac{1}{8}}\!\left(1+q+2q^{2}-2q^{\frac{5}{2}}+3q^{3}-2q^{\frac{7}{2}}+\cdots\right).
\end{equation}
A direct computation shows that \eqref{eq:I_Z3_unflavored} is annihilated by the
same fourth-order MLDE operator \eqref{eq:MLDE-order4} that governs the Schur
index of the $\mathbb{Z}_3$ S-fold theory.
Consequently, $\mathcal{I}_{\mathbb{Z}_{3}}(q;1)$ lies in the four-dimensional
solution space spanned by
$\bigl\{\operatorname{ch}_{\textrm{vac}},\,\operatorname{ch}_{\log},\,
\operatorname{ch}_{-\frac{1}{24}},\,\operatorname{ch}_{\frac{1}{8}}\bigr\}$ that are the unflavored characters of
$\mathcal{W}_{\mathbb{Z}_3}$ identified in section~\ref{sec:MLDE}.
Explicitly, the Schur index here can be written as a simple linear combination of the characters of $\mathcal{W}_{\mathbb{Z}_3}$,
\begin{equation}
\label{eq:I_Z3_decompose}
  \mathcal{I}_{\mathbb{Z}_{3} ~ \text{gauged}}(q;1)
  = \operatorname{ch}_{\frac{1}{8}}(q) + \operatorname{ch}_{\textrm{vac}}(q).
\end{equation}

Similarly, the unflavored Schur index of the parent $U(1)$ $\mathcal{N} = 4$ theory is also a linear combination of unflavored $\mathcal{W}_{\mathbb{Z}_3}$ characters. Using theta-function identities,
\begin{equation}
\label{eq:I_SYM}
  \mathcal{I}_{\mathrm{SYM}}
  = \eta(q)^{2}\,\frac{\eta(q)}{\vartheta_{4}(0|\tau)}
  = \frac{\eta(q)^{4}}{\eta(\sqrt{q})^{2}}
  = q^{\frac{1}{8}}\left(1+2q^{\frac{1}{2}}+q
    +2q^{\frac{3}{2}}+2q^2+\cdots\right),
\end{equation}
which decomposes as
\begin{equation}
\label{eq:I_SYM_decompose}
  \mathcal{I}_{\mathrm{SYM}}
  = \operatorname{ch}_{\frac{1}{8}}(q) + 3\,\operatorname{ch}_{\textrm{vac}}(q).
\end{equation}
Equations \eqref{eq:I_Z3_decompose} and \eqref{eq:I_SYM_decompose} exhibit
the same two non-logarithmic characters appearing with different multiplicities,
reflecting the distinct ways in which the $\mathbb{Z}_{3}$-gauged theory and its
$\mathcal{N} = 4$ parent populate the module spectrum of the shared VOA.

To obtain the full representation theory of the algebra, one needs to analyze in detail the null states and the flavored MLDEs. It is straightforward to compute the null states. In particular, at weight $7/2$, there are two singular vectors:
\begin{align}
\mathcal{N}_1={}&-\frac{13}{5}(JR)
+\frac{1}{10}\bigl(J(J\tilde G)\bigr)
-\frac{13}{15}(J\tilde G')
+(T\tilde G)
-\frac{1}{10}(WQ)+\frac{3}{5}(J'\tilde G)
+R'
-\frac{1}{6}\tilde G'',\cr
\mathcal{N}_2={}&\frac{13}{5}(JS)
-\frac{5}{2}\bigl(J(JG)\bigr)
-(JG')
+(TG)
-\frac{1}{10}(YP)-\frac{8}{5}(J'G)
+S'
-\frac{1}{2}G''.
\end{align}
These two states generate a 6-dimensional space of singular vectors by taking OPE with $V, P, Q$, all annihilated by the positive modes of all strong generators. These null states generate the null ideal of the VOA. 

The flavored MLDEs from the null states are more involved, since they are not equations for the character alone, but also involve $n$-point functions $\operatorname{str}\bigl[o(V)\cdots o(V)\, q^{L_0 - \frac{c_{\mathrm{2d}}}{24}}a^f\bigr]$ of the neutral operator $V$. For instance,
\begin{align}
& \Big( D_q^{(2)} - \tfrac{108}{31}\Em{1}{a^3}D_{a}D_q^{(1)} + \tfrac{9}{31}\Em{1}{a^3}D_{a}^{3}
  + \Big[\tfrac{193}{31}E_2(\tau) + \tfrac{324}{31}\Em{1}{a^3}^{2} + \tfrac{684}{31}\Em{2}{a^3}\Big]D_q^{(1)} \nonumber\\
& + \Big[-\tfrac{468}{31}E_2(\tau) - \tfrac{243}{31}\Em{1}{a^3}^{2} - \tfrac{675}{31}\Em{2}{a^3}\Big]D_{a}^{2} \nonumber\\
& + \Big[\tfrac{486}{31}\Em{1}{a^3}\Em{2}{a^3} + \tfrac{1620}{31}\Em{3}{a^3} - \tfrac{567}{31}E_2(\tau)\Em{1}{a^3}\Big]D_{a} \nonumber\\
& + \Big[\tfrac{234}{31}E_2(\tau)^2 - \tfrac{1098}{31}E_4(\tau) + \tfrac{243}{31}E_2(\tau)\Em{1}{a^3}^{2} \nonumber\\
& \quad + \tfrac{675}{31}E_2(\tau)\Em{2}{a^3} - \tfrac{162}{31}\Em{1}{a^3}\Em{3}{a^3} - \tfrac{702}{31}\Em{4}{a^3}\Big] \Big)\operatorname{ch} \nonumber\\
& + \Big( -\tfrac{23}{124}D_{a}^{2} - \tfrac{9}{31}\Em{1}{a^3}D_{a} + \tfrac{1}{6}q\partial_q \nonumber\\
& \quad + \Big[\tfrac{85}{124}E_2(\tau) + \tfrac{27}{31}\Em{1}{a^3}^{2} + \tfrac{57}{31}\Em{2}{a^3}\Big] \Big)\operatorname{str}(o[V]) \nonumber\\
& + \tfrac{1}{144}\operatorname{str}\bigl(o[V]\,o[V]\bigr) = 0 \ .
\end{align}
This falls into the more general situation discussed in \cite{Arakawa:2026nvd} where conformal blocks of the VOAs are constrained by a set of first-order flavored MLDEs.

\subsection{Gauging \texorpdfstring{$\mathbb{Z}_n$}{Zn} in the \texorpdfstring{$SU(2)$}{SU(2)} theory}

Finally, we present some results for the $\mathcal{N} = 4$ $SU(2)$ theory with a $\mathbb{Z}_n$ symmetry gauged, which we will simply refer to as the $\mathcal{N} = 3$ $SU(2)/\mathbb{Z}_n$ theory (not to be confused with a theory with ``gauge group $SU(2)/\mathbb{Z}_n$''). For general $SU(N)$, the Coulomb branch operators of the $\mathcal{N} = 4$ theory are given by $\operatorname{tr}\phi^2, \cdots, \operatorname{tr}\phi^N$, having dimension $E = 2, 3, \cdots, N$. The 4d central charge $c_{\mathrm{4d}} = a_{\mathrm{4d}} = \frac{1}{4}(N^2 - 1)$ follows from the Shapere--Tachikawa formula \cite{Shapere:2008zf}. After discrete gauging, the $a,c$ central charges remain unchanged, and ultimately the 2d central charge of the resulting $\mathcal{N} = 3$ theory is $c_{\mathrm{2d}} = -3(N^2 - 1)$.

The VOA of the $\mathcal{N} = 3$ $SU(N)/\mathbb{Z}_n$ theory is simply the $\mathbb{Z}_n$-quotient of that of the $SU(N)$ $\mathcal{N} = 4$ theory. For simplicity, we consider $N = 2$. The $SU(2)$ $\mathcal{N} = 4$ SYM is an $SU(2)_\text{gauge}$ vector multiplet coupled to an $SU(2)_\text{gauge}$-adjoint hypermultiplet in $\mathcal{N} = 2$ language. It has an associated VOA given by the small $\mathcal{N} = 4$ superconformal algebra in two dimensions, where the generators $T, J^{(AB)}, G^\pm, \tilde G^\pm$ descend from the Schur operators in four dimensions (here we are not careful about the relative coefficients and normalizations)
\begin{align}
    T = & \ [\operatorname{tr}(\epsilon_{AB}q^A D_{+\dot +} q^B + \lambda^1_+ \tilde \lambda^1_{\dot +})], 
    & J^{(AB)} = & \ [\operatorname{tr} (q^A q^B)], \\
    G^A = & \ [\operatorname{tr} q^A \lambda^1_{+}],
    & \tilde G^A = &\ [\operatorname{tr} q^A \tilde \lambda^1_{\dot +}] \ .
\end{align}
Here, $q^{A = \pm 1}$ denotes the $SU(2)_\text{gauge}$ adjoint-valued scalars in the hypermultiplet, $\lambda^I_\pm$, $\tilde \lambda^I_{\dot \pm}$ denote the fermions in the vector multiplet, and $\operatorname{tr}$ traces over the $SU(2)_\text{gauge}$-adjoint index, forming gauge singlets. The $J^{(AB)}$ generate an $\widehat{\mathfrak{su}}(2)_{-3/2}$ affine subalgebra descending from the $SU(2)_f$ flavor symmetry of the 4d theory. Note that in the free limit, $q^A$, $\lambda^1_+$, $\tilde \lambda^1_{\dot +}$ are separately Schur operators. Upon turning on gauge coupling, only gauge-invariant operators are present in the spectrum, and moreover, only the BRST cohomology of the space of all gauge-singlets survives as the Schur operators of the interacting theory, namely those generated by the generators presented above \cite{Beem:2013sza}.

By inspecting the refined single letter index of the $\mathcal{N} = 4$ vector multiplet, the $\mathbb{Z}_n$ refinement implies the following charge assignment of the free fields in Table~\ref{tab:Zn-charge-of-N=4-vector-multiplet}. As a result, we read off the $\mathbb{Z}_n$-charges of the strong generators in Table~\ref{tab:Zn-charge-small-N4-superconformal-algebra}.
\begin{table}[ht]
    \centering
    \begin{tabular}{c|c|c|c|c}
    & $q^+$ & $q^-$ &  $\lambda^1_+$ & $\tilde \lambda^1_{\dot +}$ \\
    \hline
    $\mathbb{Z}_n$ charge & $+1$ & $-1$ & $+1$ & $-1$ 
    \end{tabular}
    \caption{$\mathbb{Z}_n$ charge of fields in the vector multiplet.}
    \label{tab:Zn-charge-of-N=4-vector-multiplet}
\end{table}
\begin{table}[ht]
\centering
\begin{tabular}{c|c|c|c|c|c|c|c|c}
     & $J^- = J^{(--)}$ & $J^0 = J^{(+-)}$ & $J^+ = J^{(++)}$& $T$ & $G^+$ &  $G^-$ & $\tilde G^+$ & $\tilde G^-$\\
     \hline
     $\mathbb{Z}_n$ charge & $-2$ & $0$ & $+2$ & 
     $0$ & $+2$ & $0$ & $0$ & $-2$
\end{tabular}
\caption{The $\mathbb{Z}_n$-charges of the 2d small $\mathcal{N} = 4$ superconformal algebra.}
\label{tab:Zn-charge-small-N4-superconformal-algebra}
\end{table}
The original $\mathcal{N} = 4$ small superconformal algebra admits a free-field realization given by a single $bc\beta\gamma$ system \cite{Adamovic:2014lra,Bonetti:2018fqz},
\begin{align}\label{eq:bcbetagamma-realization-of-small-superconformal-algebra}
    J^+ = & \ \beta, \quad
    J^0 = (bc) + 2 (\beta \gamma), \quad
    J^- = - (\beta(\gamma\gamma)) - (\gamma(bc)) + \frac{3}{2} \partial \gamma\ , \nonumber\\
    T = & \ - (b\partial c) - \frac{1}{2}\partial (b c) - \beta \partial \gamma\ , \\
    G^+ = & \ b\ , \qquad
    G^- = (b\gamma)\ , \nonumber \\
    \tilde G^+ = & \ (c \partial \beta) + 2 \partial c \beta, \quad
    \tilde G^- = - (b(\partial c c)) + 2 (\beta (\gamma \partial c)) + (\partial \beta (\gamma c)) - \frac{3}{2}\partial^2 c \ . \nonumber
\end{align}
Here, $\beta(z)\gamma(w) \sim - 1/(z-w)$, $b(z)c(w)\sim 1/(z-w)$.
The $\mathbb{Z}_n$-charge of the generators discussed above can be alternatively realized by assigning the following charges in Table~\ref{tab:Zn-charge-of-ghost-system}.
\begin{table}[h]
    \centering
    \begin{tabular}{c|cccc}
         &   $\beta$& $\gamma$& $b$&$c$\\
         \hline
         $\mathbb{Z}_n$-charge& 
     $+2$& $-2$& $+2$&$-2$\end{tabular}
    \caption{$\mathbb{Z}_n$-charge assignment of the free fields.}
    \label{tab:Zn-charge-of-ghost-system}
\end{table}

\subsubsection{The \texorpdfstring{$\mathbb{Z}_n$}{Zn}-refined Schur index}

Given the $\mathbb{Z}_n$-charge of generators, the refined Schur index of the $\mathcal{N} = 4$ $SU(2)$ theory in $q$-series (to $q^5$) immediately follows\footnote{This index coincides with the twisted Schur index discussed in \cite{Maruyoshi:2026cmr}.},
\begin{align}\label{eq:Zn-refined-Schur-index}
& q^{-3/8}\mathcal{I}_{\mathcal{N} = 4}(q ; a, \epsilon) = 1
+\left(
1+\frac{1}{a^{2}\epsilon^{2}}
+a^{2}\epsilon^{2}
\right)q
-\frac{
(1+\epsilon^{2})(1+a^{2}\epsilon^{2})
}{a\epsilon^{2}}q^{3/2}
+\frac{
(1+a^{2}\epsilon^{2}+a^{4}\epsilon^{4})^{2}
}{a^{4}\epsilon^{4}}q^{2}
\nonumber\\
&\quad
-\frac{
(1+\epsilon^{2})
(1+2a^{2}\epsilon^{2}
+2a^{4}\epsilon^{4}
+a^{6}\epsilon^{6})
}{a^{3}\epsilon^{4}}q^{5/2}
\nonumber\\
&\quad
+\left(
6+\frac{1}{a^{6}\epsilon^{6}}
+\frac{2}{a^{4}\epsilon^{4}}
+\frac{5}{a^{2}\epsilon^{2}}
+5a^{2}\epsilon^{2}
+2a^{4}\epsilon^{4}
+a^{6}\epsilon^{6}
\right)q^{3}
\nonumber\\
&\quad
-\frac{
(1+\epsilon^{2})
(1+a^{2}\epsilon^{2})
(1+a^{2}\epsilon^{2}
+a^{4}\epsilon^{4})^{2}
}{a^{5}\epsilon^{6}}q^{7/2}
\nonumber\\
&\quad
+\left(
14+\frac{1+\epsilon^{-4}
+11\epsilon^{-2}}{a^{2}}
+\frac{1}{a^{8}\epsilon^{8}}
+\frac{2}{a^{6}\epsilon^{6}}
+\frac{6}{a^{4}\epsilon^{4}}
+\epsilon^{-2}
+\epsilon^{2}
\right.
\nonumber\\
&\qquad\left.
+6a^{4}\epsilon^{4}
+2a^{6}\epsilon^{6}
+a^{8}\epsilon^{8}
+a^{2}(1+11\epsilon^{2}+\epsilon^{4})
\right)q^{4}
\nonumber\\
&\quad
-\frac{
(1+\epsilon^{2})
\left(
1+3a^{2}\epsilon^{2}
+7a^{4}\epsilon^{4}
+11a^{6}\epsilon^{6}
+11a^{8}\epsilon^{8}
+7a^{10}\epsilon^{10}
+3a^{12}\epsilon^{12}
+a^{14}\epsilon^{14}
\right)
}{a^{7}\epsilon^{8}}q^{9/2}
\nonumber\\
&\quad
+\frac{
(1+a^{2}\epsilon^{2}+a^{4}\epsilon^{4})^{2}
\left(
1+3a^{4}\epsilon^{4}
+3a^{8}\epsilon^{8}
+a^{12}\epsilon^{12}
+a^{6}(\epsilon^{4}
+5\epsilon^{6}
+\epsilon^{8})
\right)
}{a^{10}\epsilon^{10}}q^{5} + \cdots.
\end{align}
We emphasize that this index \emph{cannot} be computed using the naive Molien--Weyl (unit-circle) integration,
\begin{equation}\label{eq:naive-Molien--Weyl-integration}
    q^{-3/8}\mathcal{I}_{\mathcal{N} = 4}(q ; a, \epsilon)
    \ne \oint \frac{dz}{2\pi i z} \operatorname{Haar}(z)
    \operatorname{PE}\bigg[ i_\text{s.l.}(q; a, \epsilon, z) \bigg] \ ,
\end{equation}
where $i_\text{s.l.}$ is the Schur limit of the refined superconformal single letter index of the $\mathcal{N} = 4$ vector multiplet. The reason is as follows. 

Without the $\mathbb{Z}_n$-refinement, the $\mathcal{N} = 4$ theory can be obtained by gauging the diagonal of an $SU(2) \times SU(2)$ flavor subgroup of the $SU(2)^3 \subset USp(8)$ flavor symmetry of four free hypermultiplets. At the level of chiral algebra, one needs to perform a BRST reduction of the free chiral algebra using the BRST charge \cite{Beem:2013sza}
\begin{equation}
    Q_\text{BRST} = \oint \frac{dz}{2\pi i } c_a (J^a + \frac{1}{2}J^a_\text{gh}) \ .
\end{equation}
The current $J^a$ denotes the $\widehat{\mathfrak{su}}(2)_{k_{\mathrm{2d}}}$ currents with $a$ the $SU(2)_\text{gauge}$ index, and $b, c$ are the ghost fields associated to the adjoint vector multiplet. In particular, $c_a$ carries $\mathbb{Z}_n$-charge $-1$ while $J^a, J^a_\text{gh}$ are neutral, and hence $Q_\text{BRST}$ carries a non-zero $\mathbb{Z}_n$-charge. The Molien--Weyl integration projects the product space $\mathbb{V}[\text{HM}] \otimes \mathbb{V}[\text{VM}]$ down to its gauge invariant subspace. This subspace contains $Q_\text{BRST}$-non-closed states and $Q_\text{BRST}$-exact states, which will be removed by going to the $Q_\text{BRST}$ cohomology. At the level of the integration, their contributions precisely cancel, giving the correct supertrace over the chiral algebra. However, with $\mathbb{Z}_n$ refinement, the states inside a canceling pair do not share the same $\mathbb{Z}_n$-charge, and therefore do not automatically reproduce the refined supertrace over the chiral algebra. For concreteness, we show this discrepancy at $(h = 2, f = 0)$. The gauge invariant words in this sector are listed in Table~\ref{tab:Zn-charge-of-gauge-singlets},
\begin{table}[h]
    \centering
    \begin{tabular}{c|c|c|c|c|c|c|c}
        & $\operatorname{tr}(b \beta \gamma) $
        & $\operatorname{tr}(c \beta \gamma)$
        & $\operatorname{tr}(b c)$
        & $\operatorname{tr}(\beta \partial \gamma)$
        & $\operatorname{tr}(\partial \beta \gamma)$
        & $\operatorname{tr}(\mu \mu)$
        & $\operatorname{tr}(\beta \gamma)^2$ \\
        \hline
        $\mathbb{Z}_n$-charge & $+1$ & $-1$ & $0$ & $0$ & $0$ & $0$ & 0
    \end{tabular}
    \caption{$\mathbb{Z}_n$-charges of the gauge-invariant words at $(h,f) = (2,0)$.}
    \label{tab:Zn-charge-of-gauge-singlets}
\end{table}
where $\mu^a = \epsilon^{abc} \operatorname{tr}\beta_{b} \gamma_{c}$ denotes the moment map operator of the $SU(2)_\text{gauge}$ symmetry of the free hypers, and $\operatorname{tr}$ is taken over the $SU(2)_\text{gauge}$ indices. By direct computation, we observe that
\begin{align}
    Q_\text{BRST} \operatorname{tr}(b \beta \gamma) \sim \operatorname{tr}(bc) + \operatorname{tr}(\beta \partial \gamma) 
    + \operatorname{tr}(\partial \beta \gamma)
    + \operatorname{tr}\mu \mu\,, \quad
    \operatorname{tr}(c \beta \gamma) = Q_\text{BRST} \operatorname{tr}(bc) \ .
\end{align}
We are not careful about the relative coefficients in the equalities. Therefore, two out of the seven states are $Q_\text{BRST}$-non-closed, and two are $Q_\text{BRST}$-exact, forming two canceling pairs.  Each pair contains two elements with opposite parity $(-1)^F$ and identical $(h, f) = (2, 0)$. The remaining three bosonic states form (representatives of) the $Q_\text{BRST}$-cohomology. The Molien--Weyl integral picks up contributions from all seven states; however, the contributions from the canceling pairs automatically cancel, and one is left with a $3 = \dim H^{Q_\text{BRST}}$ in the $q^2$ term
\begin{equation}
    \Big(a^4+\frac{1}{a^4}+2 a^2+\frac{2}{a^2}+3 \Big)q^2 \ .
\end{equation}
However, upon introducing the $\mathbb{Z}_n$-refinement, the states within a canceling pair no longer carry the same $\mathbb{Z}_n$-charge. Instead, they contribute $1 - \epsilon$ and $1 - \epsilon^{-1}$. These contributions are \emph{not} supposed to be present, but the naive Molien--Weyl integration \eqref{eq:naive-Molien--Weyl-integration} retains them. The correct index computation should remove these states \emph{before} collecting contributions.

\subsubsection{\texorpdfstring{$\mathbb{Z}_n$}{Zn}-gauging and VOA}

At the level of the associated VOA, the $\mathbb{Z}_n$-quotient retains only the operators with $\mathbb{Z}_n$-charge divisible by $n$. Table~\ref{tab:Zn-charge-small-N4-superconformal-algebra} immediately implies that $J^0$, $T$, $G^-$, and $\tilde G^+$ are preserved as strong generators of the $\mathcal{N} = 3$ chiral algebra. These precisely form a 2d $\mathcal{N} = 2$ superconformal algebra expected to be present in any $\mathcal{N} = 3$ chiral algebra. Note also that all the generators of the $\mathcal{N} = 4$ small superconformal algebra have even $\mathbb{Z}_n$-charge; hence, when $n = 2$, the $\mathcal{N} = 3$ chiral algebra after gauging is identical to the $\mathcal{N} = 4$ chiral algebra.

When $n = 3$, the operators preserved by the $\mathbb{Z}_3$-quotient (up to conformal weight $h = 3$) are given in Table~\ref{tab:Z3-quotient-operators}, where $W = (J^+(J^+J^+))$, $\tilde W = (J^-(J^-J^-))$. 
\begin{table}[ht]
    \centering
    \begin{tabular}{c|c}
        $h = 1$ & $J^0$ \\
        $h = 3/2$ & $G = G^-, \tilde G = \tilde G^+$ \\
        $h = 2$ & $T, \partial J^0, (J^0J^0)$ \\
        $h = 5/2$ & $\partial G, \partial \tilde G, (J^0G), (J^0\tilde G)$  \\
        $h = 3$ & $W, \tilde W, (J^0(J^0J^0)), (J^0T), \partial^2 J^0, (G\tilde G), \partial T, (\partial J^0J^0)$\\
    \end{tabular}
    \caption{Operators surviving the $\mathbb{Z}_3$ quotient of the small $\mathcal{N} = 4$ chiral algebra, up to conformal weight $h = 3$.}
    \label{tab:Z3-quotient-operators}
\end{table}
At level $h = 5/2$, one could also consider $J^+\tilde G^-$ and $J^-G^+$, which are also $\mathbb{Z}_3$-invariant; however, they are part of the null states in the small $\mathcal{N} = 4$ superconformal algebra and hence discarded,
\begin{equation}
    \partial G^- + (J^0 G^-) + 2(J^- G^+) = 0,
    \qquad
    \partial\tilde G^+ - (J^0\tilde G^+) + 2(J^+\tilde G^-) = 0 \ .
\end{equation}
Similarly, at $h = 3$, we used the null relation
\begin{equation}
\begin{aligned}
(J^+ J^0 J^-)
&=
-\frac14 (J^0 J^0 J^0)
+\frac32 (J^0 T)
+((\partial J^0) J^0)
-\frac12\partial^2 J^0
+(G^- \tilde G^+) \ ,
\\
(J^+ \partial J^-)
&=
\frac12 (J^0 T)
-\frac14 ((\partial J^0) J^0)
+\frac14\partial^2 J^0
+\frac12\partial T
+\frac12 (G^- \tilde G^+) \ ,
\\
((\partial J^+) J^-)
&=
-\frac12 (J^0 T)
-\frac14 ((\partial J^0) J^0)
+\frac14\partial^2 J^0
-\frac12 (G^- \tilde G^+) \ .
\end{aligned}
\end{equation}
$W$ and $\tilde W$ furnish new strong generators and participate in the Higgs branch relation (up to normalization, following from the original chiral ring relation $(J^+J^-) = - \frac{1}{4}(J^0)^2$)
\begin{equation}
    W  \tilde W = (J^0)^6 \ ,
\end{equation}
giving rise to the Higgs branch $\mathbb{C}^2/\mathbb{Z}_6$. In fact, the chiral algebra of the $\mathbb{Z}_3$-gauged theory and that of the $\mathbb{Z}_6$-gauged theory are the same, since all the generators of the small $\mathcal{N} = 4$ superconformal algebra carry even $\mathbb{Z}_n$-charge ($0$ or $\pm2$), so that any composite whose charge is divisible by $3$ has charge divisible by $6$ as well. This is consistent with the Coulomb/Higgs branch analysis in \cite{Bourton:2018jwb}. The operator $W$ ($\tilde W$) is chiral (anti-chiral) with respect to the $\mathcal{N} = 2$ superconformal algebra,
\begin{equation}
    \{\tilde G W\}_{n \ge 1} = 0, \qquad
    \{G \tilde W\}_{n \ge 1} = 0\ .
\end{equation}
Additional generators in the $\mathcal{N} = 2$ (anti-)chiral multiplets are given by $G_W = \{GW\}_1$ and $\tilde G_{\tilde W} = \{\tilde G \tilde W\}_1$. Moreover, there is a new long $\mathcal{N} = 2$ multiplet with primary $(J^+(J^+(J^-J^-)))$ at $h = 4$. Altogether, we find the following strong generators:
\begin{align}
    T, \quad J^0, & \quad G, \quad \tilde G, \quad W = (J^+)^3, \quad \tilde W = (J^-)^3, \quad
    G_W = \{GW\}_1, \quad \tilde G_{\tilde W} = \{\tilde G \tilde W\}_1 \ , \\
    U = & \ (J^+(J^+(J^-J^-))), \quad G_U = \{GU\}_1, \quad\tilde G_U = \{\tilde GU\}_1, \quad V = \{G\{\tilde GU\}_1\}_1 \ .
\end{align}
We refrain from writing out the OPE for this set of generators, which can be derived from the original small $\mathcal{N} = 4$ superconformal algebra OPE together with the null states.

Using the refined Schur index \eqref{eq:Zn-refined-Schur-index}, we compute the $\mathcal{N} = 3$ Schur index as a series expansion
\begin{align}\label{eq:schur-index-Z6-quotient-of-small-superconformal}
\mathcal{I}_{SU(2)/\mathbb{Z}_3} = \mathcal{I}_{SU(2)/\mathbb{Z}_6} = & \ q^{\frac{3}{8}}\Big[1+q-\frac{1+a^2}{a}q^{3/2}+3q^2
-\frac{2(1+a^2)}{a}q^{5/2}
+\left(6+\frac{1}{a^6}+a^6\right)q^3
\nonumber\\
& \quad
-\frac{1+5a^4+5a^6+a^{10}}{a^5}q^{7/2}
+\left(
14+\frac{2}{a^6}+\frac{1}{a^2}
+ a^2+2a^6
\right)q^4
\nonumber\\
&\quad
-\frac{
1+3a^2+11a^6+11a^8+3a^{12}+a^{14}
}{a^7}q^{9/2}
\nonumber\\
&\quad
+\left(
27+\frac{6}{a^6}+\frac{1}{a^4}
+\frac{2}{a^2}+2a^2+a^4+6a^6
\right)q^5 + \cdots \Big] \ .
\end{align}

When $n = 4$, the operators after the $\mathbb{Z}_4$-quotient  at low conformal weight are given in Table~\ref{tab:Z4-quotient-operators}.
\begin{table}[ht]
    \centering
    \begin{tabular}{c|c}
        $h = 1$ & $J^0$ \\
        $h = 3/2$ & $G = G^-, \tilde G = \tilde G^+$ \\
        $h = 2$ & $T, \partial J^0, (J^0J^0), W = (J^+J^+), \tilde W = (J^-J^-)$ \\
        $h = 5/2$ & $\partial G, \partial \tilde G, (J^0G), (J^0\tilde G), (J^+ G^+), (J^- \tilde G^-)$  \\
        $h = 3$ & $(TJ^0), (\partial^2 J^0), (\partial J^\pm J^\pm), (\partial J^\pm J^\mp), (G^+\tilde G^-), (J^0 W), (J^0\tilde W), (J^0(J^+J^-))$\\
    \end{tabular}
    \caption{Operators surviving the $\mathbb{Z}_4$ quotient at low conformal weight.}
    \label{tab:Z4-quotient-operators}
\end{table}
The strong generators are given by
\begin{equation}
    T, J^0, G, \tilde G,  \quad W, \tilde W,  \quad G_W = 2(J^+G^+) = \{G W\}_1, \tilde G_{\tilde W} = 2(J^-\tilde G^-) = \{\tilde G \tilde W\}_1\ .
\end{equation}
Their conformal dimension and charges are listed in Table~\ref{tab:weights-charges-Z4-quotient-of-small-superconformal}. Their OPEs include the standard $c = -9$ $\mathcal{N} = 2$ superconformal algebra generated by $T, J^0, G, \tilde G$, and some more OPEs,
\begin{align}
W(z)\tilde W(w)\sim{}&
\frac{15}{2(z-w)^4}
-\frac{10J^0(w)}{(z-w)^3}
+\frac{
-7T
+\frac{11}{2}(J^0J^0)
-5\partial J^0
}{(z-w)^2}(w)
+\frac{A_{W\tilde W}(w)}{z-w}\ , \cr
W(z)\tilde G_{\tilde W}(w)\sim{}&
\frac{10 \tilde G(w)}{(z-w)^3}+\frac{
-11(J^0 \tilde G)
+7\partial \tilde G
}{(z-w)^2}(w)
+\frac{A_{W \tilde{G}_{\tilde W}}(w)}{z-w}\ , \cr
\tilde W(z)G_W(w)\sim{}&
\frac{10G(w)}{(z-w)^3}
+\frac{
11(J^0G)
+7\partial G
}{(z-w)^2}(w)
+\frac{A_{\tilde W G_W}(w)}{z-w}\ , \cr
G(z)\tilde G_{\tilde W}(w)\sim & \
-\frac{4\tilde W(w)}{(z-w)^2}
-\frac{\partial\tilde W(w)}{z-w}\ , \cr
 \tilde G(z)G_W(w)\sim & \
-\frac{4W(w)}{(z-w)^2}
-\frac{\partial W(w)}{z-w}\ ,\cr
G_W(z)\tilde G_{\tilde W}(w)\sim{}&
    \frac{30}{(z-w)^5}
    -\frac{30J^0(w)}{(z-w)^4}
    +\frac{-24T+11(J^0J^0)-15\partial J^0}{(z-w)^3}(w)
    \cr
    &+\frac{A_1(w)}{(z-w)^2}
    +\frac{A_0(w)}{z-w} \ ,
\end{align}
where
\begin{align}
A_{W\tilde G_{\tilde W}}
={}&
3(J^0(J^0\tilde G))
-2(J^0\partial\tilde G)
-24(T\tilde G)
-11(\partial J^0\tilde G)
+11\partial^2\tilde G\ , \cr
A_{\tilde W G_W}
={}&
3(J^0(J^0G))
+2(J^0\partial G)
-24(TG )
+11(\partial J^0G)
+11\partial^2G\ , \cr
A_{W\tilde W}
={}&
11(G^-\tilde G)
-(J^0(J^0J^0))
+13(TJ^0)
+\frac{11}{2}(\partial J^0 J^0)
+2\partial T
-10\partial^2J^0\ ,\cr
A_0={}&
9(G \partial\tilde G)
+6(J^0(G \tilde G))
+24(TT)
\cr
&-3(T(J^0J^0))
+12(T\partial J^0)
-9(\partial G\tilde G)
-\frac32(\partial J^0(J^0J^0))
+15(\partial T J^0)
\cr
&+\frac{15}{2}( \partial^2J^0 J^0)
-9\partial^2T
-\frac{17}{2}\partial^3J^0\ , \cr
A_1={}&
-(J^0(J^0J^0))
+24(TJ^0)
+11((\partial J^0)J^0)-12\partial T
-17\partial^2J^0 \ .
\end{align}
The Jacobi identities of these generators are proportional to the following singular vectors:
\begin{align}
    \mathcal{N} = & \ 3(G \partial W) + 8(J^0 (G W)) - 2(J^0 (J^0 G_W)) - 3(J^0 \partial G_W) + 5(T G_W) \cr
    & \ - 12(\partial G W) + \frac{11}{2}(\partial J^0 G_W) - \partial^2G_W \ ,\cr
    \tilde{\mathcal{N}} = & \ 3\,(\tilde{G} \partial\tilde{W}) - 8\,(J^0 (\tilde{G} \tilde{W})) - 2\,(J^0 (J^0 \tilde{G}_{\tilde{W}})) + 3\,(J^0 \partial\tilde{G}_{\tilde{W}}) + 5\,(T \tilde{G}_{\tilde{W}})  \cr
    & \ - 12\,(\partial\tilde{G} \tilde{W}) - \frac{11}{2}\,(\partial J^0 \tilde{G}_{\tilde{W}}) - \partial^2\tilde{G}_{\tilde{W}} \ ,
\end{align}
which are annihilated by all positive modes of the strong generators. We expect that these two singular vectors generate the entire ideal of null states in the algebra. Note that the original $bc \beta \gamma$ system  \eqref{eq:bcbetagamma-realization-of-small-superconformal-algebra} continues to realize the irreducible (meaning all null states, such as $\mathcal{N}, \tilde{\mathcal{N}}$, vanish in the realization) version of the $\mathbb{Z}_4$ quotient VOA. We write down again the explicit realization,
\begin{align}
    T = & \ -\frac{3}{2}(b  c') - (\beta  \gamma') - \frac{1}{2}(b'  c) \ ,
    \qquad J^0 = (b  c) + 2(\beta  \gamma) \ ,\cr
    G = & \ (b  \gamma) \ ,
    \qquad 
    \tilde{G} = (c  \beta') + 2(c'  \beta) \ ,\cr
    W = & \ (\beta  \beta) \ ,\cr 
    \tilde{W} = &\ 2(b(c(\beta(\gamma(\gamma  \gamma))))) - 5(b(c(\gamma'  \gamma))) 
    + (\beta(\beta(\gamma(\gamma(\gamma  \gamma))))) - \tfrac{7}{2}(\beta(\gamma'(\gamma  \gamma))) \cr
    &\quad + 2(b'(c(\gamma  \gamma))) + 2(\beta'(\gamma(\gamma  \gamma))) + \tfrac{9}{4}(\gamma'  \gamma') \cr
    &\quad - \tfrac{3}{2}\bigl[(\beta(\gamma'(\gamma  \gamma))) - (\gamma'  \gamma') - (\gamma''  \gamma)\bigr] - \tfrac{3}{2}(\gamma''  \gamma) \ , \cr
    G_W = & \ 2(b  \beta) \ ,\cr
    \tilde{G}_{\tilde{W}} = & \ 6(b(c'(c(\beta(\gamma  \gamma))))) - 5(b(c'(c  \gamma'))) - 5(b(c''(c  \gamma))) \cr
    &\quad - 2(c(\beta'(\beta(\gamma(\gamma  \gamma))))) + 5(c(\beta'(\gamma'  \gamma))) - 2(c(\beta''(\gamma  \gamma))) \cr
    &\quad + 4(b'(c'(c  \gamma))) - 4(c'(\beta(\beta(\gamma(\gamma  \gamma))))) + 10(c'(\beta(\gamma'  \gamma))) \cr
    &\quad - 6(c'(\beta'(\gamma  \gamma))) + 5(c''(\beta(\gamma  \gamma))) - \tfrac{15}{2}(c''  \gamma') \ .
\end{align}
Note that this is different from the free-field realization for the $\mathcal{N} = 2$ VOA $\mathcal{W}_{\mathbb{Z}_4}$ discussed in \cite{Bonetti:2018fqz}.

\begin{table}[ht]
\centering
\begin{tabular}{c|cccccccc}
\hline
generators
& $T$
& $J^0$
& $G$
& $\tilde G$
& $W$
& $\tilde W$
& $G_W$
& $\tilde G_{\tilde W}$ \\
\hline
$h$
& $2$
& $1$
& $\frac{3}{2}$
& $\frac{3}{2}$
& $2$
& $2$
& $\frac{5}{2}$
& $\frac{5}{2}$ \\
$J^0\text{-charge}$
& $0$
& $0$
& $-1$
& $+1$
& $+4$
& $-4$
& $+3$
& $-3$ \\
\hline
\end{tabular}
\caption{Conformal weight and $J^0$-charge of the generators.}
\label{tab:weights-charges-Z4-quotient-of-small-superconformal}
\end{table}

With the analysis above, we are able to write down the $\mathcal{N} = 3$ Schur index $\mathcal{I} = \mathcal{I}_{SU(2)/\mathbb{Z}_4}$ to certain low $q$-order,
\begin{align}\label{eq:schur-index-Z4-quotient-of-small-superconformal}
\mathcal{I}= & \ q^{\frac{3}{8}} \bigg(1+q
-\frac{1+a^{2}}{a}q^{3/2}
+(3+\frac{1}{a^{4}}+a^{4})q^{2}
-\frac{1+2a^{2}+2a^{4}+a^{6}}{a^{3}}q^{5/2}
+(6+\frac{2}{a^{4}}+2a^{4})q^{3}
\nonumber\\
&\quad
-\frac{(1+a^{2})(1+a^{2}+a^{4})^{2}}{a^{5}}q^{7/2}
+\left(
14+\frac{1}{a^{8}}+\frac{6}{a^{4}}
+\frac{2}{a^{2}}+2a^{2}+6a^{4}+a^{8}
\right)q^{4}
\nonumber\\
&\quad
-\frac{
1+3a^{2}+7a^{4}+11a^{6}+11a^{8}
+7a^{10}+3a^{12}+a^{14}
}{a^{7}}q^{9/2}
\nonumber\\
&\quad
+\left(
27+\frac{2}{a^{8}}+\frac{13}{a^{4}}
+\frac{4}{a^{2}}+4a^{2}+13a^{4}+2a^{8}
\right)q^{5} + \cdots\bigg)\ .
\end{align}

\subsubsection{Closed-form index and non-vacuum characters}
There is a formula that allows us to write down the closed form for \eqref{eq:schur-index-Z6-quotient-of-small-superconformal} and \eqref{eq:schur-index-Z4-quotient-of-small-superconformal}. Non-vacuum characters can moreover be extracted from this closed form.

The original small $\mathcal{N} = 4$ superconformal algebra contains an $\widehat{\mathfrak{su}}(2)_{-3/2}$ subalgebra. It is therefore natural to study the branching rule proposed in \cite{Creutzig:2018ltv}. The algebra has an $SU(2)$ outer automorphism that rotates $G^\pm \leftrightarrow \tilde G^\pm$ while leaving all bosonic operators invariant. In particular, we can find the charge $o$ (which stands for \emph{outer}) of the generators under the Cartan of the diagonal $SU(2)$ in $SU(2)_\text{out} \times SU(2)_J$ given in Table~\ref{tab:o-charge-small-N4-superconformal-algebra}.
\begin{table}[ht]
\centering
\begin{tabular}{c|c|c|c|c|c|c|c|c}
     & $J^-$ & $J^0$ & $J^+$& $T$ & $G^+$ &  $G^-$ & $\tilde G^+$ & $\tilde G^-$\\
     \hline
     $o$ charge & $-2$ & $0$ & $+2$ &
     $0$ & $+2$ & $0$ & $0$ & $-2$
\end{tabular}
\caption{The charge $o$ under the Cartan of the diagonal $SU(2) \subset SU(2)_\text{out} \times SU(2)_J$, for the generators of the 2d small $\mathcal{N} = 4$ superconformal algebra.}
\label{tab:o-charge-small-N4-superconformal-algebra}
\end{table}
By inspection, this is identical to the $\mathbb{Z}_n$ charge: at the level of chiral algebra, we may view the $\mathbb{Z}_n$ symmetry as a subgroup of the diagonal $SU(2)$. See also \cite{Maruyoshi:2026cmr} for an application of this outer automorphism to non-invertible symmetries of the VOA. At the level of character, the $SU(2)_\text{out}$-refined supercharacter is given by \cite{Creutzig:2018ltv}
\begin{equation}
    \operatorname{ch}_0(q, x, y)
    = q^{3/8}\operatorname{PE}\Big[\frac{q\chi_\text{adj}(x)}{1-q}\Big] \sum_{m \ge 0}(-1)^m q^{\frac{m(m+2)}{2}} \chi_m(x)\chi_m(y) \ .
\end{equation}
Writing $x = a \epsilon$ and $y=\epsilon$, $\epsilon$ then denotes the fugacity for the $\operatorname{diag}[SU(2) \times SU(2)_\text{out}]$, and the $a$ is the standard $SU(2)$ fugacity. In terms of Jacobi theta functions,
\begin{equation}
    \mathcal{I}_{\mathcal{N} = 4}(q, a, \epsilon)
    = \frac{\vartheta_4(\mathfrak{a} + 2 \mathfrak{e}) - \vartheta_4(\mathfrak{a})}{2 \sin(2\pi \mathfrak{e}) \vartheta_1(2 \mathfrak{a} + 2 \mathfrak{e})}  \ , \qquad
    \epsilon = e^{2\pi i \mathfrak{e}} \ .
    \label{eq:SU2-refined-index-closed}
\end{equation}

With the refined Schur index in closed form, the $\mathbb{Z}_n$-quotient index can be easily written down simply by summing over $\epsilon \in \mathbb{Z}_n$ 
\begin{equation}
  \mathcal{I}_{SU(2)/\bZ_n}(q,a)
  =\frac{1}{n}\sum_{j=0}^{n-1}
  \mathcal{I}_{\mathcal{N}=4}
  \left(q,a,e^{2\pi i j/n}\right).
  \label{eq:S1-SU2-projector}
\end{equation}
Concretely, we have the flavored/unflavored Schur index
\begin{align}\label{eq:flavored-Schur-index-Zn-quotient}
    \mathcal{I}_{SU(2)/\mathbb{Z}_4}(q, a) = & \ \frac{\vartheta'_4(\mathfrak{a})}{4\pi\vartheta_1(2 \mathfrak{a})} + \frac{\vartheta_3(\mathfrak{a}) - \vartheta_4(\mathfrak{a})}{4 \vartheta_2(2 \mathfrak{a})} \ , \nonumber \\
    \mathcal{I}_{SU(2)/\mathbb{Z}_4}(q) = & \ \frac{\vartheta''_4(0)}{8\pi \vartheta_1'(0)} + \frac{\vartheta_3(0) - \vartheta_4(0)}{4 \vartheta_2(0)} \ , \nonumber \\
    \mathcal{I}_{SU(2)/\mathbb{Z}_6}(q,a) = & \ \frac{1}{18}\bigg(\frac{3}{\pi} \frac{\vartheta'_4(\mathfrak{a})}{\vartheta_1(2\mathfrak{a})}
    + \frac{2 \sqrt{3}( \vartheta_3(\frac{1}{6} + \mathfrak{a}) - \vartheta_4(\mathfrak{a}))}{\vartheta_2(2 \mathfrak{a} + \frac{1}{6})}
    + \frac{2 \sqrt{3}(-\vartheta_4(\mathfrak{a}) + \vartheta_4(\frac{1}{3} + \mathfrak{a}))}{\vartheta_1(\frac{1}{3} + 2 \mathfrak{a})}
    \bigg) \ , \nonumber \\
    \mathcal{I}_{SU(2)/\mathbb{Z}_6}(q) = & \ \frac{1}{36}\Bigg(
    \frac{4\sqrt{3}\left(
    \vartheta_3\left(\frac{1}{6}\right)-\vartheta_4(0)
    \right)}
    {\vartheta_2\left(\frac{1}{6}\right)}
    +\frac{4\sqrt{3}\left(
    -\vartheta_4(0)+\vartheta_4\left(\frac{1}{3}\right)
    \right)}
    {\vartheta_1\left(\frac{1}{3}\right)}
    +\frac{3\vartheta_4''(0)}
    {\pi\vartheta_1'(0)}
    \Bigg).
\end{align}
It is straightforward to verify that the $q$-expansions of the flavored indices reproduce the previous results \eqref{eq:schur-index-Z6-quotient-of-small-superconformal} and \eqref{eq:schur-index-Z4-quotient-of-small-superconformal}.

Next, we try to identify some non-vacuum characters. We begin with the $n = 4$ case. To verify whether an expression is a potential character, the best tools are (flavored) MLDEs. Starting from the singular vectors, we can derive additional null states. For example, $\{\tilde G \{\tilde W \mathcal{N}\}_1\}_2$ gives
\begin{align}
\{\tilde G \{\tilde W \mathcal{N}\}_1\}_2 = & \ (J^5) + 324\,(T(G\tilde{G})) - \tfrac{79}{2}\,(J(J(G\tilde{G}))) + \tfrac{11}{2}\,(G_W\tilde{G}_{\tilde{W}}) \nonumber\\
&- 16\,(J(W\tilde{W})) + 240\,(T(TJ)) - 49\,(T(J^3)) + \tfrac{187}{2}\,(J(G\tilde{G}')) \nonumber\\
&+ \tfrac{99}{2}\,(J(G'\tilde{G})) + \tfrac{241}{2}\,(T(J'J)) + 30\,(T'T) + \tfrac{185}{2}\,(T'(J^2)) \nonumber\\
&- \tfrac{33}{2}\,(J'(G\tilde{G})) - 24\,(J'(J^3)) + 6\,(W\tilde{W}') - 6\,(W'\tilde{W}) \nonumber\\
&- 126\,(G\tilde{G}'') - 120\,(G'\tilde{G}') - \tfrac{153}{2}\,(G''\tilde{G}) + \tfrac{539}{4}\,(J''(J^2)) \nonumber\\
&- \frac{601}{2}\,(TJ'') - \tfrac{151}{2}\,(T''J) - 123\,(T'J') + \tfrac{23}{4}\,(J'(J'J)) - 74\,(J''J') \nonumber\\
&- 107\,(J^{(3)}J) - 5\,T^{(3)} + \tfrac{815}{12}\,J^{(4)} \ .
\end{align}
Let $j,w,\tilde w$ denote the images of $\frac{1}{2}J,W,\tilde W$, respectively, in
$R_{\mathbb{V}}$.  Up to an overall nonzero coefficient, the $C_2$-projection
of this null state gives
\begin{equation}
    j\bigl(j^4 - w\tilde w\bigr)=0
    \qquad \text{in } R_{\mathbb{V}} \ .
\end{equation}
At weight six, additional null states give rise to the following relations in $R_{\mathbb{V}}$:
\begin{equation}
    j^2(j^4 - w \tilde w) = 0, \quad
    (j^4 - w\tilde w) w = 0, \quad
    (j^4 - w\tilde w) \tilde w = 0 \ .
\end{equation}
They imply that the radical of the generated ideal $I_{C_2}$ is
\begin{equation}
    \sqrt{I_{C_2}}=(j^4 - w\tilde w) \ .
\end{equation}
One obtains
$R_{\mathbb{V}}^{\mathrm{red}}
\cong \mathbb{C}[j,w,\tilde w]/(j^4 - w\tilde w)$ and hence
$X_{\mathbb{V}} \cong \mathbb{C}^2/\mathbb{Z}_4$. The state $\{\tilde G \{\tilde W \mathcal{N}\}_1\}_2$ also leads to a flavored MLDE annihilating the Schur index
\eqref{eq:schur-index-Z4-quotient-of-small-superconformal}, which---as well as the
equations coming from null states of higher conformal weight---is too long to
write down here. On the other hand, it is straightforward to compute the
unflavored MLDE annihilating the unflavored index, which is compact enough to
write down:
\begin{align}
0 = \Big( & D_q^{(7)} + \frac{186}{23} \Em{2}{1} D_q^{(6)} + \Big[-\frac{46645}{92} E_4(\tau) + \frac{630}{23} \Em{2}{1}^2\Big]D_q^{(5)} \nonumber\\
& + \Big[\frac{420705}{92} E_4(\tau) \Em{2}{1} - \frac{35681}{23} \Em{2}{1}^3\Big]D_q^{(4)} \nonumber\\
& + \Big[-\frac{124750}{23} E_4(\tau)^2 + \frac{1342125}{46} E_4(\tau) \Em{2}{1}^2 - \frac{177345}{23} \Em{2}{1}^4\Big]D_q^{(3)} \nonumber\\
& + \Big[\frac{863625}{46} E_4(\tau)^2 \Em{2}{1} - \frac{2220225}{46} E_4(\tau) \Em{2}{1}^3 + \frac{135180}{23} \Em{2}{1}^5\Big]D_q^{(2)} \nonumber\\
& + \Big[\frac{11495625}{92} E_4(\tau)^3 - \frac{1393875}{23} E_4(\tau)^2 \Em{2}{1}^2 \nonumber\\
& \qquad + \frac{11481675}{92} E_4(\tau) \Em{2}{1}^4 - \frac{434620}{23} \Em{2}{1}^6\Big]D_q^{(1)} \nonumber\\
& + \Big[- \frac{105249375}{92} E_4(\tau)^3 \Em{2}{1} + \frac{63149625}{46} E_4(\tau)^2 \Em{2}{1}^3 \nonumber\\
& \qquad - \frac{37889775}{92} E_4(\tau) \Em{2}{1}^5 + \frac{841995}{23} \Em{2}{1}^7\Big]\Big)\operatorname{ch} \ .
\end{align}
The indicial roots are
\begin{equation}
    \alpha = - \frac{1}{8}, \quad 0, \quad \frac{1}{8}, \quad \Big[\frac{3}{8}\Big]_2, \quad \frac{191}{184}, \quad \frac{11}{8} \ .
\end{equation}
The root $\frac{191}{184}$ corresponds to a solution 
\begin{equation}
    q^{191/184}\left(
    1
    -\frac{346657998}{295996775}q^{1/2}
    +\frac{958203650497}{316536738125}q
    -\frac{77028723876785698083046}
    {18750180698581672204375}q^{3/2}
    +\cdots
    \right) \nonumber
\end{equation}
where the denominator keeps on growing\footnote{Note that the indicial root $\frac{191}{184} = \frac{191}{8 \cdot 23}$. Denote by $a_n$ the coefficient of $q^{\alpha + n/2}$ in the solution with $a_0 = 1$, and $p_{23}(x)$ the number of occurrences of $23$ in a rational number $x$. Namely, $p_{23}(23^k \frac{u}{v}) = k$ if $u, v$ are coprime with $23$. One can show that $p_{23}(a_n) = p_{23}(a_0) - n - p_{23}(n!)$, which implies the number of factors of $23$ keeps increasing.}. This implies that the solution with root $\frac{191}{184}$ cannot be an index. In fact, there exists another equation with higher modular weight and non-zero Wronskian index, but with lower order in $q$-differential,
\begin{align}
& E_4(\tau)E_2^{-}(1)D_q^{(6)}\operatorname{ch}
  -\frac{404}{1495}E_2^{-}(1)^3D_q^{(6)}\operatorname{ch}
  +\frac{2271}{299}E_4(\tau)E_2^{-}(1)^2D_q^{(5)}\operatorname{ch}
  -\frac{1956}{1495}E_2^{-}(1)^4D_q^{(5)}\operatorname{ch}
\nonumber\\
&\quad
  -\frac{22415}{52}E_4(\tau)^2E_2^{-}(1)D_q^{(4)}\operatorname{ch}
  +\frac{3038}{23}E_4(\tau)E_2^{-}(1)^3D_q^{(4)}\operatorname{ch}
\nonumber\\
&\quad
  -\frac{10068}{1495}E_2^{-}(1)^5D_q^{(4)}\operatorname{ch}
  -\frac{45375}{598}E_4(\tau)^2E_2^{-}(1)^2D_q^{(3)}
  \operatorname{ch}
\nonumber\\
&\quad
  -\frac{210930}{299}E_4(\tau)E_2^{-}(1)^4D_q^{(3)}
  \operatorname{ch}
  +\frac{45456}{299}E_2^{-}(1)^6D_q^{(3)}\operatorname{ch}
\nonumber\\
&\quad
  -\frac{80375}{13}E_4(\tau)^3E_2^{-}(1)D_q^{(2)}
  \operatorname{ch}
  +\frac{2541200}{299}E_4(\tau)^2E_2^{-}(1)^3D_q^{(2)}
  \operatorname{ch}
\nonumber\\
&\quad
  -\frac{750105}{299}E_4(\tau)E_2^{-}(1)^5D_q^{(2)}
  \operatorname{ch}
  +\frac{66732}{299}E_2^{-}(1)^7D_q^{(2)}\operatorname{ch}
\nonumber\\
&\quad
  +\frac{5976375}{598}E_4(\tau)^3E_2^{-}(1)^2D_q^{(1)}
  \operatorname{ch}
  -\frac{11112025}{598}E_4(\tau)^2E_2^{-}(1)^4D_q^{(1)}
  \operatorname{ch}
\nonumber\\
&\quad
  +\frac{1713835}{299}E_4(\tau)E_2^{-}(1)^6D_q^{(1)}
  \operatorname{ch}
  -\frac{149052}{299}E_2^{-}(1)^8D_q^{(1)}\operatorname{ch}
\nonumber\\
&\quad
  +\frac{11694375}{52}E_4(\tau)^4E_2^{-}(1)\operatorname{ch}
  -\frac{240904125}{598}E_4(\tau)^3E_2^{-}(1)^3
  \operatorname{ch}
\nonumber\\
&\quad
  +\frac{287681625}{1196}E_4(\tau)^2E_2^{-}(1)^5
  \operatorname{ch}
  -\frac{16465680}{299}E_4(\tau)E_2^{-}(1)^7
  \operatorname{ch}
  +\frac{1272348}{299}E_2^{-}(1)^9\operatorname{ch} = 0 \ .
\end{align}
The six indicial roots of this equation precisely exclude the unwanted $\frac{191}{184}$.

The original small $\mathcal{N} = 4$ superconformal algebra naturally constitutes a module of the $\mathcal{N} = 3$ $SU(2)/\mathbb{Z}_4$ chiral algebra, and as a result, $\mathcal{I}_{\mathcal{N} = 4} = \frac{\vartheta'_4(\mathfrak{a})}{2\pi\vartheta_1(2 \mathfrak{a})}$ is a solution to all the flavored MLDEs \cite{Pan:2021mrw,Pan:2021ulr}. Consequently,
\begin{equation}
    \frac{\vartheta_3(\mathfrak{a}) - \vartheta_4(\mathfrak{a})}{4\vartheta_2(2 \mathfrak{a})}
\end{equation}
is also a solution. In fact, we can identify the following solutions/flavored characters of the $SU(2)/\mathbb{Z}_4$ VOA:
\begin{align}
    \operatorname{ch}_{\frac{3}{8}} = \mathcal{I}_{SU(2)/\mathbb{Z}_4},
    \quad
    \operatorname{ch}_{- \frac{1}{8}} = \ \frac{1}{2} \frac{\vartheta_3(\mathfrak{a}) + \vartheta_4(\mathfrak{a})}{
    \vartheta_2(2\mathfrak{a})
    }, \quad
    \operatorname{ch}_{\frac{11}{8}} = \frac{\vartheta'_4(\mathfrak{a})}{4\pi \vartheta_1(2\mathfrak{a})} + \frac{\vartheta_4(\mathfrak{a}) - \vartheta_3(\mathfrak{a})}{4\vartheta_2(2\mathfrak{a})} \ .
\end{align}
The indicial degeneracy at $\alpha = \frac{3}{8}$ implies the presence of one logarithmic character given by $STS \mathcal{I}_{SU(2)/\mathbb{Z}_4}$. We have not identified the flavored solution in closed form for $\alpha = 0, \frac{1}{8}$, nor can we rule out their correspondence to legitimate characters.

For $n = 6$, a similar pattern emerges. The unflavored Schur index $\mathcal{I}_{SU(2)/\mathbb{Z}_6}$ in \eqref{eq:flavored-Schur-index-Zn-quotient} satisfies a $10$th-order (modular weight $46$) unflavored MLDE with non-zero Wronskian index (we only display its leading term):
\begin{align}
& \Big[ -\tfrac{17140046875}{10894964032}E_4(\tau)^{6}\Em{2}{1}
  + \tfrac{89109740625}{10894964032}E_4(\tau)^{5}\Em{2}{1}^{3}\nonumber\\
& \qquad - \tfrac{40436878125}{10894964032}E_4(\tau)^{4}\Em{2}{1}^{5}
  -  \tfrac{156293898125}{10894964032}E_4(\tau)^{3}\Em{2}{1}^{7}\\
& \qquad + \tfrac{50131101625}{2723741008}E_4(\tau)^{2}\Em{2}{1}^{9}
  - \tfrac{2648747595}{340467626}E_4(\tau)\Em{2}{1}^{11}
  + \Em{2}{1}^{13} \Big]D_q^{(10)}\operatorname{ch} + \cdots = 0 \ . \nonumber
\end{align}
The indicial roots are given by
\begin{equation}
    \alpha = -\frac{1}{8},-\frac{1}{72},\frac{1}{24},\frac{5}{24},\frac{23}{72},\Big[\frac{3}{8}\Big]_2,\frac{47}{72},\frac{7}{8},\frac{11}{8} \ .
\end{equation}
Direct evaluation shows that each term in the unflavored Schur index is separately a solution to the equation. In summary, we identify three linearly independent unflavored solutions,
\begin{align}
\operatorname{ch}_{- \frac{1}{8}} = & \ \frac{\sqrt{3} \vartheta_3(\frac{1}{6})}{\vartheta_2(\frac{1}{6})} \ , \qquad
\operatorname{ch}_{\frac{3}{8}} = \frac{\vartheta''_4(0)}{4\pi\vartheta_1'(0)} \ ,
\\
\operatorname{ch}_{\frac{11}{8}} = & \ -\frac{\vartheta_3\left(\frac{1}{6}\right)}
{3\sqrt{3}\,\vartheta_2\left(\frac{1}{6}\right)}
+\frac{\vartheta_4(0)}
{3\sqrt{3}\,\vartheta_2\left(\frac{1}{6}\right)}
+\frac{\vartheta_4''(0)}
{12\pi\,\vartheta_1'(0)} \ .
\end{align}
Although we have not constructed the flavored MLDEs, we conjecture that the five terms
\begin{equation}
    \frac{\vartheta'_4(\mathfrak{a})}{\vartheta_1(2 \mathfrak{a})}, \quad
    \frac{\vartheta_3(\mathfrak{a} + \frac{1}{6})}{\vartheta_2(2 \mathfrak{a} + \frac{1}{6})}, \quad
    \frac{\vartheta_4( \mathfrak{a})}{\vartheta_2(2 \mathfrak{a} + \frac{1}{6})}, \quad
    \frac{\vartheta_4( \mathfrak{a})}{\vartheta_1(2 \mathfrak{a} + \frac{1}{3})}, \quad
    \frac{\vartheta_4(\mathfrak{a} + \frac{1}{3})}{\vartheta_1(2 \mathfrak{a} + \frac{1}{3})}, \ .
\end{equation}
in the flavored Schur index are separately flavored solutions, in complete analogy with the $n = 4$ case.

\section{\texorpdfstring{$S^1$}{S1} reduction}\label{sec:S1-reduction}

In this section, we use the closed-form expressions of the Schur indices to test the expected reduction of these 4d $\mathcal{N} = 3$ theories to 3d ABJM theories.\footnote{We are grateful to Takahiro Nishinaka for suggesting the possibility.}
 After matching the 4d fugacities with the 3d mass and squashing parameters, we compare the divergent small-circle behavior of the Schur index with that of the ABJM $S^3$ partition function, first for the $\mathbb{Z}_3$ S-fold theory and then for the discretely gauged $SU(2)/\mathbb{Z}_n$ theories.

The Schur index can be viewed as the partition function of the 4d theory on
$S^1\times S^3$ \cite{Pan:2019bor,Jeong:2019pzg,Oh:2019bgz,Dedushenko:2019yiw}.
As the circle shrinks, a 4d $\mathcal{N} = 2$ theory reduces to a 3d
$\mathcal{N} = 4$ theory, and the index is expected to reduce to the $S^3$
partition function of the latter
\cite{Dolan:2011rp,Gadde:2011ia,Imamura:2011uw,Razamat:2014pta,Buican:2015hsa,Minahan:2021pfv}.
For the $\mathcal{N} = 3$ theories of interest here, the 3d theory is expected
to be an ABJM theory, and this reduction was analyzed in
\cite{Nakanishi:2022fvr}.

The identification of the two sides is non-trivial at the level of
symmetries. For 4d $\mathcal{N}=3$ SCFTs on $S^1\times S^3_{b}$, it was shown
in \cite{Nakanishi:2022fvr} that the Schur limit reaches a
squashing-independent locus of the ABJM partition function when the 4d
$R$-symmetry is mixed with the baryonic symmetry of the 3d theory,
\begin{equation}
    R_I^{\rm 4d}=R_I^{\rm 3d}- J_{U(1)_B}^{\rm 3d}~,
  \label{eq:R-mixing}
\end{equation}
where $J_{U(1)_B}^{\rm 3d}$ generates the baryonic $U(1)_B$ flavor symmetry of
the ABJM theory. On this locus, with a rescaling of the remaining mass parameter, the partition function becomes independent of the squashing parameter $b$, so
that the comparison can be performed on the round sphere.

Before turning to the individual theories, we record the dictionary between the
4d fugacities and the 3d parameters, and review the origin of the divergence
that both sides of the comparison will turn out to have \cite{Nakanishi:2022fvr}.

Consider the $\mathcal{N} = 2$ superconformal index $\mathcal{I}(p,q,t; a)$, with $a$ the $U(1)_f$ flavor fugacity \cite{Gadde:2011uv}. The small-circle limit  is implemented by the identifications
\begin{equation}
  p=e^{-\beta b},\qquad
  q=e^{-\beta b^{-1}},\qquad
  \frac{t}{\sqrt{pq}}=e^{-i\beta m},\qquad
  a=e^{-i\beta M},
  \label{eq:small-circle}
\end{equation}
where $\beta\to0^{+}$ parameterizes the vanishing circle radius, while $b$, $m$ and $M$ are held fixed. In 3d, $M$ is a real mass. The Schur specialization $t=q$ imposes
$m=\frac{i}{2}(b-b^{-1})$; on this locus, the residual
squashing dependence is removed by $M\to b^{-1}M$.  Taking the round sphere limit $b=1$, one has $m=0$. It is essential that $a\to1$ and $q\to1$ be
correlated as in \eqref{eq:small-circle}: holding $a$ fixed at a generic
value as $q\to1$ sends $M=i\beta^{-1}\log a$ to infinity and produces a 3d
large-mass limit rather than a fixed-mass reduction.

The $U(N)_k \times U(N)_{-k}$ ABJM theory contains four $\mathcal{N} = 2$ bifundamental/anti-bifundamental chiral multiplets, whose masses are parameterized by the three mass parameters $\mu_{1,2,3}$. Under the circle reduction from 4d, $\mu_i$ are identified with the 4d parameters by
\begin{equation}
  \mu_1=-\Bigl(M+\frac{m}{2}+\frac{iQ}{4}\Bigr),\qquad
  \mu_2=m,\qquad
  \mu_3=M-\frac{m}{2}-\frac{iQ}{4},\qquad
  Q=b+b^{-1},
\end{equation}
where $\mu_1$ couples to $U(1)_B$, while $\mu_2$ and $\mu_3$ couple to the
$SO(2)^2\subset SO(6)_R$ Cartan.  They satisfy
\begin{equation}
  \mu_1+\mu_2+\mu_3=-\frac{iQ}{2}.
  \label{eq:accidental-plane}
\end{equation}
The combination \eqref{eq:accidental-plane} is a mass direction for a purely 3d
accidental global symmetry, with no counterpart among the 4d fugacities.  On the round-sphere Schur locus $b=1$,
$m=0$, this gives
\begin{equation}
  \mu_1=-M-\frac{i}{2},\qquad
  \mu_2=0,\qquad
  \mu_3=M-\frac{i}{2}.
  \label{eq:masses-Schur}
\end{equation}
The four chiral multiplets of the ABJM theory therefore carry the masses
\begin{equation}
  \begin{aligned}
    \nu_1&=\frac{\mu_1+\mu_2+\mu_3}{2}=-\frac{i}{2},
    &\qquad
    \nu_2&=\frac{\mu_1-\mu_2-\mu_3}{2}=-M,\\
    \nu_3&=\frac{-\mu_1-\mu_2+\mu_3}{2}=M,
    &\qquad
    \nu_4&=\frac{-\mu_1+\mu_2-\mu_3}{2}=\frac{i}{2}.
  \end{aligned}
  \label{eq:four-chiral-masses}
\end{equation}
Two of them sit at the imaginary values $\pm i/2$.  At these values, the
corresponding one-loop determinants no longer suppress large eigenvalues, so the
matrix integral develops flat directions and diverges as a power of the cutoff,
\begin{equation}
    Z_{S^3} \sim \Lambda^{N_{\rm C}} \ ,
\end{equation}
where $N_{\rm C}=\dim_{\mathbb{C}}\mathcal{C}_{\mathrm{4d}}$ is the complex dimension of the 4d ($\mathcal{N}=2$) Coulomb branch and $\Lambda$ is a cutoff on the VEV of the 3d
Coulomb branch operator. The 4d side must then be singular as
well, and the comparison has to be made between the coefficients of the two
divergences rather than between the partition functions themselves.

\subsection{The \texorpdfstring{$\mathbb{Z}_3$}{Z3} S-fold theory}

On the grounds of supersymmetry and of the moduli space, the rank-one
$\mathbb{Z}_3$ S-fold theory of section~\ref{section:3} is expected to reduce on the
circle to the $\mathcal{N}=6$ $U(1)_3\times U(1)_{-3}$ ABJM theory.  We can now
test this expectation quantitatively.  Since the limit is singular, a
truncated $q$-series carries no information about it; the closed form
\eqref{eq:flavored-Schur-index-WZ3}, on the other hand, allows the small-circle
asymptotics, and in particular the residue, to be extracted analytically.

The Schur index of the 4d $\mathcal{N} = 3$ theory is given by \eqref{eq:flavored-non-vacuum-characters-WZ3-1}--\eqref{eq:flavored-non-vacuum-characters-WZ3-3} and \eqref{eq:flavored-Schur-index-WZ3}:
\begin{equation}
  \operatorname{ch}_{\textrm{vac}}
  =-\frac{\operatorname{ch}^+_{\frac18}+\operatorname{ch}^-_{\frac18}}{6}
  +\frac{\operatorname{ch}^+_{\frac18}-\operatorname{ch}^-_{\frac18}}{3}\,
  \mathcal{E},
  \qquad
  \mathcal{E}=\Em{1}{a}
  +\Ep{1}{a^2} .
\end{equation}
It is convenient to set $\ell=\beta/(2\pi)$ and $x=\pi M$, so that $\tau=i\ell$
and the 4d $U(1)_f$ chemical potential is $\mathfrak a=-\ell M$; the
small-circle limit is then $\ell\to0^{+}$ at fixed $x$.

Because $\ell\to0$ is the cusp $\tau\to0$, the asymptotics are governed by the
modular $S$ transformation of the $\eta$ and Jacobi $\vartheta$ functions.
Concretely, as $\ell \to 0$,
\begin{equation}
  \operatorname{ch}^{\pm}_{\frac18}
  \longrightarrow
  \frac{1}{\sqrt3\,\bigl[\cosh x\pm i\sqrt3\sinh x\bigr]},
  \qquad
  \operatorname{ch}^+_{\frac18}-\operatorname{ch}^-_{\frac18}
  =-\frac{2i\sinh x}{\cosh^2x+3\sinh^2x}+\cO(1)\  ,
\end{equation}
while the twisted Eisenstein series have asymptotic behavior \cite{Pan:2024bne}
\begin{equation}
  \Em{1}{a}
  =\frac{i}{2\ell}\tanh x-iM+\cO(e^{-2\pi/\ell}),
  \qquad
  \Ep{1}{a^2}
  =\frac{i}{2\ell}\coth(2x)-2iM+\cO(e^{-2\pi/\ell}),
\end{equation}
so that $\mathcal{E}=\frac{i}{2\ell}[\tanh x+\coth(2x)]+\cO(1)$.  The symmetric
combination $\operatorname{ch}^+_{\frac18}+\operatorname{ch}^-_{\frac18}$
contributes only at $\cO(1)$, so the entire $1/\ell$ divergence comes from the
term containing $\mathcal{E}$.  Using
\begin{equation}
  \tanh x+\coth(2x)=\frac{\cosh^2x+3\sinh^2x}{2\sinh x\cosh x},
\end{equation}
we obtain the leading Laurent expansion in $\beta$,
\begin{equation}
  \operatorname{ch}_{\textrm{vac}}\bigl(e^{-\beta},e^{-i\beta M}\bigr)
  =\frac{\pi}{3\beta\cosh(\pi M)}+\cO(1),
  \quad
  \log\operatorname{ch}_{\textrm{vac}}
  =\log\frac{2\pi}{\beta}-\log\bigl(6\cosh(\pi M)\bigr)+\cO(\beta).
  \label{eq:S1-4d-asymptotics}
\end{equation}
For a 4d $\mathcal{N}=2$ SCFT with $a_{\mathrm{4d}}\neq c_{\mathrm{4d}}$ and a finite
$S^3$ partition function, the small-circle behavior would instead take the
Cardy-like form
\begin{equation}
  \log Z_{S^1\times S^3}
  \sim\frac{8\pi^2}{\beta}\bigl(c_{\mathrm{4d}}-a_{\mathrm{4d}}\bigr)
  +\log Z_{S^3}+\cO(\beta),
  \label{eq:generic-cardy}
\end{equation}
in which the exponentially growing prefactor is controlled by $c_{\mathrm{4d}}-a_{\mathrm{4d}}$ and the finite part is $\log Z_{S^3}$.  Neither ingredient is available in the present case. Any $\mathcal{N}=3$ theory has $a_{\mathrm{4d}}=c_{\mathrm{4d}}$, so the first term is absent, while the $S^3$ partition function diverges, as we saw above. The expected behavior is logarithmic instead \cite{Nakanishi:2022fvr},
\begin{equation}
        \log Z_{S^1\times S^3}\sim N_{\rm C}\log\frac{2\pi}{\beta} + \cO(1) \ .
\end{equation}
Our result \eqref{eq:S1-4d-asymptotics} reproduces precisely this form with $N_{\rm C}=1$, as it should for a rank-one theory, and it moreover fixes the constant term to the simple function $[6\cosh(\pi M)]^{-1}$.  We now show that this constant is the residue of the divergent ABJM $S^3$ partition function at its pole.

The $S^3$ partition function of the mass-deformed ABJM matrix model for
$U(N)_k\times U(N)_{-k}$ with two mass parameters $\mu_A,\mu_B$\footnote{We have renamed the parameters $m_{1,2}$ in  \cite{Nosaka:2024gle} into $\mu_{A,B}$.} was computed in
\cite{Nosaka:2024gle}. In the rank-one case $N=1$, Eq.~(A.1) in \cite{Nosaka:2024gle} gives the
exact result
\begin{equation}
  Z_{k}(\mu_A,\mu_B)=\frac{1}{4k\cosh(\mu_A/2)\cosh(\mu_B/2)}.
  \label{eq:ABJM-rank-one}
\end{equation}
Comparing the localization integrands of \cite{Nakanishi:2022fvr} and \cite{Nosaka:2024gle} in the round-sphere limit $b \to 1$, where the
paired double-sine factors satisfy
$s_1\bigl(\tfrac{i}{2}-u\bigr)s_1\bigl(\tfrac{i}{2}+u\bigr)=[2\cosh(\pi u)]^{-1}$, fixes the dictionary
\begin{equation}
  \mu_A=\pi(\mu_1+\mu_2+\mu_3),\qquad
  \mu_B=\pi(\mu_3-\mu_1),
\end{equation}
up to the sign flip $\mu_{A,B}\to-\mu_{A,B}$ and the exchange of the two mass
parameters. On the locus
\eqref{eq:masses-Schur}, we have
\begin{equation}
  \mu_A=-i\pi,\qquad \mu_B=2\pi M.
  \label{eq:ABJM-mass-map}
\end{equation}
The exact result \eqref{eq:ABJM-rank-one} has a pole precisely at
$\mu_A=-i\pi$, since $\cosh(-i\pi/2)=0$.  This is the divergence anticipated
above from the flat direction of the matrix integral at the masses
\eqref{eq:four-chiral-masses}.  Setting
$\mu_A=-i\pi+\epsilon$ and $\mu_B=2\pi M$, the generic rank-one result has the
Laurent expansion
\begin{equation}
  Z_k(-i\pi+\epsilon,2\pi M)
  =\frac{i}{2k\epsilon\cosh(\pi M)}+\cO(\epsilon),
  \qquad
  \lim_{\epsilon\to0}\frac{\epsilon}{i}Z_k
  =\frac{1}{2k\cosh(\pi M)}.
  \label{eq:S1-ABJM-residue-general}
\end{equation}

For $k=3$, this divergence is identical to that of the Schur index in the small-circle limit, once the two deformation parameters are identified as
\begin{equation}
    2\pi \epsilon = i \beta .
\end{equation}
Concretely, we have
\begin{equation}
  \lim_{\beta\to0^{+}}\frac{\beta}{2\pi}\,
  \mathcal{I}\bigl(e^{-\beta},e^{-i\beta M}\bigr)
  =\lim_{\epsilon\to0}\frac{\epsilon}{i}\,
  Z_{3}(-i\pi+\epsilon,2\pi M)
  =\frac{1}{6\cosh(\pi M)}.
  \label{eq:S1-residue-match}
\end{equation}
We emphasize that the equality only holds in the correlated limit $\beta = \frac{2\pi \epsilon}{i} \to 0$. Away from this limit, the Schur index and the $S^3$ partition function are not equal.

\subsection{The discretely gauged \texorpdfstring{$SU(2)/\mathbb{Z}_n$}{SU(2)/Zn} theories}

The same analysis can be repeated for the theories of
section~\ref{sec:discrete-gauging}, starting from the closed-form index
\eqref{eq:SU2-refined-index-closed}.  In this case, the residue can be traced to the individual holonomy sectors of
the projection. Discrete gauging projects onto invariant operators by averaging
over the $n$ holonomy sectors \eqref{eq:S1-SU2-projector}
and, in the fixed-mass limit $q=e^{-\beta}$, $a=e^{-i\beta M}$, only some of
these sectors are singular.  Indeed, \eqref{eq:SU2-refined-index-closed} becomes singular exactly when 
\begin{equation}
  2j\equiv0\pmod n
  \label{eq:S1-singular-sectors}
\end{equation}
in \eqref{eq:S1-SU2-projector}, while every other
sector stays $\cO(1)$. Each singular sector then contributes the same
pole as the parent $\mathcal{N}=4$ $SU(2)$ index, whose asymptotics read
\begin{equation}
  \mathcal{I}_{\rm par}(e^{-\beta},e^{-i\beta M})
  =\frac{\vartheta'_4(\mathfrak a)}
  {2\pi\vartheta_1(2\mathfrak a)}
  =\frac{\pi}{2\beta\cosh(\pi M)}+\cO(1).
  \label{eq:S1-SU2-parent-asymptotics}
\end{equation}

The counting of singular sectors is naturally expressed in terms of
\begin{equation}
  k_n=\operatorname{lcm}(2,n) \ ,
  \label{eq:S1-kn-definition}
\end{equation}
which is also the order of the rank-one orbifold obtained by combining the
Weyl $\mathbb{Z}_2$ quotient of the parent theory with the additional
$\mathbb{Z}_n$ action.  Condition~\eqref{eq:S1-singular-sectors} has
\begin{equation}
  N_{\rm pole}=\gcd(2,n)=\frac{2n}{k_n}
\end{equation}
solutions, and each sector enters \eqref{eq:S1-SU2-projector} with weight
$1/n$, so the singular sectors carry total weight $2/k_n$.  The pole of the
parent index is thus rescaled by this factor,
\begin{equation}
  \mathcal{I}_n(e^{-\beta},e^{-i\beta M})
  =\frac{\pi}{k_n\beta\cosh(\pi M)}+\cO(1) \ ,
  \label{eq:S1-SU2-Zn-asymptotics}
\end{equation}
and hence
\begin{equation}
  \lim_{\beta\to0^+}\frac{\beta}{2\pi}\,
  \mathcal{I}_n(e^{-\beta},e^{-i\beta M})
  =\frac{1}{2k_n\cosh(\pi M)}
  =\lim_{\epsilon\to0}\frac{\epsilon}{i}
  Z_{k_n}(-i\pi+\epsilon,2\pi M).
  \label{eq:S1-SU2-Zn-residue-match}
\end{equation}
The factor $q^{3/8}$ in the vacuum-character convention tends to unity and does
not affect this leading residue.  The $\mathbb{Z}_n$-gauged $SU(2)$ theory thus has
the small-circle residue of the rank-one $\mathrm{ABJM}_{k_n}$ theory.  Together with the $\mathbb{Z}_3$ S-fold case above, this
gives the correspondences collected below, where $\mathcal{M}_{\mathrm{4d}}$ denotes
the moduli space of the 4d theory and $\mathcal{M}_{\rm 3d}$ that of the
candidate 3d theory:
\begin{center}
\begin{tabular}{c c c c}
  \toprule
  4d theory & $\mathcal{M}_{\mathrm{4d}}$ & $\mathcal{M}_{\rm 3d}$ & rank-one 3d candidate \\
  \midrule
    $\mathbb{Z}_3$ S-fold & $\mathbb{C}^3/\mathbb{Z}_3$ & $\mathbb{C}^4/\mathbb{Z}_3$ & $\mathrm{ABJM}_3$ \\
  $SU(2)\ \mathcal{N}=4$ & $\mathbb{C}^3/\mathbb{Z}_2$ & $\mathbb{C}^4/\mathbb{Z}_2$ & $\mathrm{ABJM}_2$ \\
  $SU(2)/\mathbb{Z}_3$ & $\mathbb{C}^3/\mathbb{Z}_6$ & $\mathbb{C}^4/\mathbb{Z}_6$ & $\mathrm{ABJM}_6$ \\
  $SU(2)/\mathbb{Z}_4$ & $\mathbb{C}^3/\mathbb{Z}_4$ & $\mathbb{C}^4/\mathbb{Z}_4$ & $\mathrm{ABJM}_4$ \\
  $SU(2)/\mathbb{Z}_6$ & $\mathbb{C}^3/\mathbb{Z}_6$ & $\mathbb{C}^4/\mathbb{Z}_6$ & $\mathrm{ABJM}_6$ \\
  \bottomrule
\end{tabular}
\end{center}
In particular, the $n=3$ and $n=6$ quotients share the same Schur index, as
already observed in
\eqref{eq:schur-index-Z6-quotient-of-small-superconformal}, and accordingly
both reduce to the $k_n=6$ residue.  

\section{Discussion}

We have determined protected character data for several rank-one
$\mathcal{N}=3$ SCFTs.  For the interacting $\mathbb{Z}_3$ S-fold, the
$\mathcal{N}=1$ UV description gives the vacuum character in closed form, while
formal Wilson-line insertions give the unflavored non-vacuum characters.
Combining the unflavored and flavored MLDEs with Zhu's algebra and the
free-field realization, we study four highest-weight modules and obtain their
flavored characters in closed form.  A logarithmic solution completes the
modular orbit but is not an additional ordinary module.  The flavored
null-state equations also form a quasi-modular system under $\Gamma^0(2)$.
We further conjecture that these modules are counted by a reduced PBW basis
built from only five of the eight strong generators, whose state count
reproduces the flavored characters.  The corresponding monomials are not
linearly independent beyond the lowest levels, however, and the systematic
construction of the corrected normal forms remains open.

For theories obtained by discrete gauging, we derived closed-form indices from
free fields in the $U(1)$ case and from the branching rule of
\cite{Creutzig:2018ltv} in the $SU(2)$ case.  The $\mathbb{Z}_3$ quotient of the
$U(1)$ theory shares the unflavored MLDE solution space of the interacting
$\mathbb{Z}_3$ S-fold.  We also found a free-field realization of the $n=4$
quotient VOA and identified several of its characters.  For $n=6$, the
flavored equations and the classification of ordinary modules remain open.
Their small-circle residues nevertheless follow directly from the holonomy
projector: they are controlled by $k_n=\operatorname{lcm}(2,n)$ and agree with
the mass-deformed rank-one $\mathrm{ABJM}_{k_n}$ residues, as summarized in
\eqref{eq:S1-SU2-Zn-residue-match}.

The closed form also permits a direct small-circle test.  In the correlated
limit $q,a\to1$, the vacuum character has a $\beta^{-1}$ divergence whose
residue matches that of the corresponding ABJM
$S^3$ partition function at its pole, as shown in
\eqref{eq:S1-residue-match} and \eqref{eq:S1-SU2-Zn-residue-match}.  This equality of residues supports the proposed
circle reduction.

It would be important to extend the study of the representation theory of VOAs associated with the other rank-one
S-folds and to higher rank.  One accessible case may be the theory with moduli
space $(\mathbb{C}^3)^3/\mathsf{G}(3,3,3)$ discussed in \cite{Zafrir:2020epd}, whose
construction involves the $E_6$ Minahan--Nemeschansky (MN) theory
\cite{Minahan:1996cj}.  The closed-form $E_6$ Schur index of
\cite{Pan:2021mrw} could serve as input for the orbifold index.  It would then
be possible to test whether contour integrals with formal Wilson-line
insertions again recover non-vacuum characters and whether the modular
solution space is sufficient to determine the flavored characters.

There is also a geometric question suggested by the modular representation.
The 4d mirror symmetry program \cite{Fredrickson:2017yka,Fredrickson:2017jcf,Kozcaz:2018usv,Dedushenko:2018bpp,Gukov:2022gei,Shan:2023xtw,Shan:2024yas,Xie:2026xxg} has studied the Argyres--Douglas theories \cite{Argyres:1995jj} extensively. In particular, it has been shown in detail that the representation theory of the boundary admissible Kac--Moody algebras and the $\mathcal{W}$-algebras from Drinfeld--Sokolov reduction is closely related to the $\mathbb{C}^\times$ fixed points of the corresponding Hitchin moduli space.
It has recently been extended to examples from class $\mathcal{S}$ and
$\mathcal{N}=4$ theories \cite{Pan:2024epf,Pan:2024hcz,Li:2025nhc}.  The
circle-compactified Coulomb branch of the $\mathbb{Z}_3$ S-fold is a
Liouville integrable system with a singular fiber of Kodaira type $IV^*$
\cite{Argyres:2016xua,Nishinaka:2016hbw}.  The same fiber occurs for the
 $E_6$ MN theory, but the available resolution differs
because the S-fold has only $U(1)$ flavor symmetry.  Under the conjectural
dictionary of \cite{Pan:2024epf,Pan:2024hcz}, the Jordan form
\eqref{eq:full-Jordan-form} predicts three isolated fixed points and one
one-dimensional connected fixed component under the $\mathbb{C}^\times$-action.  Studying $\mathbb{C}^\times$-fixed components of the Coulomb branch, including its mixed-Hodge refinement, directly would test this prediction and could
extend the 4d mirror symmetry to
$\mathcal{N}=3$ theories.

Finally, the high-temperature limit can be applied to observables beyond the
vacuum character.  Taking correlated limits of flavored Schur operator
correlation functions should give protected correlators in the compactified
theory.  These data may be compared with deformation quantizations of the
common Higgs branch coordinate ring, whose classical limit is a Poisson
algebra
\cite{Dedushenko:2016jxl,Beem:2016cbd,Dedushenko:2019mzv,Dedushenko:2019mnd,Pan:2020cgc}.

\acknowledgments

We would like to express our gratitude to Jirui Guo and Hao Zou for inviting us to the workshop \href{https://sites.google.com/view/math-physics-2025/}{Algebra and Geometry in Quantum Field Theory and String Theory} at Tongji University, where this project was initiated. S.N. thanks G. Zafrir for explanations of his work \cite{Zafrir:2020epd}. We are grateful to T. Nishinaka and T. Nosaka for intriguing discussions.
This work is supported by the Shanghai Municipal Science and Technology Major Project (No. 24ZR1403900) and the National Natural Science Foundation of China (NSFC) under Grant No. 11905301.
\paragraph{Statement on use of AI:}  We acknowledge the use of OpenAI Codex (GPT 5.5/5.6) and Anthropic Claude Code (Opus 4.8/5) for assistance with
drafting, editing, consistency checks, and manuscript organization. Nevertheless, the ideas came from the authors, and all results were author-verified.

\appendix
\section{Special functions}\label{app:special-functions}

\subsection*{Nome variables and modular parameter}

Throughout the paper, we use the standard nome variables associated with the
modular parameter $\tau$ and the elliptic variables $\mathfrak{z}$, $\mathfrak{y}$:
\begin{equation}
  q := e^{2\pi i\tau}, \qquad
  z := e^{2\pi i\mathfrak{z}}, \qquad
  y := e^{2\pi i\mathfrak{y}}.
\end{equation}
We pass freely between the multiplicative variables $(z,y,q)$ and their
additive counterparts $(\mathfrak{z},\mathfrak{y},\tau)$; for instance, we write
$\vartheta_{i}(z,q)$ and $\vartheta_{i}(\mathfrak{z}|\tau)$ interchangeably.
When the dependence on $q$ or $\tau$ is clear from context, it is omitted and
we write simply $\vartheta_{i}(z)$ or $\vartheta_{i}(\mathfrak{z})$.

\subsection*{Jacobi theta functions and Dedekind eta function}

The Jacobi theta functions $\vartheta_{i}(\mathfrak{z}|\tau)$ and the Dedekind
eta function $\eta(\tau)$ (sometimes denoted as $\eta(q)$) are defined by the series
\begin{align}
	\vartheta_1(\mathfrak{z}|\tau) \coloneqq & \ -i \sum_{r \in \mathbb{Z} + \frac{1}{2}} (-1)^{r-\frac{1}{2}} e^{2\pi i r \mathfrak{z}} q^{\frac{r^2}{2}} ,
	& \vartheta_2(\mathfrak{z}|\tau) \coloneqq & \sum_{r \in \mathbb{Z} + \frac{1}{2}} e^{2\pi i r \mathfrak{z}} q^{\frac{r^2}{2}} \ ,\cr
	\vartheta_3(\mathfrak{z}|\tau) \coloneqq & \ \sum_{n \in \mathbb{Z}} e^{2\pi i n \mathfrak{z}} q^{\frac{n^2}{2}},
	& \vartheta_4(\mathfrak{z}|\tau) \coloneqq & \sum_{n \in \mathbb{Z}} (-1)^n e^{2\pi i n \mathfrak{z}} q^{\frac{n^2}{2}} \ .
\end{align}

Alternatively, they are given by the following infinite products:
\begin{equation}
\begin{aligned}
  \vartheta_{1}(\mathfrak{z}|\tau)
  &= -i\,z^{\frac{1}{2}}q^{\frac{1}{8}}
    \prod_{n=1}^{\infty}(1-q^{n})
    \prod_{n=0}^{\infty}(1-z\,q^{n+1})(1-z^{-1}q^{n}),\\[4pt]
  \vartheta_{2}(\mathfrak{z}|\tau)
  &= z^{\frac{1}{2}}q^{\frac{1}{8}}
    \prod_{n=1}^{\infty}(1-q^{n})
    \prod_{n=0}^{\infty}(1+z\,q^{n+1})(1+z^{-1}q^{n}),\\[4pt]
  \vartheta_{3}(\mathfrak{z}|\tau)
  &= \prod_{n=1}^{\infty}(1-q^{n})
    \prod_{r\in\mathbb{Z}_{\ge 0}+\frac{1}{2}}(1+z\,q^{r})(1+z^{-1}q^{r}),\\[4pt]
  \vartheta_{4}(\mathfrak{z}|\tau)
  &= \prod_{n=1}^{\infty}(1-q^{n})
    \prod_{r\in\mathbb{Z}_{\ge 0}+\frac{1}{2}}(1-z\,q^{r})(1-z^{-1}q^{r}),\\[4pt]
  \eta(\tau)
  &= q^{\frac{1}{24}}\prod_{n=1}^{\infty}(1-q^{n}).
\end{aligned}
\end{equation}
The zeros of $\vartheta_{i}(\mathfrak{z}|\tau)$ are manifest in these product formulas.

\subsection*{Elliptic Gamma function}

The elliptic Gamma function $\Gamma(z;p,q)$ is the meromorphic function of
$(z,p,q)$ defined by
\begin{equation}
  \Gamma(z;p,q)
  \coloneqq \prod_{m,n\geq 0}
  \frac{1-z^{-1}p^{m+1}q^{n+1}}{1-z\,p^{m}q^{n}}.
\end{equation}
In the Schur limit $p = \sqrt{q}$ used throughout the main text, we adopt the
shorthand $\Gamma(z) \coloneqq \Gamma(z;\sqrt{q},q)$, which specializes to
\begin{equation}
  \Gamma(z)
  = \Gamma\!\left(z;\sqrt{q},q\right)
  = \prod_{m,n\geq 0}
  \frac{1-z^{-1}\,q^{\frac{2m+n+3}{2}}}{1-z\,q^{\frac{2m+n}{2}}}.
\end{equation}

\noindent The shorthand $\Gamma(z)$ satisfies the following identities, which are
used repeatedly in the main text.
The shift relations
\begin{equation}
\label{eq:Gamma-shift-sqrtq}
  \Gamma\!\left(q^{\frac{1}{2}}z\right)
  = \frac{-i\,q^{-\frac{1}{12}}\sqrt{z}\;\vartheta_{1}(\mathfrak{z}|\tau)}{\eta(\tau)}\,\Gamma(z),
\end{equation}
\begin{equation}
\label{eq:Gamma-shift-q}
  \Gamma(qz)
  = \frac{-i\,q^{-\frac{1}{24}}\sqrt{z}\;\vartheta_{1}\!\left(\mathfrak{z}\big|\tfrac{\tau}{2}\right)}
         {\eta\!\left(\tfrac{\tau}{2}\right)}\,\Gamma(z),
\end{equation}
express the effect of shifting $z$ by a power of $q$, and together imply the
reflection formula
\begin{equation}
\label{eq:Gamma-reflection}
  \Gamma(z)\,\Gamma(z^{-1})
  = \frac{q^{\frac{1}{8}}\,\eta(\tau)\,\eta\!\left(\tfrac{\tau}{2}\right)}
         {\vartheta_{1}\!\left(\mathfrak{z}\big|\tfrac{\tau}{2}\right)\,\vartheta_{1}(\mathfrak{z}|\tau)}.
\end{equation}
Finally, the product formula makes the inversion relation
$\Gamma(z;p,q)\,\Gamma\!\left(pq/z;p,q\right)=1$ manifest, the two factors being
reciprocal term by term.  At $p=\sqrt{q}$ this reads
$\Gamma(z)\,\Gamma\!\left(q^{\frac{3}{2}}/z\right)=1$, and setting $z=\sqrt{q}$ yields
the normalization identity
\begin{equation}
\label{eq:Gamma-normalization}
  \Gamma(q)\,\Gamma\!\left(\sqrt{q}\right) = 1.
\end{equation}

\subsection*{Twisted Eisenstein series}

For a pair of characteristics $(\phi,\theta)$, the twisted Eisenstein series
of weight $k\geq 1$ is defined by the $q$-series
\begin{equation}
\label{eq:twisted-Eisenstein-def}
  E_{k}\!\begin{bmatrix}\phi\\\theta\end{bmatrix}(\tau)
  := -\frac{B_{k}(\lambda)}{k!}
    + \frac{1}{(k-1)!}\sum_{r\geq 0}^{\prime}
      \frac{(r+\lambda)^{k-1}\theta^{-1}q^{r+\lambda}}
           {1-\theta^{-1}q^{r+\lambda}}
    + \frac{(-1)^{k}}{(k-1)!}\sum_{r\geq 1}
      \frac{(r-\lambda)^{k-1}\theta\,q^{r-\lambda}}
           {1-\theta\,q^{r-\lambda}},
\end{equation}
where $\lambda\in[0,1)$ is determined by $\phi = e^{2\pi i\lambda}$, $B_{k}(x)$
is the $k$-th Bernoulli polynomial, and the prime on the first sum indicates
that the $r=0$ term is omitted when $\phi=\theta=1$.
Since every characteristic occurring in the main text has upper entry $\pm1$, we abbreviate
\begin{equation}\label{eq:Epm-shorthand}
  \Ep{k}{x} \coloneqq E_{k}\!\begin{bmatrix}+1\\ x\end{bmatrix} \ , \qquad
  \Em{k}{x} \coloneqq E_{k}\!\begin{bmatrix}-1\\ x\end{bmatrix} \ .
\end{equation}
The untwisted series are defined by specialization to $\lambda=0$ and
$\theta=1$, with the $r=0$ term omitted as prescribed above:
\begin{equation}\label{eq:untwisted-Eisenstein-specialization}
 E_{2n}(\tau):=E_{2n}\!\begin{bmatrix}1\\1\end{bmatrix}(\tau)
 =-\frac{B_{2n}}{(2n)!}
 +\frac{2}{(2n-1)!}\sum_{r\ge1}\frac{r^{2n-1}q^r}{1-q^r},
 \qquad n\ge1.
\end{equation}
Thus $E_2=-1/12+2q+\cdots$ in our normalization. This specialization is
not a simultaneous limit of the characteristics. For odd weights $k\ge3$,
$E_k\!\begin{bmatrix}\phi\\\theta\end{bmatrix}=0$ at $\phi,\theta\in\{1,-1\}$.
At weight one the same holds except at $(\phi,\theta)=(1,1)$, where the
omission prescription gives $E_1\!\begin{bmatrix}1\\1\end{bmatrix}=1/2$ (we take $0^0=1$).
The meromorphic function $\Ep{1}{z}$ instead has a pole at $z=1$:
\begin{equation}\label{eq:E1-theta-identities}
 \Ep{1}{z}=\frac{1}{2\pi i}
 \frac{\vartheta_1'(\mathfrak z)}{\vartheta_1(\mathfrak z)}.
\end{equation}
For use in generating series we also set $E_0\!\begin{bmatrix}\phi\\\theta\end{bmatrix}=-1$.

The relation to theta derivatives can be written without a formal differential
operator. With an auxiliary variable $\xi$, define
\begin{equation}
 P_2(\xi):=-\sum_{n\ge1}\frac{E_{2n}(\tau)}{2n}\xi^{2n}.
\end{equation}
Then, away from the zeros of $\vartheta_1(\mathfrak z)$,
\begin{equation}\label{EisensteinToTheta}
 \Ep{k}{z}
 =-[\xi^k]\left(
 e^{-P_2(\xi)}
 \frac{\vartheta_1(\mathfrak z-\xi/(2\pi i))}
      {\vartheta_1(\mathfrak z)}\right),\qquad k\ge1.
\end{equation}
Here $[\xi^k]$ denotes the coefficient of $\xi^k$ in the Taylor expansion
at zero, and primes on theta functions are derivatives with respect to the
additive argument. In particular, $\vartheta_i^{(n)}/\vartheta_i$ means a
normalized $n$th derivative, not an iterated derivative of a ratio.
More explicitly, for any weight $k\geq1$, we have $\Ep{k}{z}$ in terms of $\vartheta_1^{(n)}(\mathfrak{z})$ and $E_{2n}(\tau)$:
\begin{equation}\label{EisensteinToTheta-2}
	\Ep{k}{z} = \sum_{\ell = 0}^{\lfloor k/2 \rfloor}  \frac{(-1)^{k + 1}}{(k - 2\ell)!}\left(\frac{1}{2\pi i}\right)^{k - 2\ell} \mathbb{E}_{2\ell} \frac{\vartheta_1^{(k - 2\ell)}(\mathfrak{z})}{\vartheta_1(\mathfrak{z})} \ ,
\end{equation}
where
\begin{align}
  \mathbb{E}_{2\ell} \coloneqq \sum_{\substack{\{n_p\} \cr \sum_{p \ge 1} (2p)n_p = 2\ell}} \prod_{p\ge 1} \frac{1}{n_p !} \left(\frac{1}{2p}E_{2p}\right)^{n_p} \ .
\end{align}
For every $k\geq1$, the remaining twisted Eisenstein series of weight $k$ are
given by the same formula, with $\vartheta_1$ replaced by $\vartheta_{2,3,4}$
according to
\begin{equation}\label{eq:theta-Eisenstein-dictionary}
  \vartheta_1 \leftrightarrow \Ep{k}{z} \ , \quad
  \vartheta_4 \leftrightarrow \Em{k}{z} \ , \quad
  \vartheta_2 \leftrightarrow \Ep{k}{-z} \ , \quad
  \vartheta_3 \leftrightarrow \Em{k}{-z} \ .
\end{equation}
At $k=1$, for instance, this reads $\Em{1}{z} = \frac{1}{2\pi i}\vartheta_4'(\mathfrak{z})/\vartheta_4(\mathfrak{z})$,
and similarly for $\vartheta_{2,3}$.

Using the representation in terms of Jacobi theta functions, it is straightforward to obtain the modular transformations of the twisted Eisenstein series under the generators
\begin{align}
  S: \tau \to - \frac{1}{\tau}, \ \mathfrak{z} \to \frac{\mathfrak{z}}{\tau}, \qquad
  \qquad
  T: \tau \to \tau + 1 , \ \mathfrak{z} \to \mathfrak{z} \ .
\end{align}
Concretely, under $S$, the series $E^{\pm}_{k}\!\left(\pm z\right)$ transform as
\begin{align}
  \Ep{n}{z} \xrightarrow{S} &
  \left(\frac{1}{2\pi i}\right)^n\left[\bigg(\sum_{k \ge 0}\frac{1}{k!}(- \log z)^k y^k\bigg)
  \bigg(\sum_{\ell \ge 0}(\log q)^\ell y^\ell \Ep{\ell}{z}\bigg)\right]_n\ ,\cr
  \Em{n}{z} \xrightarrow{S} &
  \left(\frac{1}{2\pi i}\right)^n\left[\bigg(\sum_{k \ge 0}\frac{1}{k!}(- \log z)^k y^k\bigg)
  \bigg(\sum_{\ell \ge 0}(\log q)^\ell y^\ell \Ep{\ell}{-z}\bigg)\right]_n\ ,\cr
  \Ep{n}{-z} \xrightarrow{S} &
  \left(\frac{1}{2\pi i}\right)^n\left[\bigg(\sum_{k \ge 0}\frac{1}{k!}(- \log z)^k y^k\bigg)
  \bigg(\sum_{\ell \ge 0}(\log q)^\ell y^\ell \Em{\ell}{z}\bigg)\right]_n\ ,\cr
  \Em{n}{-z} \xrightarrow{S} &
  \left(\frac{1}{2\pi i}\right)^n\left[\bigg(\sum_{k \ge 0}\frac{1}{k!}(- \log z)^k y^k\bigg)
  \bigg(\sum_{\ell \ge 0}(\log q)^\ell y^\ell \Em{\ell}{-z}\bigg)\right]_n\ .
\end{align}
Here, $[f(y)]_n$ extracts the coefficient of $y^n$. 
Under the $T$-action,
\begin{align}
  \Ep{n}{z} \xrightarrow{T}& \ \Ep{n}{z}, & 
  \Em{n}{z} \xrightarrow{T}& \
  \Em{n}{-z}\ , \cr
  \Ep{n}{-z} \xrightarrow{T}& \ \Ep{n}{-z}, & 
  \Em{n}{-z} \xrightarrow{T}& \ 
  \Em{n}{z} \ .
\end{align}
We may also combine the two and obtain that under $STS$,
\begin{align}
  \Em{n}{z} \xrightarrow{STS}
  \left(\frac{1}{2\pi i}\right)^n\left[\bigg(\sum_{k \ge 0}\frac{1}{k!}(- \log z)^k y^k\bigg)
  \bigg(\sum_{\ell \ge 0}(\log q - 2\pi i)^\ell y^\ell \Em{\ell}{z}\bigg)\right]_n\ . 
\end{align}

\subsection*{Serre derivative and modular differential operators}

Given a (quasi-)modular form $f(\tau)$ of weight $k$, the
\emph{Serre derivative} is defined by
\begin{equation}
\label{eq:serre-derivative}
  \partial_{(k)} f
  \;:=\;
  \Bigl(q\,\partial_{q} + k\,E_{2}(\tau)\Bigr) f,
\end{equation}
where $\partial_{q} := d/dq$ and $E_{2}(\tau)$ is the weight-two Eisenstein
series (which is quasi-modular).
The operator $\partial_{(k)}$ raises the modular weight by two: it maps
weight-$k$ forms to weight-$(k+2)$ forms.
As representative examples, the Eisenstein series satisfy
\begin{equation}
\label{eq:serre-examples}
  \partial_{(2)}E_{2} = E_{2}^{2} + 5E_{4}, \qquad
  \partial_{(4)}E_{4} = 14\,E_{6}, \qquad
  \partial_{(6)}E_{6} = 20\,E_{8},
\end{equation}
where, in the normalization \eqref{eq:twisted-Eisenstein-def},
$E_{8} = \tfrac{3}{7}E_{4}^{2}$.

Iterating the Serre derivative with successive weight shifts yields the
modular differential operators
\begin{equation}
\label{eq:Dqk-def}
  D_{q}^{(k)}
  \;:=\;
  \partial_{(2k-2)} \circ \cdots \circ \partial_{(2)} \circ \partial_{(0)},
\end{equation}
so that $D_{q}^{(k)}$ maps a weight-zero (quasi-)modular form to one of
weight $2k$.
These are precisely the operators $D_{q}^{(n)}$ appearing in the MLDE
\eqref{eq:MLDE-order4}.

Under the standard $SL(2,\mathbb{Z})$ action, these operators transform
covariantly.
For $\gamma = \begin{pmatrix}a & b \\ c & d\end{pmatrix}\in SL(2,\mathbb{Z})$
with
\begin{equation}
\label{eq:tau-transform}
  \tau' = \gamma\cdot\tau := \frac{a\tau+b}{c\tau+d}, \qquad
  q' := e^{2\pi i\tau'},
\end{equation}
the corresponding operators satisfy
\begin{equation}
\label{eq:Dqk-covariance}
  D_{q'}^{(k)} \;=\; (c\tau+d)^{2k}\,D_{q}^{(k)}.
\end{equation}
This covariance property ensures that the MLDE \eqref{eq:MLDE-order4},
whose coefficients are modular forms for $\Gamma^0(2)$, transforms homogeneously under
$\Gamma^0(2)$, and hence that its solution space carries a well-defined
modular representation.

\section{Useful identities}\label{app:identities}

In this appendix, we collect the identities used to bring the contour
integrals of the main text into closed form.  They fall into three groups:
theta--eta relations at shifted arguments, used to reduce the $\mathbb{Z}_3$ S-fold Schur
integral \eqref{eq:Schur-integral} to \eqref{eq:SchurClosedForm}; ratio
identities at third-integral arguments, used for the $\mathbb{Z}_3$-quotient
index \eqref{eq:I_Z3_flavored}; and duplication and shift properties of the
twisted Eisenstein series, used throughout the flavored analysis.
Throughout, $\eta$ and $\vartheta_i$ take the modulus as argument, as in
appendix~\ref{app:special-functions}, and a suppressed modulus means $\tau$.

\paragraph{Theta--eta relations at shifted arguments.}
\begin{align}
    \vartheta_1(\mathfrak{z}|2\tau)\vartheta_4(0|2\tau) = & \ - \vartheta_1(\frac{\mathfrak{z}}{2}|\tau)
    \vartheta_1(\frac{\mathfrak{z} }{2} - \frac{1}{2}|\tau) \ ,\\
    \frac{\vartheta_1(\mathfrak{z}|\tau)}{
    \vartheta_1(\mathfrak{z}|2\tau)
    \vartheta_1(\mathfrak{z} + \tau|2\tau)
    } = & \  - i q^{\frac{1}{4}} z^{\frac{1}{2}} \frac{\eta(\tau)}{\eta(2\tau)^2} \ ,\\
    \vartheta_1(\frac{1}{6} + \frac{\tau}{2}|\tau) 
    = & \ \frac{i \eta(2\tau)^2\eta(3\tau) \vartheta_1(\frac{1}{6}|\tau) \vartheta_1(\frac{1}{3}|\tau)}{
    \sqrt{3} q^{\frac{1}{4}}
    \eta(\tau)
    \eta(6\tau)^2
    \vartheta_1(\frac{1}{3} + \frac{\tau}{2}|\tau)
    } \ , \\
    \vartheta_3\left(\frac{1}{6}\right)
     = & \ 
     \frac{\eta(\tau)^{2}\,\eta\left(\frac{3\tau}{2}\right)}
          {\eta\left(\frac{\tau}{2}\right)\eta(3\tau)}\ , \\
    \vartheta_1(\frac{1}{6}|\tau)^4
    = & \ \frac{3 \eta(\tau)^2 \eta(6\tau)^4 \vartheta_4(0|2\tau)^3}{
    \eta(2\tau) \eta(3\tau)^2
    \vartheta_1(\frac{1}{3}|\tau)^2
    } \ ,\\
    \vartheta_1(\frac{1}{2} + \frac{\tau}{2}|\tau)
    \vartheta_1(\frac{\tau}{2}|\tau) = & \ - i q^{\frac{1}{4}}\vartheta_1(\tau|2\tau)^2\ , \\
    \vartheta_1\left(\frac{2}{3}+\frac{\tau}{3}\right)
   \vartheta_1\left(\frac{1}{6}+\frac{\tau}{3}\right)^{2}
   = & \ e^{-\frac{\pi i}{3}}\,
   \vartheta_1 \left(\frac{\tau}{3}\right)
   \vartheta_1 \left(\frac{\tau}{3}+\frac{1}{2}\right)^{2} \nonumber \\
   & \ +e^{-\frac{2\pi i}{3}}\,
   \vartheta_1\left(\frac{\tau}{3}+\frac{1}{3}\right)
   \vartheta_1\left(-\frac{1}{6}+\frac{\tau}{3}\right)^{2}\ , \\
   \vartheta_4(0) = & \ \frac{\eta(\frac{1}{2}\tau)^2}{\eta(\tau)} \ .
\end{align}

\paragraph{Ratio identities at third-integral arguments.}
\begin{align}
    \frac{\vartheta_1\!\left(\tfrac{1}{6}+\tfrac{\tau}{3}\right)^{2}\vartheta_1\!\left(\tfrac{2}{3}+\tfrac{\tau}{3}\right)}{\eta(\tau)\,\eta(2\tau)\,\vartheta_1(\tau|2\tau)}
    = & \ \frac{q^{1/12}}{\eta(2\tau)}\left[\frac{\eta(3\tau)^{2}}{\eta(6\tau)}-e^{\frac{i\pi}{3}}\frac{\eta(\tau)\,\eta(6\tau)^{2}}{\eta(2\tau)\,\eta(3\tau)}\right]\ ,\\
    \frac{\vartheta_1 \left(-\tfrac{1}{6}+\tfrac{\tau}{3}\right)^{2}\vartheta_1 \left(\tfrac{1}{3}+\tfrac{\tau}{3}\right)}{\eta(\tau)\,\eta(2\tau)\,\vartheta_1(\tau|2\tau)}
    = & \ \frac{ - q^{1/12}}{\eta(2\tau)}\left[\frac{\eta(3\tau)^{2}}{\eta(6\tau)}-e^{\frac{2\pi i}{3}}\frac{\eta(\tau)\,\eta(6\tau)^{2}}{\eta(2\tau)\,\eta(3\tau)}\right]\ ,\\
    \frac{\vartheta_1 \left(\tfrac{1}{6}\right)^{3}\vartheta_1 \left(\tfrac{1}{3}\right)\vartheta_1 \left(-\tfrac{1}{6}+\tfrac{\tau}{2}\right)\vartheta_1 \left(\tfrac{1}{3}+\tfrac{\tau}{2}\right)}{\eta(2\tau)^{3} \vartheta_1(\tau|2\tau)^{3}} = & \ -\sqrt{3} e^{\frac{i\pi}{3}} q^{1/2} \frac{\eta(\tau) \eta(6\tau)^{2}}{\eta(2\tau)^{2} \eta(3\tau)}\ ,\\
    \frac{\eta(2\tau) \vartheta_1\left(\tfrac{1}{6}\right)\vartheta_1\left(\tfrac{1}{3}\right)\vartheta_1(\tau|2\tau)}{\eta(\tau)^{2} \vartheta_1\left(-\tfrac{1}{6}+\tfrac{\tau}{2}\right)\vartheta_1\left(\tfrac{1}{3}+\tfrac{\tau}{2}\right)}
    = & \ \sqrt{3} e^{-\frac{i\pi}{3}} \frac{\eta(\tau) \eta(6\tau)^{2}}{\eta(2\tau)^{2} \eta(3\tau)}\ , \\
    \frac{\vartheta_1 \left(\tfrac{1}{2}+\tfrac{\tau}{3}\right)^{2}\vartheta_1 \left(\tfrac{\tau}{3}\right)}{\eta(\tau)\,\eta(2\tau)\,\vartheta_1(\tau|2\tau)}
    = & \ \frac{q^{1/12}}{\eta(2\tau)} \left[\frac{\eta(3\tau)^{2}}{\eta(6\tau)}+\frac{\eta(\tau)\,\eta(6\tau)^{2}}{\eta(2\tau)\,\eta(3\tau)}\right]\ ,\\
    \frac{\vartheta_1 \left(-\frac{1}{6}+\frac{\tau}{3}\right)^{2}
      \vartheta_1 \left(\frac{1}{3}+\frac{\tau}{3}\right)}
     {\vartheta_1 \left(\frac{1}{2}+\frac{\tau}{3}\right)^{2}
      \vartheta_1 \left(\frac{\tau}{3}\right)}
      = & \ -
   \frac{\eta \left(\frac{\tau}{3}\right)\eta(2\tau)\eta(3\tau)^{2}}
        {\eta \left(\frac{2\tau}{3}\right)^{2}\eta(\tau)\eta(6\tau)}
        -
   e^{\frac{2\pi i}{3}}
   \frac{\eta \left(\frac{\tau}{3}\right)\eta(6\tau)^{2}}
        {\eta \left(\frac{2\tau}{3}\right)^{2}\eta(3\tau)}\ , \\
    \frac{\eta(\tau)^{4}}{3\eta\!\left(\frac{\tau}{2}\right)^{2}}
     +
     \frac{2\eta\!\left(\frac{\tau}{2}\right)\eta(3\tau)^{2}}
          {3\eta(\tau)^{2}\eta\!\left(\frac{3\tau}{2}\right)}
    = &\ +
     \frac{2\eta(3\tau)}{3\vartheta_3\!\left(\frac{1}{6}\right)}
     +
     \frac{\eta(\tau)^{3}}{3\vartheta_4(0)} \ .
\end{align}

\paragraph{Duplication and shift properties.}
The twisted Eisenstein series satisfy the duplication formulas
\begin{align}
    \sum_{\pm \pm} E^{\pm}_{k}\!\left(\pm z\right)
    = \frac{4}{2^k} \Ep{k}{z^2}, \qquad
    \sum_\pm \Ep{k}{\pm z}(\tau) = 2 \Ep{k}{z^2}(2\tau) \ ,
\end{align}
and some special shift properties
\begin{align}
    \Em{1}{e^{\frac{2\pi i}{3}}q^{\frac{2}{3}}}
    + \Em{1}{e^{- \frac{2\pi i}{3}}q^{\frac{1}{3}}} = - 1, \qquad
    \Em{1}{e^{\frac{\pi i}{3}}q^{\frac{2}{3}}}
    + \Em{1}{e^{- \frac{\pi i}{3}}q^{\frac{1}{3}}} = - 1 \ .
\end{align}

\section{Details of the residue computation}\label{app:residues}

In this appendix, we collect the explicit expressions for the quantities $S_{i}$
and $W_{i}$ entering the evaluation of $\mathcal{I}_{r,s}$ in \eqref{eq:Imn}.  Throughout,
$r,s\in\mathbb{Z}$ are the monomial exponents of \eqref{eq:Imn} and we abbreviate
\begin{equation}
  \omega \coloneqq e^{\frac{2\pi i}{3}} \ ,
\end{equation}
so that every root of unity below is written as $\pm\omega^{k}$ or
$\pm e^{\pm i\pi/3}$; all fractional powers of $q$ are taken on the principal
branch.  Each quantity is
expressed directly in terms of the two non-vacuum characters
$\operatorname{ch}_{-\frac{1}{24}}$ \eqref{eq:nonvac-characters-1} and $\operatorname{ch}_{\frac{1}{8}}$
 \eqref{eq:nonvac-characters-2}, which makes the decomposition
\eqref{eq:Wilson-jz} manifest.
We further abbreviate the recurring denominators as
\begin{equation}
\begin{aligned}
  \Delta_{1} &= \left(1+q^{\frac{r}{6}}\right)
               \left(1-q^{\frac{r}{6}}+q^{\frac{r}{3}}\right)
               \left(q^{\frac{r}{6}}+q^{\frac{s}{3}}\right)
               \left(q^{\frac{r}{3}}-q^{\frac{1}{6}(r+2s)}+q^{\frac{2s}{3}}\right) \ ,\\
  \Delta_{2} &= \left(1+q^{\frac{r}{2}}\right)\left(1+q^{\frac{1}{2}(r+2s)}\right) \ ,\qquad
  \Delta_{3}  = \left(1+q^{\frac{r}{2}}\right)\left(q^{r+s}-1\right) \ ,\\
  \Delta_{4} &= \left(1+q^{\frac{r}{2}}\right)\left(q^{s}-q^{r}\right) \ .
\end{aligned}
\end{equation}

\begin{equation}
\label{S1}
\begin{aligned}
    S_1=&\frac{q^{-\frac{5}{8}+\frac{r}{4}}}{3(1+q^\frac{r}{2})}\Bigg[
    \frac{\eta(\tau)^4}{\eta(\frac{\tau}{2})^2}
    \\
    &
    +\frac{1}{4}\operatorname{ch}_{-\frac{1}{24}}\Big\{(-1)^{r}\Big[
    -\Em{1}{-q^{\frac{1}{6}}}
    +e^{\frac{i\pi}{3}}\Em{1}{e^{-\frac{i\pi}{3}}q^{\frac{1}{6}}}
    +\omega \Em{1}{e^{\frac{i\pi}{3}}q^{\frac{1}{6}}}\\ 
    &\qquad \qquad \qquad
    -\Em{1}{-q^{\frac{1}{3}}}
    -\omega \Em{1}{e^{-\frac{i\pi}{3}}q^{\frac{1}{3}}}
    +e^{\frac{i\pi}{3}}\Em{1}{e^{\frac{i\pi}{3}}q^{\frac{1}{3}}}
    \Big] \\
    &\qquad \qquad \qquad
     -\Em{1}{-q^{\frac{1}{6}}}
     -\omega \Em{1}{e^{-\frac{2i\pi}{3}}q^{\frac{1}{6}}}
     +e^{\frac{i\pi}{3}}\Em{1}{e^{\frac{2i\pi}{3}}q^{\frac{1}{6}}}
     \\
     &\qquad \qquad \qquad
     -\Em{1}{q^{\frac{1}{3}}}
     +e^{\frac{i\pi}{3}}\Em{1}{e^{-\frac{2i\pi}{3}}q^{\frac{1}{3}}}
     -\omega \Em{1}{e^{\frac{2i\pi}{3}}q^{\frac{1}{3}}}
    \Big\}\\
    &
    +\frac{1}{4}\operatorname{ch}_{\frac{1}{8}}
    \Big\{
    (-1)^{r}
    \Big[ 
    i\sqrt{3}\Em{1}{e^{\frac{i\pi}{3}}}
    -\Em{1}{-q^{\frac{1}{6}}}
    -\Em{1}{e^{-\frac{i\pi}{3}}q^{\frac{1}{6}}}
    -\Em{1}{e^{\frac{i\pi}{3}}q^{\frac{1}{6}}}
    \\
    &\qquad \qquad \quad
    -\Em{1}{-q^{\frac{1}{3}}}
    -\Em{1}{e^{-\frac{i\pi}{3}}q^{\frac{1}{3}}}
    -\Em{1}{e^{\frac{i\pi}{3}}q^{\frac{1}{3}}}
    -i\sqrt{3}\Ep{1}{e^{\frac{i\pi}{3}}}
    \Big]\\
    &\qquad \qquad  \quad
    -i\sqrt{3}\Em{1}{e^{\frac{2i\pi}{3}}}
    -\Em{1}{q^{\frac{1}{6}}}
    -\Em{1}{e^{-\frac{2i\pi}{3}}q^{\frac{1}{6}}}
    -\Em{1}{e^{\frac{2i\pi}{3}}q^{\frac{1}{6}}}
    \\
    &\qquad \qquad \quad
    -\Em{1}{q^{\frac{1}{3}}}
    -\Em{1}{e^{-\frac{2i\pi}{3}}q^{\frac{1}{3}}}
    -\Em{1}{e^{\frac{2i\pi}{3}}q^{\frac{1}{3}}}
    +i\sqrt{3}\Ep{1}{e^{\frac{2i\pi}{3}}}
    \Big\}
    \Bigg] \ ,
\end{aligned}
\end{equation}

\begin{equation}
\begin{aligned}
    S_2&=\frac{q^{-\frac{5}{8}+\frac{r}{4}}}{3(1+q^\frac{r}{2})}\left[\frac{\eta(\tau)^4}{\eta(\frac{\tau}{2})^2}-\frac{1}{4}\operatorname{ch}_{\frac{1}{8}}\right] \ ,
\end{aligned}
\end{equation}

\begin{equation}
\begin{aligned}
W_{1} = & \ \frac{\omega^{s-r}\, q^{\frac{1}{24}(10 r+8 s-15)}}{12\,\Delta_{1}} \cr
& \qquad \times \Big[\,q^{\frac{r}{6}}\big(\omega^{1-r}-e^{\frac{i\pi}{3}}\omega^{s}+\omega^{r-s}\big)
     +q^{\frac{s}{3}}\big(\omega^{1+s}-e^{\frac{i\pi}{3}}\omega^{-r}+\omega^{r-s}\big)\Big]\,\operatorname{ch}_{-\frac{1}{24}} \cr
& -\frac{i\,\omega^{r}\, q^{\frac{1}{24}(6 r+8 s-15)}}{24\,\Delta_{1}} \cr
& \qquad \times \Big[\,2i\big(\omega^{s}+\omega^{2r-s}+\omega^{s-r}\big)\big(q^{\frac{r}{3}}+q^{\frac{1}{6}(r+2s)}\big)
     +\sqrt{3}\,\big(\omega^{r-s}-2\omega^{s}+\omega^{2r-s}\big)q^{\frac{2s}{3}}\Big]\,\operatorname{ch}_{\frac{1}{8}} \ ,
\end{aligned}
\end{equation}

\begin{equation}
\begin{aligned}
W_{2} = & \ \frac{\omega^{s-r}\, q^{\frac{1}{24}(10 r+8 s-15)}}{12\,\Delta_{2}} \cr
& \qquad \times \Big[\,\omega-e^{\frac{i\pi}{3}}q^{\frac{1}{6}(r+2s)}
     +\omega^{r-s}\big(1+q^{\frac{1}{6}(r+2s)}\big)
     +e^{\frac{i\pi}{3}}\omega^{s-r}\big(e^{\frac{i\pi}{3}}q^{\frac{1}{6}(r+2s)}-1\big)\Big]\,\operatorname{ch}_{-\frac{1}{24}} \cr
& +\frac{\omega^{s-r}\, q^{\frac{1}{24}(10 r+8 s-15)}}{12\,\Delta_{2}} \cr
& \qquad \times \Big[\,\big(1+\omega^{r-s}\big)\big(1+q^{\frac{1}{6}(r+2s)}\big)+i\sqrt{3}\,q^{\frac{1}{3}(r+2s)} \\
& \qquad\qquad +\omega^{s-r}\big(1+q^{\frac{1}{6}(r+2s)}-i\sqrt{3}\,q^{\frac{1}{3}(r+2s)}\big)\Big]\,\operatorname{ch}_{\frac{1}{8}} \ ,
\end{aligned}
\end{equation}

\begin{equation}
\begin{aligned}
W_{3} = & -\frac{\omega^{r+s}\big(\omega^{r+s}-1\big)\, q^{\frac{1}{24}(14 r+8 s-15)}}{24\,\Delta_{3}} \cr
& \qquad \times \Big[\,i\sqrt{3}-1+\big(1+i\sqrt{3}\big)q^{\frac{1}{3}(r+s)}
     +2\omega^{r+s}\big(q^{\frac{1}{3}(r+s)}-1\big)\Big]\,\operatorname{ch}_{-\frac{1}{24}} \cr
& -\frac{\omega^{r+s}\, q^{\frac{1}{24}(14 r+8 s-15)}}{12\,\Delta_{3}} \cr
& \qquad \times \Big[\,q^{\frac{1}{3}(r+s)}-1-i\sqrt{3}\,q^{\frac{2}{3}(r+s)}
     +\omega^{-r-s}\big(q^{\frac{1}{3}(r+s)}-1\big) \cr
& \hspace{4.2cm}
     +\omega^{r+s}\big(q^{\frac{1}{3}(r+s)}-1+i\sqrt{3}\,q^{\frac{2}{3}(r+s)}\big)\Big]\,\operatorname{ch}_{\frac{1}{8}} \ ,
\end{aligned}
\end{equation}

\begin{equation}
\begin{aligned}
W_{4} = & \ \frac{\omega^{r+s}\, q^{\frac{1}{24}(14 r+8 s-15)}}{12\,\Delta_{4}} \cr
& \qquad \times \Big[\,\omega^{-r-s}\big(q^{\frac{r}{3}}-q^{\frac{s}{3}}\big)
     +e^{\frac{i\pi}{3}}\omega^{r}\big(e^{\frac{i\pi}{3}}q^{\frac{r}{3}}+q^{\frac{s}{3}}\big)
     -e^{\frac{i\pi}{3}}\omega^{s}\big(q^{\frac{r}{3}}+e^{\frac{i\pi}{3}}q^{\frac{s}{3}}\big)\Big]\,\operatorname{ch}_{-\frac{1}{24}} \cr
& +\frac{\omega^{r+s}\, q^{\frac{1}{24}(6 r+8 s-15)}}{12\,\Delta_{4}} \cr
& \qquad \times \Big[\,\omega^{-r-s}q^{\frac{r}{3}}\big(q^{\frac{r}{3}}-q^{\frac{s}{3}}\big)
     +\omega^{s}\big(q^{\frac{2r}{3}}-q^{\frac{1}{3}(r+s)}-i\sqrt{3}\,q^{\frac{2s}{3}}\big) \cr
& \hspace{3.4cm}
     +\omega^{r}\big(q^{\frac{2r}{3}}-q^{\frac{1}{3}(r+s)}+i\sqrt{3}\,q^{\frac{2s}{3}}\big)\Big]\,\operatorname{ch}_{\frac{1}{8}} \ .
\end{aligned}
\end{equation}

\bibliographystyle{JHEP}
\bibliography{references}

\end{document}